\documentclass{aa}  
\bibpunct{(}{)}{;}{a}{}{,} 
\usepackage{graphicx}
\usepackage{txfonts}
\usepackage{xcolor}
\usepackage{amsmath}
\usepackage{tablefootnote}
\usepackage{hyperref}
\usepackage{subfig}
\usepackage[normalem]{ulem} 
\hypersetup{
    colorlinks=true,
    linkcolor=blue,
    citecolor=blue
    }
\newcommand{\mdy}[1]{#1}
\newcommand{\kms}{km s$^{-1}$}                  
\defcitealias{de2020alma}{DV20}                 
\defcitealias{de2022modeling}{DV22}             

\begin{document} 

    \title{JWST study of the DG Tau B disk wind candidate: \\
    \textbf{I - Overview and Nested H$_2$/CO outflows}  
    } 

   \author{V.~Delabrosse\inst{\ref{inst1}} \and
          C.~Dougados \inst{\ref{inst1}} \and
          S.~Cabrit \inst{\ref{inst2},\ref{inst1}} \and
          B.~Tabone \inst{\ref{inst3}} \and
          L.~Tychoniec \inst{\ref{inst4}} \and 
          T.~Ray \inst{\ref{inst5}} \and
          L.~Podio \inst{\ref{inst6}} \and
          M.~McClure \inst{\ref{inst4}}
          }

   \institute{Univ. Grenoble-Alpes, CNRS, IPAG, 38000 Grenoble, France\label{inst1} \and
             Observatoire de Paris, PSL University, Sorbonne University, CNRS, LERMA, 75014 Paris, France\label{inst2} \and
             Univ. Paris-Saclay, CNRS, Institut d'Astrophysique Spatiale 91405 Orsay, France\label{inst3} \and
             \mdy{Leiden Observatory, Leiden University, PO Box 9513, NL–2300 RA Leiden, The Netherlands}\label{inst4} \and
             School of Cosmic Physics, Dublin Institute for Advanced Studies, Dublin, Ireland\label{inst5} \and
             INAF, Osservatorio Astrofisico di Arcetri, 50125 Firenze, Italy\label{inst6} 
             }

    \date{Received date / Accepted date}

 
  \abstract
   {The origin of outflows and their impact on proto-planetary disk evolution and planet formation processes are still crucial open questions. DG Tau B is a Class I protostar associated with a structured disk and a rotating conical CO outflow, recently identified by ALMA as one of the best CO disk wind candidate. It is therefore a perfect forming star to study these questions. 
   }
   {We aim to map and study any outflow component intermediate between the axial jet and the CO outflow, in order to constrain the origin \mdy{(irradiated/shocked} disk wind or swept-up material) of the red-shifted molecular outflows in DG Tau B.
   }
   {We analyze observations obtained with JWST NIRSpec-IFU and NIRCam, supplemented by IFU data from the SINFONI/VLT instrument. We investigate the morphology, kinematics and excitation conditions of the ro-vibrational H$_2$ emission lines and their relation with the atomic jet and CO outflow. We focus our analysis on the red-shifted outflow lobe.}
   {We observe a global layered structure of the red-shifted outflows in DG Tau B, with the atomic jet inside the H$_2$ cavity which in turn is nested inside the CO conical outflow and with temperature, velocity and collimation increasing towards the axis of the flow. The red-shifted H$_2$ emission traces a narrow conical cavity (semi-opening angle of $9.4^\circ$) wider than the axial jet but nested just inside the CO outflow. Both
   the jet and the H$_2$ cavity originate from the innermost regions of the disk ($r_0$ $<$ 6 au). 
   The red-shifted H$_2$ cavity  flows with a constant vertical velocity $V_z = 22.5 \pm 0.8$ \kms, twice faster than the conical CO flow. The excitation conditions imply a hot H$_2$ gas ($T_{ex} \simeq 2200$~K) with an average mass flux of $\dot{M}(\text{H}_2) = 3 \times 10^{-11} \text{M}_{\odot} \text{yr}^{-1}$ significantly lower than the jet and CO values. 
   }
   {The global layered H$_2$/CO structure in temperature, velocity and collimation in the DG Tau B red-shifted lobe is consistent with an MHD disk wind scenario. The hot H$_2$  could trace the inner dense photodissociation layer in the wind.
    An H$_2$ launching region at disk radii \mdy{0.2-0.4~au} combined with a large ejection efficiency ($\xi \simeq 1$) would account for the mass flux and kinematics. Alternatively, the near-IR ro-vibrational H$_2$ could be emitted in the interaction layer driven by successive jet bow-shocks into an outer disk wind or envelope. Further constraints on both scenarios will be obtained from the analysis of MIRI observations. 
    }


   \maketitle
%

\section{Introduction}

Star formation is accompanied by striking and powerful bipolar ejections. Two prominent components are observed: collimated and fast axial jets, and wider low-velocity outflows. These ejection signatures are detectable throughout the star formation process, from Class 0 onwards. They may have a major impact on the final mass of the star and disk accretion process, by potentially carrying a significant fraction of the system mass and angular momentum \citep[see for a recent review][]{pascucci2023}.

Slow molecular outflows ($< 10$~\kms) observed in CO lines are traditionally interpreted as swept-up material, tracing the interaction between the inner jet and/or wide-angle wind with the infalling envelope or parent core. However, recent interferometric observations probing CO outflows on smaller scales from the driving source (less than a few thousand au)  suggest an additional contribution. They have revealed outflowing conical 'cavities' anchored well inside the disk, also detected in more evolved Class II sources with no evidence for a residual envelope \citep[e.g. HH30:][]{louvet2018hh30}. 
Rotation signatures consistent with an origin in the inner regions of the disk (r$_{0} \simeq 0.1-40$ au) have been detected in most cases \citep{launhardt2009, zapata2015kinematics, bjerkeli2016resolved, tabone2017alma, louvet2018hh30, de2020alma, de2022modeling, lopez2023HH30}. The inferred mass fluxes are very significant, on the order of the mass accretion rate onto the star, and largely exceed the levels achievable by disk photoevaporation \citep{pascucci2023}. 

These observations suggest that part of the small-scale emission in these slow molecular outflows could trace matter directly ejected from the disk, 
by magnetic processes \citep[e.g.][]{pudritz2006disk,bai2016magneto}.
Such magnetic disk winds, if fully confirmed, would have a crucial impact on proto-planetary disks. They would solve the problem of angular momentum transport in the low-ionization dead zones of disks where MRI turbulence is quenched \citep[e.g.][]{bai2016magneto, riols2018} and they would significantly affect the disk properties, in particular the surface density profile, affecting  planet formation and migration \citep{kimmig2020effect}.
 
However, the swept-up interpretation cannot be fully excluded at this stage. To test this scenario, we are critically missing resolved spatial information on any gas component filling the gap between the fast axial jets and the slow and cold molecular outflows. In the disk wind paradigm, such a component would trace streamlines with intermediate velocities (a few 10 \kms) and temperatures (from a few 100 K to a few 1000 K), expected to emit preferentially in the near and mid-infrared domain. With ground-based IFUs, e.g. SINFONI/VLT \mdy{and NIFS/Gemini}, it is possible to probe the gas component at 2000~K with H$_2$ ro-vibrational emission lines \citep[e.g.][]{takami2007micro, Beck2008, agra2014origin}. 
However, these ground-based observations are limited in sensitivity and resolution, especially close to the central object or in embedded sources too faint for adaptive optics correction.
 
JWST offers a unique opportunity to detect and map this intermediate temperature wind component with increased angular resolution (comparable to ALMA) and unprecedented sensitivity. Furthermore, the numerous H$_2$ transitions probed with JWST allow to constrain with great accuracy the gas excitation conditions.

The present article is the first in a series describing and analysing JWST observations of the prototypical MHD disk wind candidate in DG Tau B. DG Tau B is a low-luminosity ($\simeq$ 1 L$_{\odot}$) source in the Taurus cloud ($d=140$~pc) embedded in a thick dark lane in optical and near-infrared images \citep{stapelfeldt1997,padgett1999}, suggesting an inclined disk orientation and classified as Class I  based on its spectral energy distribution \citep{furlan2009disk, luhman2009disk}. 
ALMA observations reveal a disk in the millimetric continuum viewed at an inclination of $\text{i}=63^{\circ} \pm 2^{\circ}$ \citep[herefater DV20]{de2020alma}. The DG Tau B disk is one of the few known young ($\le$ 0.5 Myrs) disks with bright rings identified in millimetric continuum emission, possibly indicative of early stages of planetesimal formation \citep[\citetalias{de2020alma},][]{garufi2020alma}. The disk is also detected in CO out to $R=700$~au with a Keplerian rotation curve constraining the central stellar mass at $1.1 \pm 0.2 \text{ M}_{\odot}$ \citepalias{de2020alma}.

In addition, DG Tau B is associated with a bipolar jet \citep{eisloffel1998imaging} and an asymetric molecular CO outflow \citep{mitchell1997dg}. The bright red-shifted CO outflow displays a striking narrow conical morphology at its base originating from well within the disk, and with rotation signatures in the same sense as the disk \citep[\citetalias{de2020alma}]{zapata2015kinematics}. The conical outflow also shows a radial stratification in velocities $V_p$ and specific angular momentum $J$ \citep[hereafter DV22]{de2022modeling} compatible with a magnetic disk wind of constant magnetic lever arm ($\lambda \simeq$ 1.6) and launched from $r_0 = 0.7 - 3.4$~au in the disk, i.e. corresponding to the expected dead zone location. The large mass flux of the CO conical flow ($2 \times 10^{-7} \text{M}_\odot \text{yr}^{-1}$), a factor 3 larger than the mass accretion rate onto the central protostar, is expected for MHD disk winds dominating extraction of angular momentum in the dead zone region \citep[][\citetalias{de2022modeling}]{lesur2021}. DG Tau B is therefore currently one of the most convincing candidates for  an MHD disk wind reported so far. 
 
We discuss here the first results from Cycle I JWST observations of the inner 5$^{\prime\prime}$ ($=700$~au) of the prototypical DG Tau B system in ro-vibrational H$_2$ and selected atomic emission lines, obtained with NIRSpec-IFU \& NIRCam. JWST observations are complemented with H$_2$ spectro-imaging SINFONI/VLT observations at $\sim 70$~\kms ~resolution to constrain the kinematics. 
 
The outline of the article is as follows. The JWST and SINFONI observations and data reduction process are presented in \S 2. We describe the main results in \S 3. In \S 4 we analyze the morphology, kinematics and excitation conditions of ro-vibrational H$_2$ in the red-shifted lobe, and its relation with the jet and CO conical outflow. We discuss in \S 5 the implications for the various scenarios proposed for the origin of small-scale rotating molecular cavities, and we conclude in \S 6.

\section{Observations and data reduction}

\subsection{NIRCam/JWST images}
NIRCam data of DG Tau B were obtained on October 12, 2022 within the JWST Cycle 1 program 1644 (PI: C. Dougados), in four narrow-band filters: F164N, F212N, F323N and F405N \mdy{in order to observe respectively the [Fe {\sc ii}] $\lambda$1.64$\mu$m line, a hot atomic gas tracer to image jets, the  H$_2$ 1-0 S(1) $\lambda$2.12$\mu$m and  1-0 O(5) $\lambda$3.23$\mu$m lines, to map the warm molecular outflow component and the Br$\alpha$ $\lambda$4.05$\mu$m emission line as an accretion tracer. 
} 
\mdy{The angular resolutions (FWHM) for each filter in increasing order of wavelength are: 56 mas, 72 mas, 108 mas and 136 mas.}
The F164N and F212N images were  acquired with NIRCam's \textit{Short Wavelength Channel} and have a final field of view (FoV) of 50.7"x50.4" for a square pixel size of $\sim$0.03". The two images at longer wavelengths were obtained using the \textit{Long Wavelength Channel}, with a FoV size of 46.6" x 46.7" and a pixel size of 0.06$^{\prime\prime}$. They are all centered on DG Tau B, at R.A. 04:27:02.42 and Dec. +26:05:31.31 (J2000.0). Each of the images is the result of 4-point dithering, which is necessary to compensate for bad pixels and reduce flat field uncertainties.
Individual dither exposures are the result of four integrations of five groups each, giving a total effective exposure time of 1205s per filter.

The observations were reduced using version \textit{1.12.3} of the JWST calibration pipeline and the CRDS context \textit{jwst\_1138.pmap}. 
\mdy{The raw products (uncal files) were generated with the 2022\_3a version of the Science Data Processing (SDP) software.}
The settings for \texttt{Detector1}, \texttt{Image2} and \texttt{Image3} have all been left to default values. Following the \texttt{Detector1} level, the reduction products show a structured background of horizontal bands known as $1/f$ correlated noise \citep{rauscher2011}. We have built a procedure\footnote{The routine is available at: https://github.com/delabrov/JWST-Background-Noise-Removal}, described in Appendix \ref{appendix:noise_bgd}, to correct for this noise.

The absolute pointing accuracy\footnote{https://jwst-docs.stsci.edu/jwst-observatory-characteristics/jwst-pointing-performance} of JWST is 0.1$^{\prime\prime}$. There are no stars in the NIRCam field of view to improve the astrometric solution. In the following, the (0,0) position for all NIRCam images is defined as the position of the ALMA 230GHz continuum intensity peak of the disk from \citetalias{de2020alma} ($\alpha_{J2000}$: 04$^{h}$27$^m$02$^s$.573, $\delta_{J2000}$: +26$^\circ$05$^\prime$30$^{\prime\prime}$.170). It agrees within 0.05$^{\prime\prime}$ with the positions of the jet axis and the apex of the \mdy{line and continuum} scattered light cavity in NIRCam images, indicating good absolute astrometry. 

\subsection{NIRSpec/JWST IFU data}
NIRSpec IFU data of DG Tau B were obtained on September 05, 2022 within the JWST Cycle 1 program 1644 (PI: C. Dougados). Two different pointings were observed, one in each outflow lobe. Each pointing covers a FoV of 3$^{\prime\prime}$ x 3$^{\prime\prime}$, sampled with spaxels of 0.1$^{\prime\prime}$ and was observed in two different filter/grating combinations: F100LP/G140H covering from $\lambda=0.97\mu$m to $1.89\mu$m and F170LP/G235H from $\lambda=$1.66$\mu$m to 3.17$\mu$m, both with a spectral resolving power of R$\sim$2700 ($\simeq 110$ \kms). The spectral sampling step is 2.35$\AA$ in the first grating, 3.96$\AA$ in the second, which corresponds to a velocity sampling of $\sim$49 km/s. \mdy{The NIRSpec angular resolution is close to the diffraction limit of the telescope ($\sim \lambda/D$), i.e. $\sim$30 mas FWHM at 1$\mu$m and $\sim$90 mas FWHM at 3$\mu$m.} On each pointing, we used a 4-Point-Dither strategy with an offset of 0.4$^{\prime\prime}$. For each dither position, we used two integrations of twenty groups in the F100LP/G140H configuration, which corresponds to an effective exposure time of 4668 sec per pointing. The F170LP/G235H configuration has only one integration with twenty groups, giving an effective exposure time of 2334 sec per pointing. 

The observations also included four "leakage" exposures intended to correct for possible leaks of sky light through the micro-shutter array of detectors (MSA). The leakage correction works correctly for the observations of the blue-shifted lobe. However, residual leakage emission remains in the processed datacubes for the observations of the red-shifted lobe. 
Only a few spaxels are affected by these leakage residuals, most probably coming from DG~Tau located less than 1.5$^{\prime}$ from DG~Tau~B. These spaxels are located outside of the DG Tau B source emission and are not affecting the spectra we are interested in. 

The data were reduced using version \textit{1.12.3} of the JWST calibration pipeline and the CRDS context \textit{jwst\_1138.pmap}. 
\mdy{The raw products were generated with the 2022\_3a version of the SDP software.} 
The first level of reduction, \texttt{Detector1}, is used with the default settings, except for the \textit{jump} stage. As the detection of snowballs and showers artifacts caused by cosmic rays is turned off by default, we have activated the \textit{expand\_large\_events} parameter. \texttt{Spec2}, the second level of reduction is used without modifying the parameters. At this stage, the files are corrected for MSA leakage. A file specifying the associations between the science observations and the MSA observations is provided as an input to \texttt{Spec2}. The last level of the reduction pipeline, \texttt{Spec3}, is used by choosing the "emsm" weighting parameter in the \textit{cube\_build} reconstruction step, in order to reduce the fringing observed in the spectra. It is also important to note that previous versions had an issue with the \textit{outlier\_detection} step. We noticed that this step was too aggressive, modifying the line fluxes and the continuum of the spectra. The problem has been completely corrected in the version of the pipeline used here.

We also used the method described in Appendix \ref{appendix:noise_bgd} on the \texttt{Detector1} stage 1 reduction product files to correct for residual 1/f noise. 
The absolute flux calibration of NIRSpec data is accurate to $\sim$10\% \citep{sturm2023}. Similarly to the NIRCam images, the \mdy{center offset position (0,0)} of the NIRSpec spectro-images is defined as the position of the ALMA 230 GHz continuum emission peak of the disk.

\begin{figure*}[ht]
\begin{minipage}{.5\hsize}
  \centering
  \includegraphics[width=0.75\hsize]{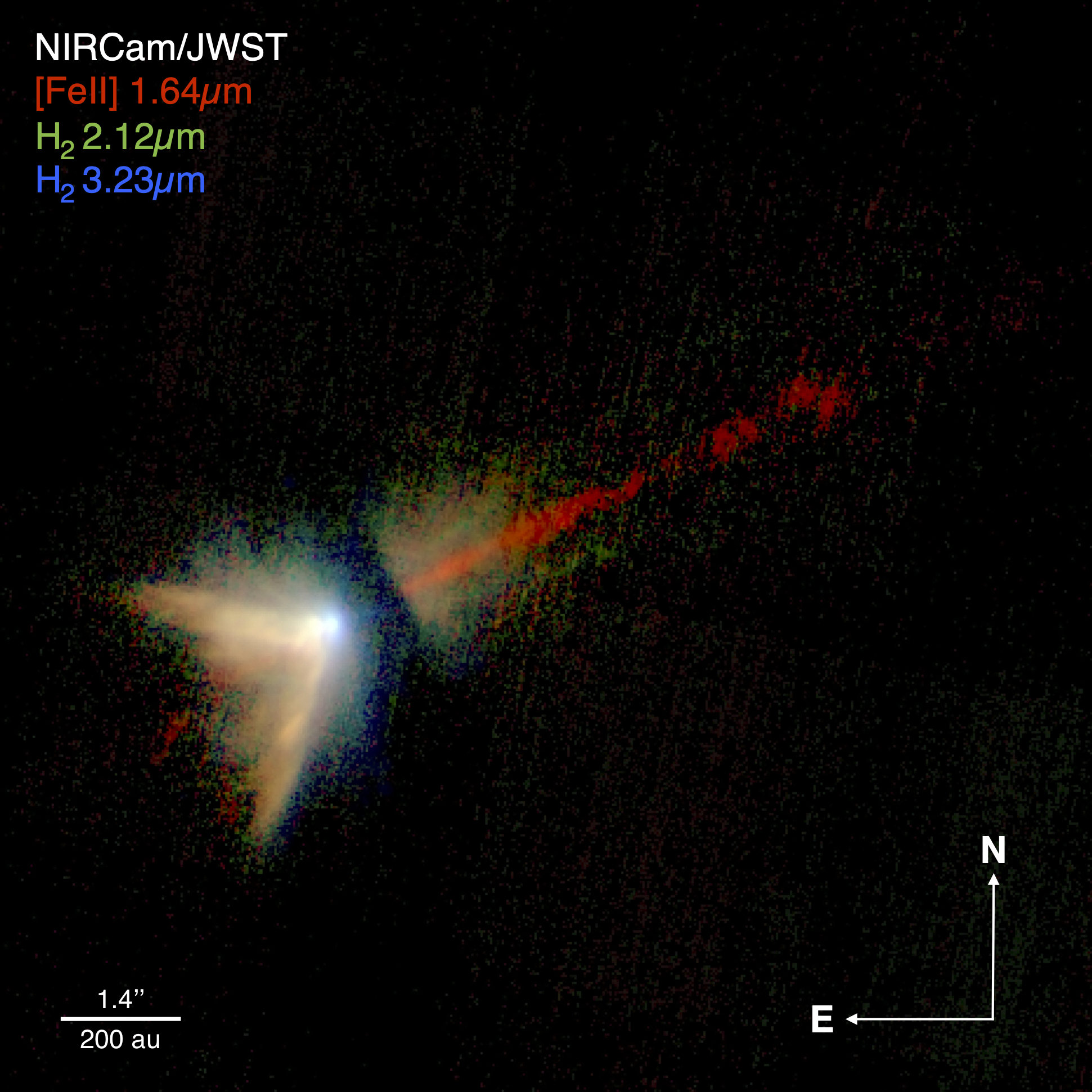}
\end{minipage}%
\begin{minipage}{.5\hsize}
  \centering
  \includegraphics[width=0.8\hsize]{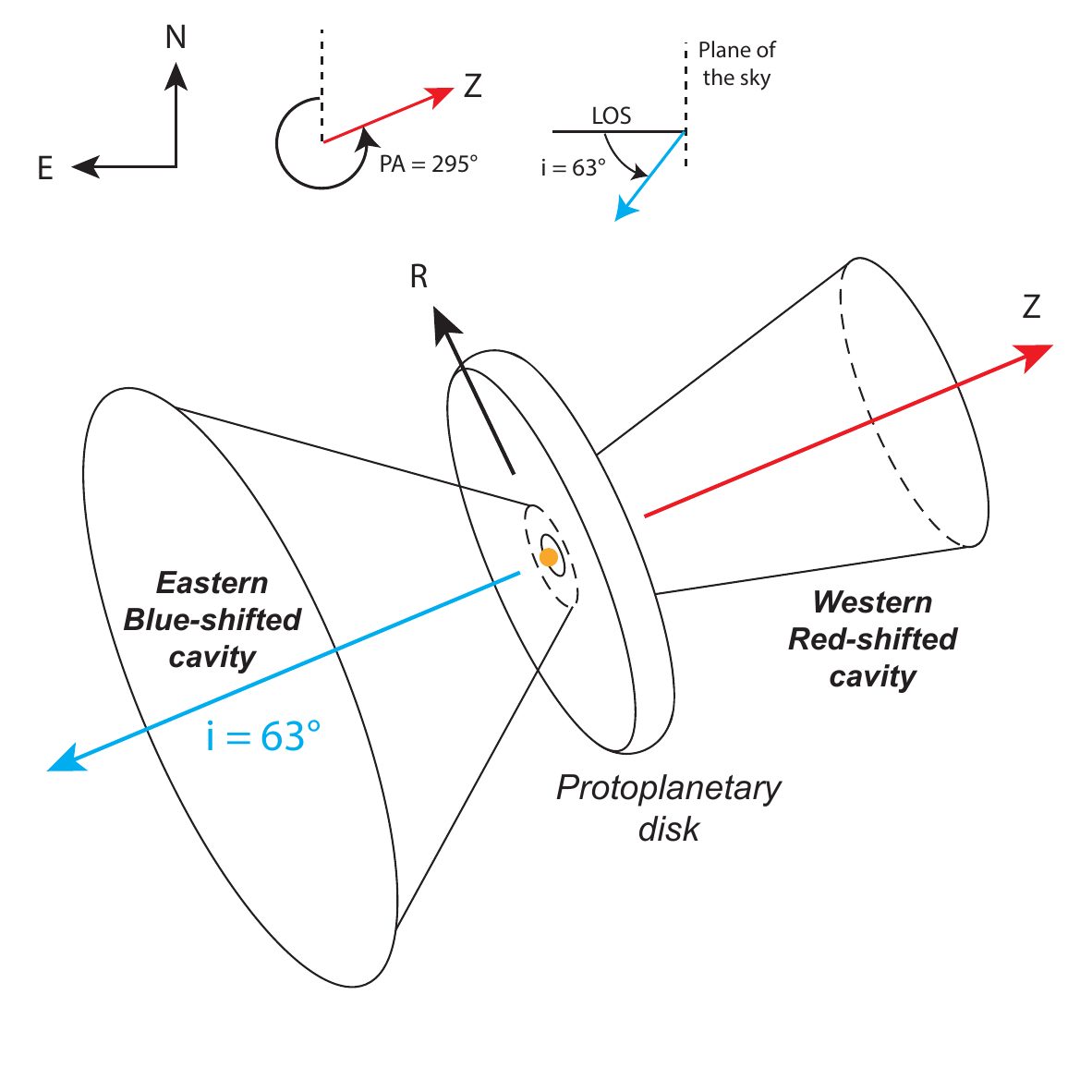}
\end{minipage}
    \caption{RGB NIRCAM images and sketch of the DG Tau B outflow geometry. Left: RGB image of DG Tau B combining three NIRCam exposures in F164N (red) for the [Fe{\sc ii}] 1.64$\mu$m line + continuum, F212N (green) for the H$_2$ 1-0S(1) 2.12$\mu$m line + continuum, and F323N (blue) for the 1-0O(5) 3.23$\mu$m line + continuum. Right: Schematic representation of the DG Tau B disk and outflows showing their orientation and inclination in space.}
    \label{fig:RGB_NIRCam}
\end{figure*}

\subsection{SINFONI/VLT IFU data}
\label{sect:SINFONI_data}
DG Tau B was observed on November 29, 2018 with the SINFONI/VLT Integral Field Spectrograph (program 0102.C-0615 PI: C. Dougados). 
The instrument was used in the seeing limited mode with a spaxel size of 125mas x 250mas. The smallest spaxel dimension (along the slicing mirrors) was set perpendicular to the jet. The source was observed in the K-band grating with a spectral sampling of 0.25nm/pixel ($\sim 30$~\kms) corresponding to a spectral resolution $R \sim$4500 (67~\kms) and covering the wavelength range 1.94$\mu$m to 2.45$\mu$m. Five dithering positions were used for a total on source integration time of 1500s and providing a final FoV of $9^{\prime\prime}$x$9^{\prime\prime}$. Sky exposures were also obtained for a total integration time of 750s.

The reduction was performed with the SINFONI Reflex package (version 3.3.2) using standard parameters and included dark and flat correction, wavelength calibration, sky subtraction and dither combination. The reduction procedure was repeated without sky subtraction, to refine the wavelength calibration (see below and Appendix \ref{appendix:wvs_calibration}). The final combined datacubes are resampled to a square pixel grid of size 125mas.

Flux calibration followed a standard procedure using the reference star Hip028649 observed on the same night, under similar airmass and seeing conditions. To derive the final flux calibration, we estimated the intrinsic K-band spectrum of Hip028649 by fitting a power law to its 2MASS magnitudes ($\text{J} = 7.700\pm0.021$, $\text{H} = 7.789\pm0.016$, $\text{K} = 7.808\pm0.020$). Using this procedure, we estimated an absolute flux calibration uncertainty of $\sim$10\%. Because the standard star was observed close in time and under the same conditions as DG Tau B, the flux calibration procedure also corrects for telluric absorption.

The primary goal of the SINFONI observations is to constrain the kinematics of the H$_2$ outflow in DG Tau B, in particular to search for radial velocity gradients across the flow indicative of rotation. It is therefore necessary to measure the peak radial velocities of the H$_2$ emission lines with a high degree of accuracy (better than 1 km/s in principle). We have improved the wavelength calibration performed by the SINFONI reduction pipeline using the atmospheric OH lines present in the data (see Appendix \ref{appendix:wvs_calibration}). After recalibration, there are however still systematic residual radial velocity variations across the outflow of $ \Delta V \simeq$ 3~\kms (3$\sigma$) which limit our ability to detect velocity gradients. 
The H$_2$ centroid velocities reported in the present paper are corrected for the earth motion (Barycentric Earth radial velocity BERV) and for the star velocity estimated by \citetalias{de2020alma} (V$_{LSR} = 6.35 \pm 0.05$~\kms).

\begin{figure*}
    \centering
    \includegraphics[width=0.75\textwidth]{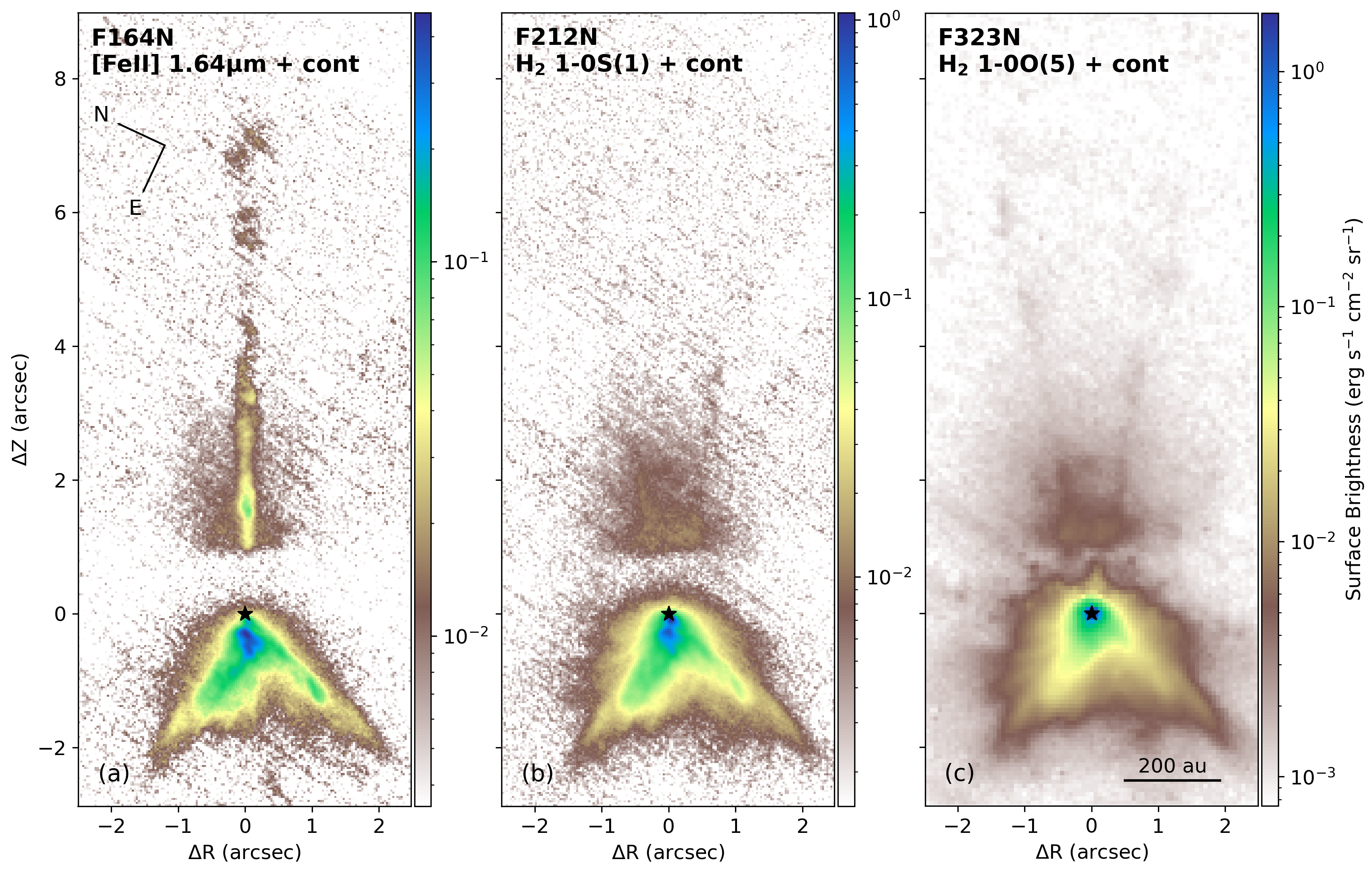}
    \caption{NIRCam narrow-band images of DG Tau B: (a) F164N = continuum + [Fe {\sc ii}], (b) F212N = continuum + H$_2$ 1-0 S(1), (c) F323N = continuum + H$_2$ 1-0 O(5). 
    The black star at (0,0) marks the position of the disk continuum emission peak observed with ALMA at 232GHz \citepalias{de2020alma}, considered to be the source position.
    The orientation on the sky is shown in panel a. The $\Delta Z$ axis points towards the red-shifted jet at $PA = 295^\circ$.}
    \label{fig:NIRCam_images}
\end{figure*}

\begin{figure*}
    \centering
    \includegraphics[width=1.0\textwidth]{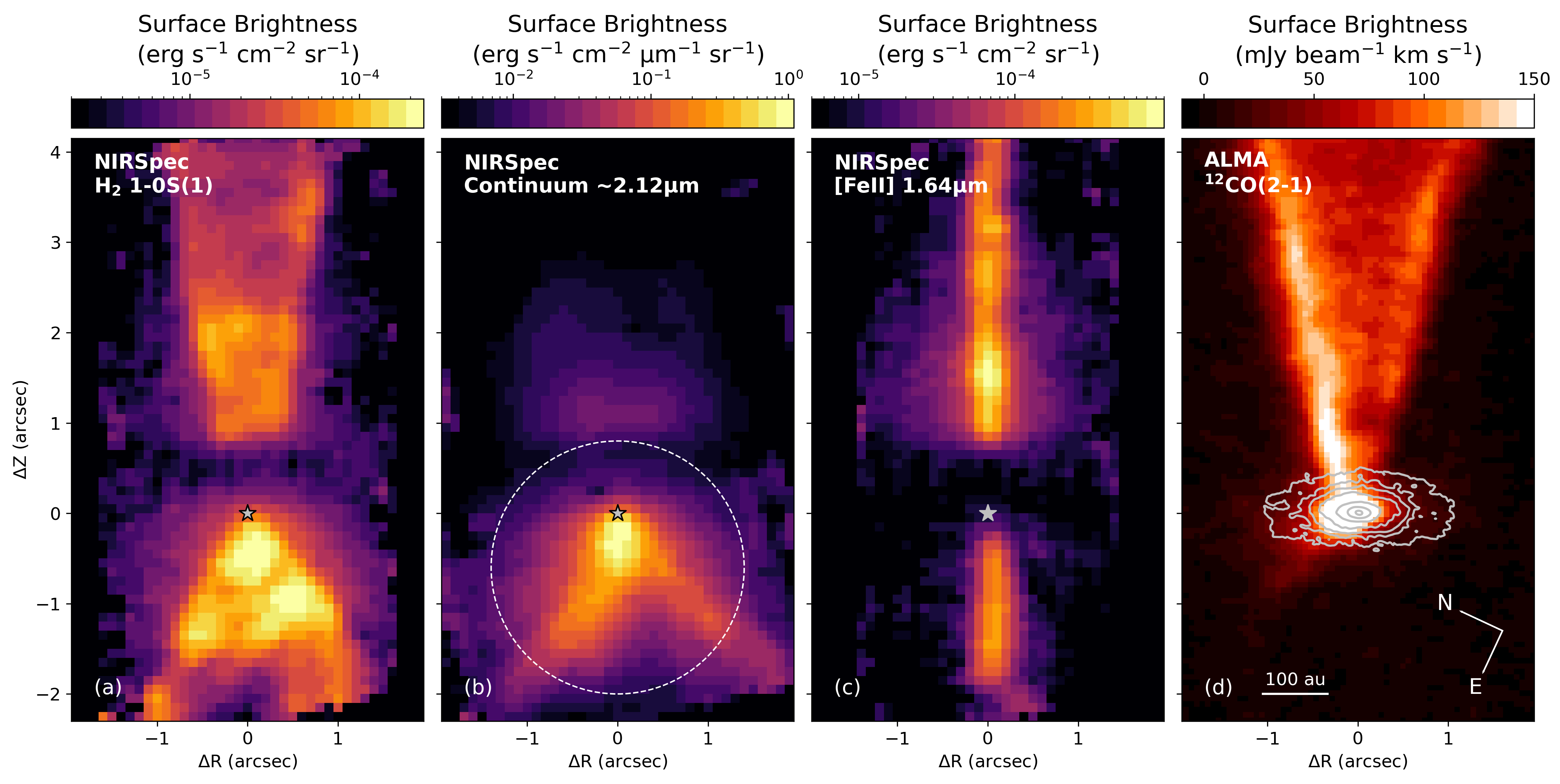}
    \caption{NIRSpec and ALMA maps of DG Tau B outflows: (a) NIRSpec H$_2$ 1-0S(1) $\lambda$2.12$\mu$m integrated surface brightness map, continuum subtracted. (b) NIRSpec average continuum map estimated over a spectral width $1.8 \times 10^{-2}$ $\mu$m centered on the H$_2$ 1-0S(1) line. \mdy{The white dashed circle corresponds to the aperture used to extract the spectra from Fig~\ref{fig:NIRSpec_spectra}.} (c) NIRSpec [Fe {\sc ii}] $\lambda$1.64$\mu$m integrated surface brightness map, continuum subtracted. (d) Moment 0 map of the $^{12}$CO(2-1) emission observed with ALMA, integrated between ($V-V_{sys}$) = +2.15 km s$^{-1}$ and 8 km s$^{-1}$. The contours of the ALMA disk continuum at 232GHz are superimposed. Adapted from \citetalias{de2020alma}. The grey star at offset (0,0) marks the position of the disk continuum emission peak, considered to be the source position. The $\Delta Z$ axis points towards the red-shifted jet at $PA = 295^\circ$.}
    \label{fig:NIRSpec_images}
\end{figure*}


\begin{figure*}[ht!]
    \centering
     \includegraphics[width=1.0\textwidth]{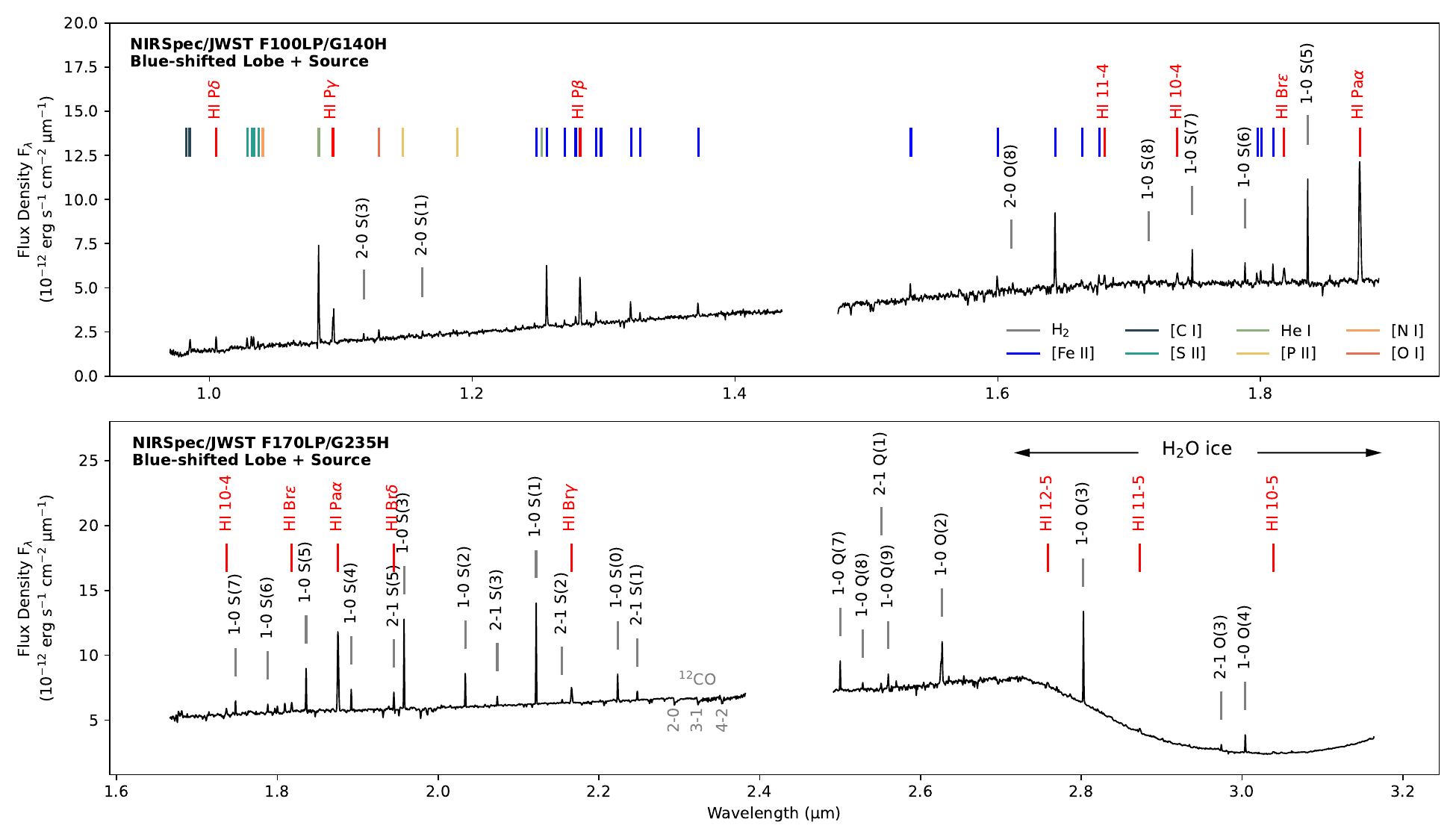}
     \caption{NIRSpec-IFU spectra of DG Tau B. Top panel: spectrum in the F100LP/140H grating integrated over a \mdy{circular aperture with a radius of $1.4^{\prime\prime}$ centered in the blue-shifted lobe containing the central source position (Fig.~\ref{fig:NIRSpec_images}b).} Many atomic transitions are detected (in particular for [Fe {\sc ii}]). The mid-band value-free gap is associated with the physical separation between the two NIRSpec detectors. Bottom panel: spectrum in the F170LP/G235H grating integrated in \mdy{the same circular aperture.} Many ro-vibrational H$_2$ transitions are detected, as well as a few [Fe {\sc ii}] and H{\sc i} atomic transitions. Several photospheric absorption lines, including the CO overtone bands, are observed. The water ice absorption band around 3$\mu$m is clearly visible. The mid-band instrumental gap is also present.}
     \label{fig:NIRSpec_spectra}
\end{figure*}

\section{Results}
In this section, we present an overview of the results obtained from the NIRCam/JWST, NIRSpec/JWST, and SINFONI/VLT observations. We first discuss in detail the H$_2$ ro-vibrational lines tracing the outflow, then we present the detected signatures in the bipolar atomic jet and the central star, and finally we discuss Hydrogen line fluxes and the large uncertainty affecting inferred accretion rate. 

\subsection{H$_2$ Outflows}

\subsubsection{Cavity morphology}
\label{sect:morphology_results}

A three-colour combination of the F164N, F212N and F323N narrow-band NIRCam images is shown in Fig.~\ref{fig:RGB_NIRCam} together with a sketch illustrating the geometry of the system.
Figure~\ref{fig:NIRCam_images} displays the individual NIRCam images in the F164N, F212N and F323N filters, centered on the [Fe II] 1.64$\mu$m, ro-vibrational H$_2$ 2.1218$\mu$m 1-0 S(1), and 3.2349$\mu$m 1-0 O(5) transitions, respectively. Each image is obtained in a narrow spectral band including continuum emission, and is rotated so that the red-shifted lobe axis points upward.

The F212N and F323N images show two V-shaped lobes, separated by a dark lane corresponding to the circumstellar obscuration (disk and envelope), as already shown in the HST NICMOS images of \citet{padgett1999}. 
The eastern blue-shifted lobe downward in Fig.~\ref{fig:NIRCam_images}) shows a similar V-shaped morphology in all images, including F164N, suggesting that the emission is predominantly \mdy{continuum} scattered light. As the wavelength increases the central peak becomes brighter and a point source begins to be distinguished near the origin. 

The receding and much fainter red-shifted lobe (upward in Fig.~\ref{fig:NIRCam_images}) shows a narrow conical geometry in the F212N and F323N filters, not seen in F164N, suggesting that it is tracing in-situ H$_2$ emission (as confirmed below by NIRSpec). Around the base of this cone, faint extended emission about 2$^{\prime\prime}$ wide, \mdy{with a similar morphology as in F164N, hence probably due to scattered light (as confirmed by NIRSpec data, Sect.~\ref{sect:scattering})}. The base of the lobe is extincted by the disk/envelope. The edges of the receding conical cavity appear brighter than its center and are marginally resolved in the NIRCam images. We estimate their deconvolved thickness at $\Delta Z =1.5^{\prime\prime}$ from the central source as $\simeq 0.16^{\prime\prime}\pm 0.04^{\prime\prime}$ (see Appendix \ref{appendix:cavity_thickness}), corresponding to $22 \pm 5$~au. The limb-brightening observed suggests a hollow conical geometry for H$_2$, as already observed in CO with ALMA (\citetalias{de2020alma, de2022modeling}). 

Turning now to NIRSpec results, \mdy{we show in Fig.~\ref{fig:NIRSpec_spectra} two spectra in the F100LP/G140H and F170LP/G235H bands obtained in a circular aperture of 1.4$^{\prime\prime}$ radius centered in the approaching lobe ($\Delta Z = -0.6^{\prime \prime}$).} \mdy{Twenty-four} ro-vibrational H$_2$ transitions are detected, all between \mdy{1.1$\mu$m} and 3.1$\mu$m. The H$_2$ lines all share the same morphology, which is illustrated in Fig.~\ref{fig:NIRSpec_images}a for the continuum-subtracted H$_2$ 2.12$\mu$m emission map. H$_2$ emission stands out more clearly in the NIRSpec observations than in narrow-band NIRCam images due to the combination of spectral resolution and accurate continuum subtraction, demonstrating the performance of NIRSpec-IFU for outflow studies. 

The red-shifted western lobe (pointing upward) shows a narrow conical morphology in H$_2$, strikingly similar to the red-shifted CO cavity mapped with ALMA by \citetalias{de2020alma}, shown for comparison in Fig.~\ref{fig:NIRSpec_images}d. Substructures can also be seen as internal arc-shaped enhancements and brightness asymetries along the cavity edges. The blue-shifted eastern lobe (pointing downward) shows a wider conical morphology in H$_2$, with multiple bright peaks along the cone walls and down the cone axis. Both H$_2$ lobes appear brighter in their inner $\sim1.5-2^{\prime\prime}$ region, with a sharp brightness decrease (factor 2--3) beyond. Both are surrounded by a fainter and broader halo of H$_2$ signal.

For comparison, we plot in Fig.~\ref{fig:NIRSpec_images}b an image of the local continuum level in the vicinity of the H$_2$ 2.12$\mu$m transition. 
Very extended emission is observed, tracing photons from the central continuum source that are scattered by surrounding dust grains.
In the red-shifted lobe, the continuum map morphology is strikingly different from the bright H$_2$ cone: it  traces a broader nebulosity of width $\Delta R \simeq 2^{\prime\prime}$ emerging from the dark lane and detected out to $\Delta Z = 2^{\prime\prime}$, similar to the nebulosity observed in the narrow-band images in Fig.~\ref{fig:NIRCam_images}. In the blue-shifted lobe, in contrast, the V-shaped continuum map morphology is very similar to the region of bright H$_2$ in the continuum-subtracted map of Fig.~\ref{fig:NIRSpec_images}a, indicating a stronger contribution of scattered H$_2$ in that lobe. In the next section we attempt to quantify the contribution of scattered versus in-situ H$_2$ emission.

\subsubsection{Scattered versus in-situ H$_2$}
\label{sect:scattering}

As discussed in Sect.~\ref{sect:morphology_results}, the similar morphology of the blue-shifted lobe in all NIRCam images, and in the continuum and H$_2$ NIRSpec maps, indicates a significant contribution of scattering \mdy{to the total observed intensities}. We investigate below the relative importance of scattering and in-situ H$_2$ emission using the NIRSpec emission maps.  We concentrate on the H$_2$ 1-0 S(1) transition at 2.12$\mu$m, which shows the best SNR. 

\begin{figure}[h!]
    \resizebox{\hsize}{!}{\includegraphics{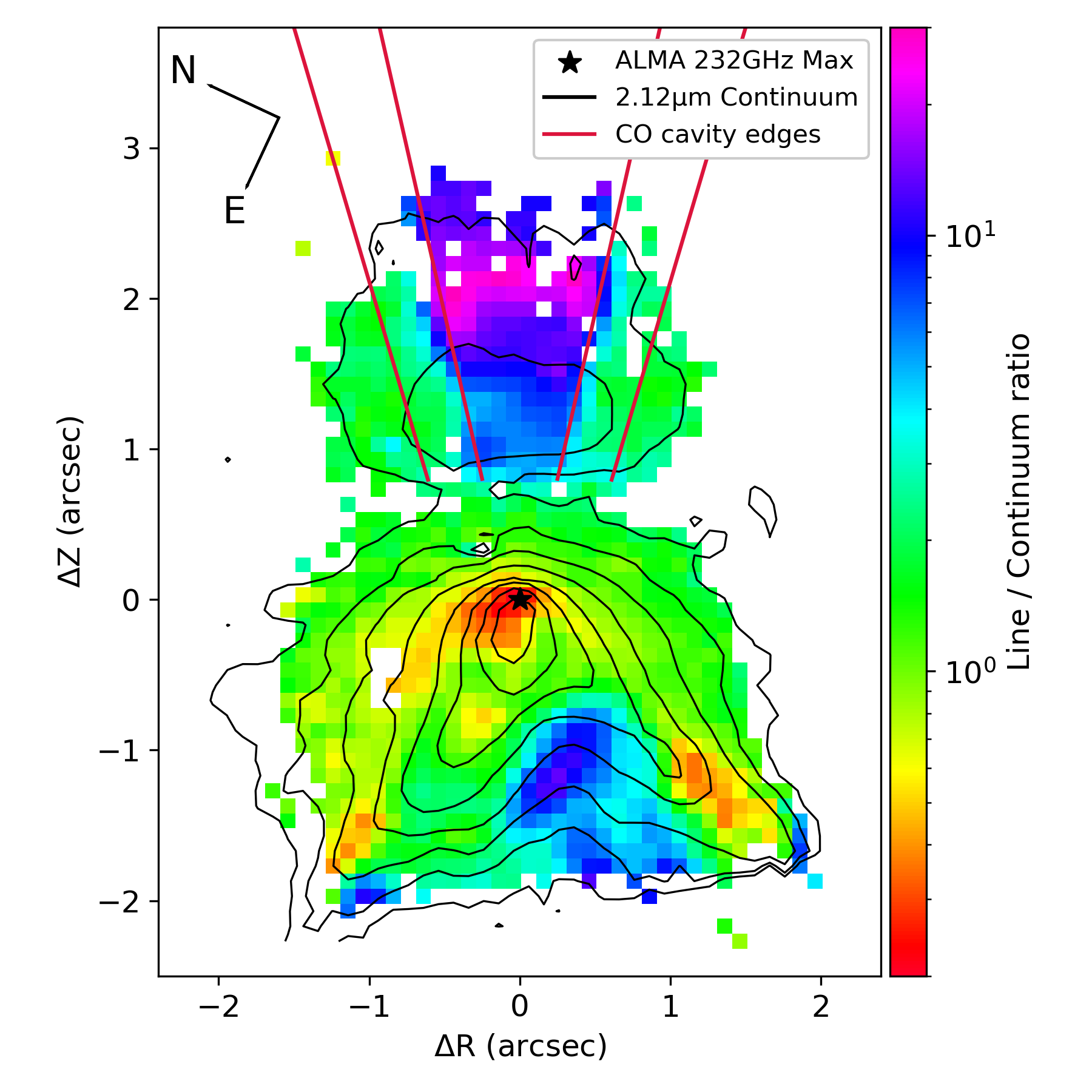}}
    \caption{NIRSpec H$_2$ line to continuum map. The color map shows the flux ratio between the 1-0S(1) H$_2$ line and the local continuum. The ratio is computed for a SNR $> 3$ in both line and continuum. Black contours plot the lcontinuum map. Levels start at 4
    $\sigma$ (with $\sigma = 1.4 \times 10^{-3} \text{ erg} \text{ s}^{-1} \text{cm}^{-2} \mu \text{m}^{-1} \text{ sr}^{-1}$) and increase by a factor of 2. Solid red lines show the projected inner and outer limits of the bright CO cone derived by \citetalias{de2022modeling}.
    }
    \label{fig:Line_continuum_map}
\end{figure}

We quantify the effect of scattering on H$_2$ by estimating the ratio between the line and adjacent continuum for each spaxel in the NIRSpec data cube. The resulting ratio map is shown as background colors in Fig.~\ref{fig:Line_continuum_map}. Only spaxels where both the line and continuum are detected with a SNR $\geq 3$ are considered. If the extended H$_2$ "emission" is dominated by scattering by surrounding dust of bright H$_2$ in the central source spectrum, the line-to-continuum ratio should be uniform over the image. In the case of in-situ H$_2$ line emission, the ratio should increase locally. To guide the eye, we superpose in Fig.~\ref{fig:Line_continuum_map} a contour map of the local continuum level (from Fig.~\ref{fig:NIRSpec_images}b).

Towards the red-shifted lobe, a wide component with constant line to continuum ratio $\sim 1-2$ (green shades) is detected \mdy{on both sides of the} conical flow, coincident with the nebulosity seen in the continuum map, and with a faint halo in the H$_2$ map (Fig.~\ref{fig:NIRSpec_images}a). It appears unlikely that this component traces in-situ H$_2$ emission as its spatial distribution coincides with the continuum. This region may trace scattering of H$_2$ from the inner disk and/or from the strong in-situ H$_2$ emission associated with the conical red lobe.
Interestingly, this nebula is wider than the limits of the bright CO conical outflow as determined by \citetalias{de2022modeling} (shown as red lines in Fig.~\ref{fig:Line_continuum_map}) and extends into the lower velocity wide CO outflow mapped by \citetalias{de2020alma}. In contrast, inside the bright cone of H$_2$, the line to continuum ratio is distinctively higher than in the surrounding nebula, with values of 5--10 that increase away from the source. From this, we estimate that the scattering contribution in H$_2$ is at most 30\% at the base of the red-shifted conical lobe and less than 15\% beyond projected distances of $\Delta Z =1.3^{\prime\prime}$ ($=180$~au). 

The  H$_2$ emission in the blue, eastern lobe is much more affected by scattering. The regions with substantial continuum emission (see black contours in Fig. \ref{fig:Line_continuum_map}) exhibit a surprisingly broad range of H$_2$ line-to-continuum ratios. A large fraction of the spaxels have ratio values around $1-2$ (green shades) similar to those observed in the nebula around the red lobe, suggesting a large contribution from scattered H$_2$. However, some regions  near the apex and along the V-shaped cavity walls have much smaller line-to-continuum ratios of $0.2-0.3$, complicating the interpretation. One possibility might be that the illuminating H$_2$ source is offset from and/or more extended than the continuum, so that it does not illuminate the innermost cavity walls in the same way. Until the origin of this spatial variation in line-to-continuum ratio is fully understood and modeled, however, only the (small) patches in the blue lobe with large ratio values $> 5$ can be unambiguously considered as dominated by intrinsic in-situ H$_2$ emission.

In the remainder of this paper, we will therefore focus our analysis on the red-shifted (western) lobe, where the contribution of scattering is negligible inside the bright H$_2$ cone. This will enable us to compare quantitatively H$_2$ in-situ emission properties with the red-shifted CO cavity studied by \citetalias{de2022modeling}.

\subsubsection{SINFONI Velocity map}
\label{sect:velocitymap}

Thanks to wavelength recalibration on OH sky emission lines (Sect. \ref{sect:SINFONI_data}, Appendix \ref{appendix:wvs_calibration}) and a higher spectral resolution ($\sim 67$ \kms), the SINFONI data can be used to constrain the radial velocity of H$_2$ more precisely than with NIRSpec. 

We first attempt to correct the SINFONI H$_2$ 1-0S(1) spectra for scattering based on the line to continuum ratio. We take as reference the spectrum with the minimum value of this ratio in the SINFONI data. This is our best estimate of the central source spectrum. At each spaxel position, this reference spectrum is then rescaled to the local continuum level and subtracted from the current spectrum. This method only corrects for the contribution of scattering of the central reference spectrum at each spatial position and may underestimate the scattering contribution in the blue lobe from a more extended H$_2$ source (see discussion in previous section).

\begin{figure}[ht!]
    \resizebox{\hsize}{!}{\includegraphics{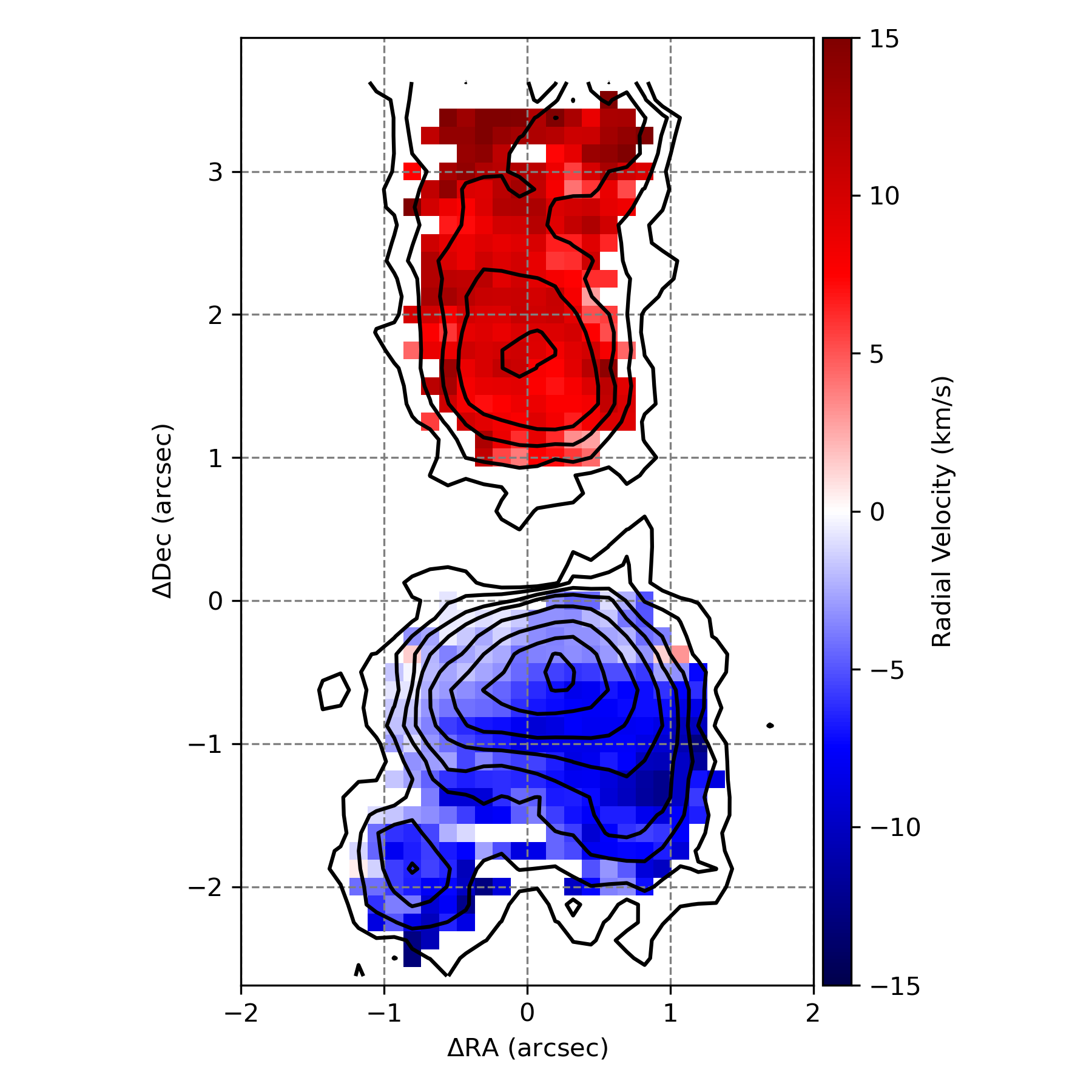}}
    \caption{Map of H$_2$ 1-0S(1) centroid velocities from SINFONI observations. Only values for which the SNR $>10$ at the line peak are shown. The black contours represent the SINFONI integrated, continuum subtracted H$_2$ 1-0S(1) surface brightness map. Contour levels are 3, 5, 8, 11, 17, 26, 38 and 50$\sigma$ with $\sigma = 4 \times 10^{-6} \text{ erg} \text{ s}^{-1} \text{cm}^{-2} \text{ sr}^{-1}$.}
    \label{fig:velocity_map}
\end{figure}

Figure~\ref{fig:velocity_map} shows the intensity and centroid velocity maps of the H$_2$ 1-0S(1) transition, derived from the SINFONI observations corrected (in part) for scattering as described above. Centroid velocities are estimated from a Gaussian fit of the line profiles. Only values corresponding to a line peak SNR $>10$ are considered. The blue-shifted lobe exhibits radial velocities between -3 and -13~\kms. However, the light blue regions at low radial velocities close to the source may still be affected by scattering, which biases the estimate of centroid velocities. The red-shifted lobe shows more uniform radial velocities with an average value of $\sim 10$~\kms and a dispersion of 3~\kms~rms.  
In both lobes, no transverse velocity gradient are detected at the 3$\sigma$ level above $\simeq$ 3~\kms, i.e. at the level of the residual calibration systematic effects (see Appendix~\ref{appendix:wvs_calibration}). We therefore take $\Delta V = 3$~\kms as a conservative upper limit for transverse radial velocity gradients. 
We discuss in Sect.~\ref{sect:kinematics} the constraints on H$_2$ kinematics (vertical and rotational motions) derived from these measurements. 

\subsection{Atomic Jets}

The top panel of Fig.~\ref{fig:NIRSpec_spectra} shows that the 0.97-1.81$\mu$m wavelength range in NIRSpec observations is dominated by atomic lines, tracing the bipolar jets. We detect transitions of [C {\sc i}], [S {\sc ii}], [N {\sc i}], [P {\sc ii}], \mdy{[O {\sc i}]}, He {\sc i} as well as 17 lines of [Fe {\sc ii}]. Figure~\ref{fig:NIRSpec_images}c shows an example of a NIRSpec continuum-subtracted map in the [Fe {\sc ii}] 1.64$\mu$m  transition. The local continuum was removed through a baseline fit to the individual spectra. 
The red-shifted jet morphology is the same as in the F164N NIRCam image (Fig.~\ref{fig:NIRCam_images}a). The blue-shifted jet is much more visible in \mdy{the NIRSpec image (Fig.~\ref{fig:NIRSpec_images}c)}, however, thanks to an accurate subtraction of the strong continuum emission in that lobe. The two jet lobes are collimated and well aligned. The inner regions of the receding jet are obscured by the dark lane.
The emission peak at 1.64$\mu$m \mdy{in the [Fe {\sc ii}] NIRSpec map (Fig.~\ref{fig:NIRSpec_images}c)} corresponds to an emission knot in the receding jet located at $\Delta Z \simeq 1.4^{\prime\prime}$. Wider and fainter nebulosity remains visible, even after continuum subtraction, around the base of the receding jet. 
This nebulosity likely traces scattering of in-situ [Fe {\sc ii}] jet emission by the dusty nebula surrounding the base of the jet/outflow, similarly to what is seen in H$_2$ (Fig. \ref{fig:Line_continuum_map}).

We estimate the position angle (PA) of the red-shifted jet by fitting a Gaussian function to each transverse intensity profile in Fig.~\ref{fig:NIRCam_images}a, at $PA_{jet,red} = 295.0^{\circ} \pm 0.2^{\circ}$. This value is identical to the $PA_{CO,red} = 295.0^{\circ} \pm 1^{\circ}$ of the red-shifted CO outflow derived by \citetalias{de2022modeling} and perpendicular to the large scale disk mapped with ALMA at $PA_{disk} = 25.7^{\circ} \pm 0.3^{\circ}$  \citepalias{de2020alma}. Our result is also consistent with the $PA$ values between 273$^{\circ}$ and 302.5$^{\circ}$ determined at large distance from the star by \citet{eisloffel1998imaging} for the red knots observed in [S {\sc ii}]6716,6731\AA. 

\subsection{Central source}

In the NIRCam F164N image, the peak of the disk millimetric continuum is located near the apex of the approaching lobe, consistent with the expected projection effects. As wavelength increases, the NIR emission peak approaches the millimetric intensity maximum. This drift suggests that the source is not seen directly at wavelengths shorter than 3$\mu$m but through \mdy{continuum} scattering by surrounding dust. \mdy{At short wavelengths, the central source emission is completely absorbed by the dust grains in the envelope and upper layers of the flared disk atmosphere, producing the dark lane. The large scale continuum emission is dominated by scattered light. As the wavelength increases, the opacity of the dust decreases and so does the width of the occulting dark lane, thereby revealing emission closer to the source position. This effect is illustrated in the Class I model images computed by \citet{whitney2003} (see their Fig. 11).}



The NIRSpec spectrum of the blue lobe in the bottom panel of Fig.~\ref{fig:NIRSpec_spectra} shows photospheric absorption lines compatible with an early K type (e.g. Si I, Ca I transitions at $\sim$1.98$\mu$m and $\sim$2.26$\mu$m, Mg I transitions at $\sim$2.28$\mu$m). The first three overtone ro-vibrational bands of $^{12}$CO (2-0, 3-1 and 4-2) are also detected in absorption between 2.29 and 2.38$\mu$m.  A more detailed study of the central star spectrum will be conducted in a forthcoming publication.

\subsection{Hydrogen lines and accretion rate}
\label{sect:BrGamma_emission}

\begin{figure}[t!]
    \resizebox{\hsize}{!}{\includegraphics{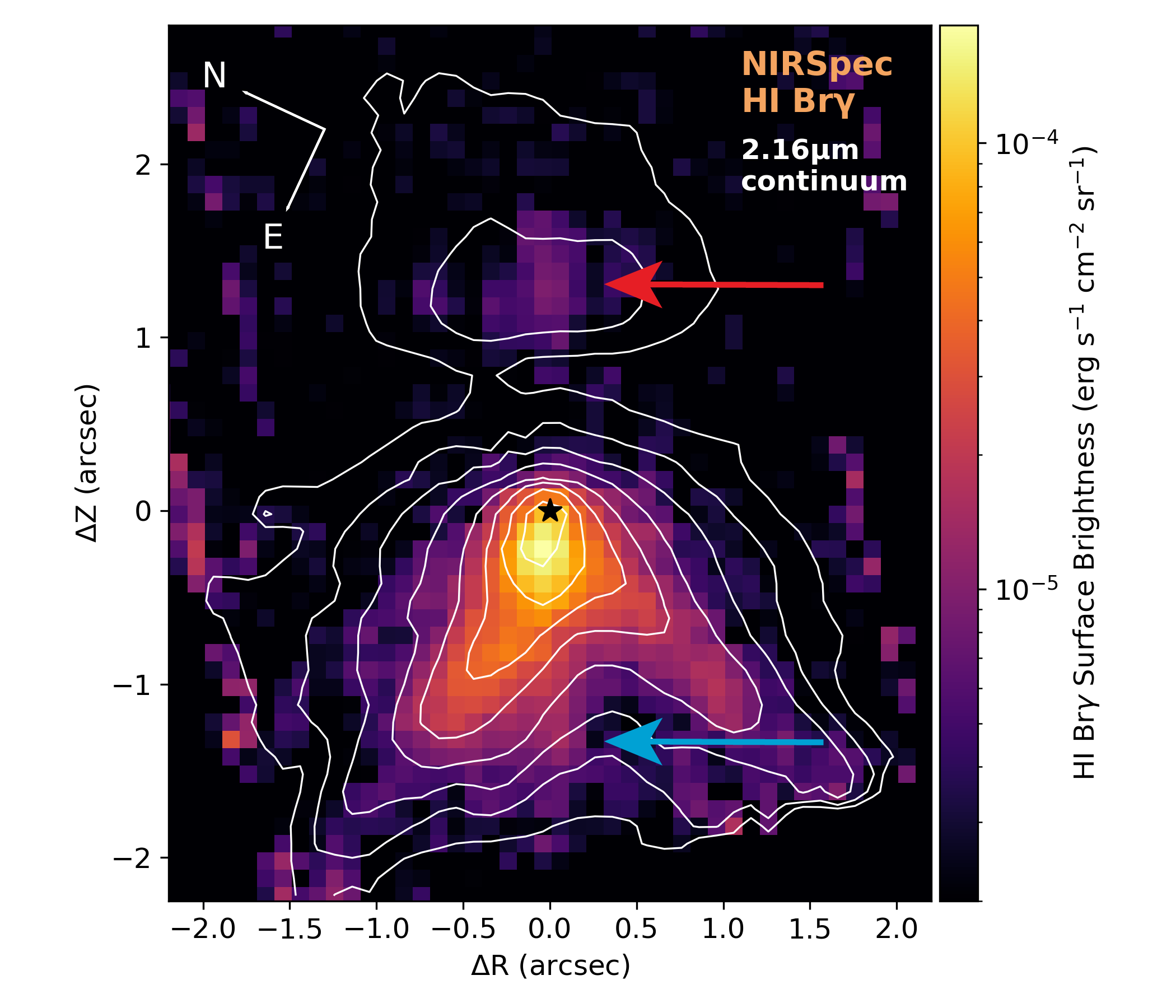}}
    \caption{NIRSpec H{\sc i} Br$\gamma$ surface brightness map, continuum subtracted (color map). White contours plot the continuum near 2.16$\mu$m. Contour levels start at 3$\sigma$ ($\sigma$ = 2.3 $\times 10^{-3} \text{ erg} \text{ s}^{-1} \text{ cm}^{-2} \mu\text{m}^{-1} \text{ sr}^{-1}$) and increase by factors of 2. The ALMA 232GHz peak continuum position is represented with a black star. \mdy{The red and blue arrows show the base of the red-shifted and blue-shifted jets respectively.}}
    \label{fig:Map_BrG}
\end{figure}

In addition to the main jet and outflow tracers, 11 H{\sc i} transitions, considered to be accretion tracers, are detected across the 0.97$\mu$m --- 3.17$\mu$m spectral range (see Fig~\ref{fig:NIRSpec_spectra}). All these transitions share a similar extended morphology, illustrated in Fig.~\ref{fig:Map_BrG} with the continuum-subtracted Br$\gamma$ emission map.  The contours of the adjacent continuum are also shown for comparison.

In the approaching lobe, the Br$\gamma$ emission mostly follows the spatial distribution of the continuum, except for some faint component along the blue-shifted jet \mdy{(see blue arrow, Fig.~\ref{fig:Map_BrG})}. In particular, the Br$\gamma$ emission peak overlaps with the continuum intensity maximum. This suggests that most of the H{\sc i} emission is formed close to the source and scattered along the walls of the V-shaped, blue-shifted cavity. In the red-shifted lobe, the H{\sc i} emission traces the base of the receding jet \mdy{(see red arrow, Fig.~\ref{fig:Map_BrG})}. 


As matter accretes, energy is released in the form of accretion luminosity $L_{acc}$. To estimate it, we use the empirical relations between $L_{acc}$ and (dereddened) luminosity of near-infrared H{\sc i} emission lines first established by \citet{muzerolle1998} and recently updated by \citet{alcala2017x}. In the same way as \citet{fiorellino2021kmos}, we assume that the relations remain valid for Class I stars.
We use the following 4 H{\sc i} transitions: Pa$\delta$ 7-3 at 1.00$\mu$m, Pa$\gamma$ 6-3 at 1.09$\mu$m, Pa$\beta$ 5-3 at 1.28$\mu$m and Br$\gamma$ at 2.16$\mu$m. We derive their fluxes in an aperture of radius 2.2$^{\prime\prime}$ centered on the continuum peak position.
We assume here that all the flux within the aperture comes from an unresolved central source exclusively as the jet contribution is negligible (see Fig.~\ref{fig:Map_BrG}). We then derive the best estimate of reddening that yields the same $L_{acc}$ (and accretion rate) for all dereddened line fluxes, using the empirical relations from \citet{alcala2017x}.
Details on the method can be found in Appendix~\ref{appendix:accretion_rate}. We find an extinction $A_V = 13 \pm 4$~mag and an accretion rate $\dot{M}_{acc}$ $  = 1.0 \pm 0.5 \times 10^{-10}$ $\text{M}_\odot \; \text{yr}^{-1}$.
Assuming that the accretion rate onto the star is ten times the red jet mass flux derived by \citet{podio2011tracing} ($\dot{M}_{jet} = 6.4 \times 10^{-9} \text{M}_\odot \; \text{yr}^{-1}$) we expected an accretion rate of $\sim 6 \times 10^{-8} \text{M}_\odot \; \text{yr}^{-1}$. Our $\dot{M}_{acc}$ estimate is 600 times smaller, \mdy{suggesting} that we are massively underestimating the intrinsic luminosity of H{\sc i} transitions. Such a discrepancy is unlikely to be due to time variability. 
Indeed, \citet{podio2011tracing} show that the red-shifted jet mass-loss rate has varied by less than a factor 3 over a period of $\sim 60$ years, suggesting small variations in accretion rate as well. 

As discussed before, the bulk of the H{\sc i} Br$\gamma$ emission follows the same spatial distribution as the continuum. Therefore, a likely explanation for the discrepancy is that a large fraction of the H{\sc i} emission coming from the source is occulted, and only a small fraction ($\simeq 1/600$) is reaching us via scattering into the line of sight. Our method only gives the extinction between us and the last scattering surface of the H{\sc i} transitions, and therefore severely underestimates the intrinsic line flux. This result shows that extreme care should be taken when estimating accretion rates from H{\sc i} line fluxes in embedded sources especially at large system inclinations. Similar conclusions have been reached in the recent JWST study of TMC1A by \citet{harsono2023jwst}. 
The self-consistent approach presented by \citep{antoniucci2008, fiorellino2021kmos, fiorellino2023} should provide a more reliable estimate of the accretion rate, as it uses isochrones (or a birthline) to fully correct for occultation of the central K-band source, and of the associated Br$\gamma$ emission. 
 
\section{Analysis of the red-shifted lobe structure}
\label{sect:analysis}

\subsection{Jet and H$_2$ opening angle and relation to CO}
\label{sect:morphology}

\begin{figure}[ht!]
    \resizebox{\hsize}{!}{\includegraphics{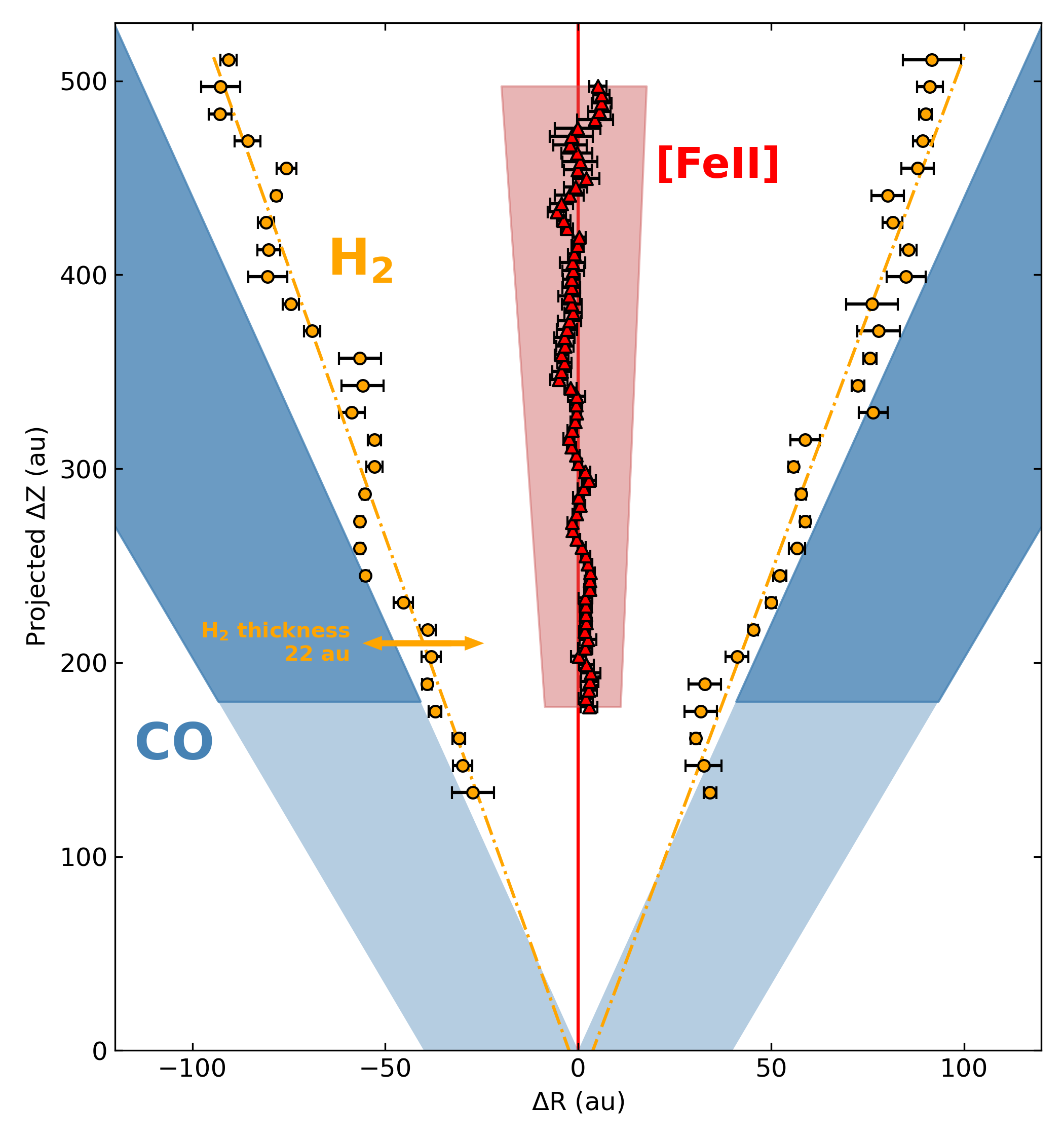}}
    \caption{Collimation of the DG Tau B red-shifted outflows. The orange dots show the derived positions of the H$_2$ edges from the NIRSpec map. The red filled area indicates the full jet opening angle and the red dots trace the jet axis positions both derived from the F164N NIRCam image. The filled blue conical area show the projected outlines of the outer and inner limits 
    derived by \citetalias{de2022modeling}
     for the bright conical CO outflow. The edges of the CO cavity are extrapolated below $\sim180$ au, where no measurement are available, in light blue.  All error bars are 3$\sigma$. The double yellow arrow shows the thickness of the H$_2$ cavity estimated in Sect.~\ref{sect:morphology_results}.}
    \label{fig:widening}
\end{figure}

The H$_2$ emission of the receding lobe appears to be very conical, preserving its opening angle along its axis (Figs. \ref{fig:NIRCam_images}b,c \& \ref{fig:NIRSpec_images}a). This morphology is very similar to the bright conical outflow mapped with ALMA on similar scales, where CO traces a hollow conical layer of deprojected inner and outer semi-opening angles of 12.8$^{\circ}$ and 16.5$^{\circ}$ respectively (see DV22 and Figure~\ref{fig:widening}). 
We estimated the opening angle of the H$_2$ cavity from the NIRSpec data.
In order to improve SNR, we used a multiplex H$_2$ map constructed from the 7 most intense transitions: \mdy{1-0 S(5) 1.83$\mu$m, 2-1 S(5) 1.94$\mu$m, 1-0 S(3) 1.95 $\mu$m, 1-0 S(2) 2.03$\mu$m, 1-0 S(1) 2.12 $\mu$m, 1-0 S(0) 2.22 $\mu$m, 1-0 O(3) 2.80$\mu$m}. As the edges of the H$_2$ cavity are brighter than the center, we estimated their position accurately using a Gaussian fit to the transverse intensity profile, independently on the two sides. 
By fitting the derived edge positions with a linear function, we estimate the H$_2$ projected semi-opening angle to be $10.4^{\circ} \pm 0.5^{\circ}$ (deprojected semi-opening angle equal to $9.4^{\circ} \pm 0.4^{\circ}$, corresponding to $Z/R \simeq$ 6). These results are shown in Fig.~\ref{fig:widening}. The H$_2$ emitting layer is nested just inside the CO conical cavity. Extrapolating the H$_2$ edge positions to the source, we derive an upper limit on the H$_2$ cavity anchoring radius in the disk of $r_{0,H_2} \leq$ 6~au.

From the NIRCam F163N image in Fig.~\ref{fig:NIRCam_images}a we can also precisely constrain the position of the jet axis and its width as a function of distance from the star. Each transverse intensity cut is fitted with a Gaussian profile to estimate an FWHM, which we consider to be the width of the jet, and a center position used to estimate the jet axis position. Small amplitude jet axis wiggling is visible. However, there is no clear increase of axis displacements with distance from the source as would be expected for a jet axis precession signature \mdy{and/or orbital motion of the jet source \citep[see for e.g.][]{masciadri2002}.}
We estimate a maximum precession angle of $1^{\circ}$. 
The measured jet FWHM are corrected for the PSF FWHM (0.056$^{\prime\prime}$ for the F164N filter\footnote{https://jwst-docs.stsci.edu/jwst-near-infrared-camera/nircam-performance/nircam-point-spread-functions}). We estimate a jet projected semi-opening angle of $1.6^{\circ} \pm 0.3^{\circ}$ (deprojected semi-opening angle of $1.45^{\circ} \pm 0.3^{\circ}$), similar to that estimated by \citet{podio2011tracing} for the optical jet.This corresponds to a jet deprojected $Z/R$ ratio of $\simeq$ 40, which is around 7 times larger than the ratio estimated for the H$_2$ cavity. Extrapolation at the source gives an upper limit on the jet origin (radius at $\Delta Z = 0$) of $r_{0,jet} \leq 4$~au. 

The western red-shifted DG Tau B ejecta form a layered structure. The [Fe {\sc ii}] jet is significantly more collimated than the H$_2$ emission, itself nested inside the CO outflow. 
The H$_2$ layer thickness, estimated in Sect.~\ref{sect:morphology_results} (cf horizontal orange arrow in Fig.~\ref{fig:widening}), suggests that 
there is no gap between the H$_2$ and CO flows. Indeed, the transverse Position-Velocity diagrams in Fig.~1 of \citetalias{de2022modeling} show that the highest velocity CO component is aligned with the H$_2$ opening angle. In contrast, no clear connection is observed between the jet and the H$_2$ lobe on large scale ($Z > 150$~au). However, if we extrapolate the conical morphologies at the origin, the jet, the H$_2$ emission and the inner CO surface all originate from within radial distances $R=6$~au of the central star. The connection between these different components could be established close to the disk mid-plane.

\subsection{Constraints on H$_2$ flow velocity and rotation}
\label{sect:kinematics}

Using the SINFONI centroid velocity map from Fig.~\ref{fig:velocity_map}, we can estimate the vertical velocity $V_Z$ and the specific angular momentum $J =RV_{\phi}$ of the red-shifted lobe. Assuming that the flow is axisymmetric, these quantities are given by:
\begin{equation}
\label{eq:Vz}
    V_z \simeq - \frac{V_{los}(R) + V_{los}(-R)}{2\cos{i}}
\end{equation}
\begin{equation}
\label{eq:RVphi}
    RV_\phi \simeq R \times \frac{V_{los}(R) - V_{los}(-R)}{2\sin{i}} \text{ ,}
\end{equation}
where $V_{los}$ is the projected velocity along the line of sight (peak centroid velocities in Fig.~\ref{fig:velocity_map}), $R$ the cavity edge radius at a given $Z$ and $i$ the inclination of the flow axis to the line of sight. 

Figure~\ref{fig:mass_flux}a shows the inferred axial velocities $V_z$ obtained, using Equation~\ref{eq:Vz}. The values of $R$ are those estimated in Fig.~\ref{fig:widening} from the NIRSpec H$_2$ datacube, which more precisely locates the cavity edges thanks to its higher angular resolution. The vertical velocity appears to be constant with height at $V_z = 22.5 \pm 0.8$ \kms. Assuming that the gas flows along the cavity wall, we then estimate a poloidal velocity $V_p = 22.8 \pm 2.5$ \kms. The derived $V_p$ for H$_2$ is larger than estimated for the conical CO outflow, ranging between 6 and 14 \kms (Fig.~9 of \citetalias{de2022modeling}). \citet{podio2011tracing} estimate the velocity of the red-shifted jet at $\sim$140 \kms. The layered morphology of the outflow is accompanied by a velocity stratification: the more internal the flow, the more collimated and the larger its vertical velocity. 

Regarding rotation signatures in the H$_2$ outflow, we derive in Sect.~\ref{sect:velocitymap}  upper limits of $\Delta V \le 3$~\kms~for $\Delta Z \geq 200$~au. Using Equation \ref{eq:RVphi} and the H$_2$ cavity radius $R(Z)$, we estimate an upper limit for the specific angular momentum beyond $\Delta Z=250$ au of $J = R V_{\phi} \le 90$ au \kms. We will discuss in Section~\ref{sec:discussion} the implications brought by the hot H$_2$ kinematics on the MHD disk wind scenario. 

\subsection{H\texorpdfstring{\textsubscript{2}}{2} Excitation conditions: Extinction, Temperature, column density and mass flux}
\label{sect:excitation_conditions}

Numerous ro-vibrational H$_2$ transitions are detected in both lobes. To estimate a column density and excitation temperature in the red-shifted lobe, we use the excitation diagram method, assuming optically thin emission (a good assumption for H$_2$ emission in outflows). 
Thanks to the very good SNR of the lines, the excitation diagrams can be studied at different positions along the flow axis (using typically 18 H$_2$ ro-vibrational lines at each position \mdy{, see Appendix~\ref{appendix:excitation_diagram}}). 

\begin{figure}[ht!]
    \resizebox{\hsize}{!}{\includegraphics{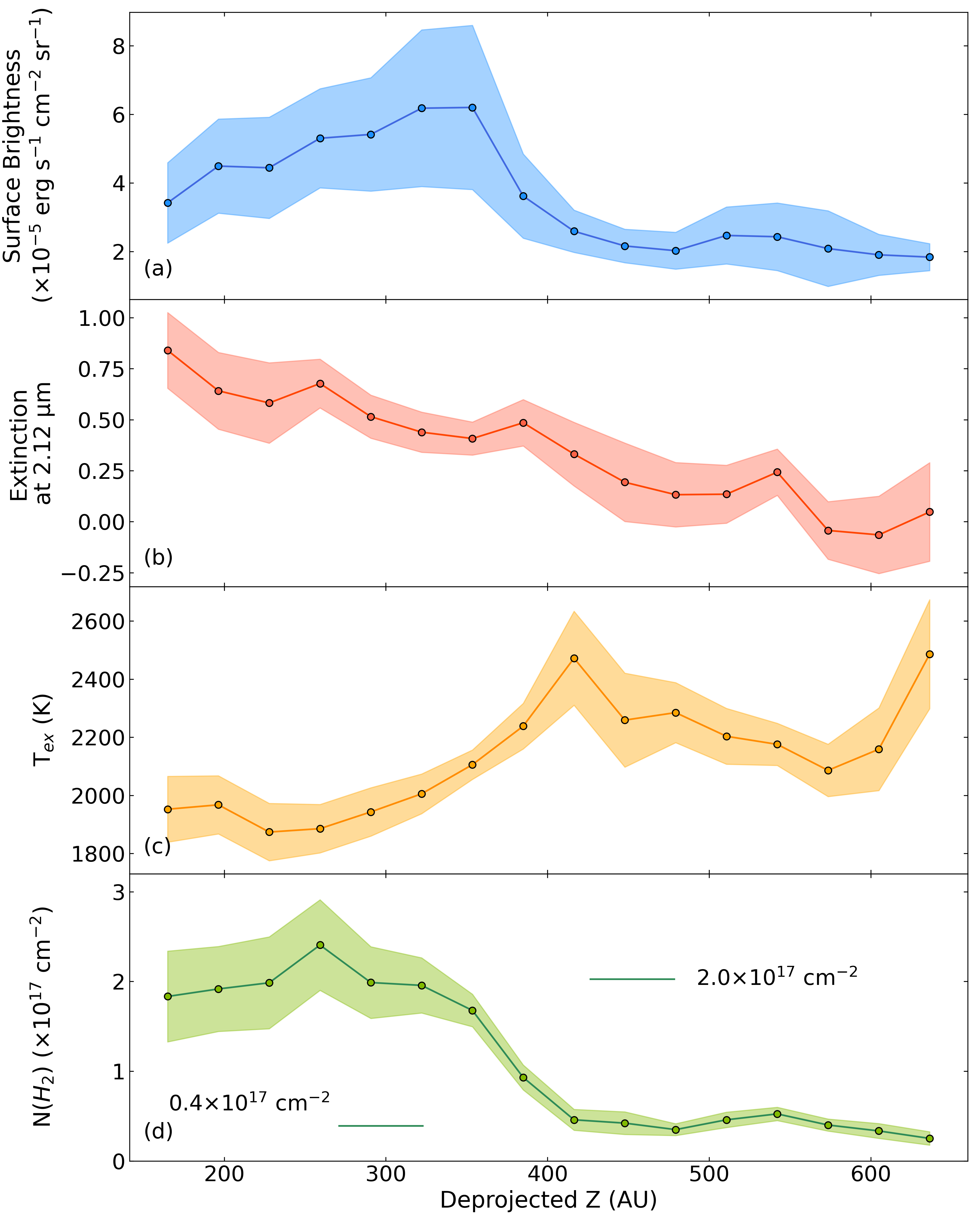}}
    \caption{H$_2$ excitation conditions along the red-shifted lobe axis. (a) Surface brightness of H$_2$ 1-0S(1) averaged across the lobe, not corrected from extinction, as a function of deprojected distance from the source. (b) Average extinction at 2.12$\mu$m obtained from the line ratio H$_2$ 1-0Q(7)/1-0S(5) (see text). (c) H$_2$ excitation temperature obtained by fitting the excitation diagram. (d) H$_2$ column density calculated from the same excitation diagrams (see Appendix~\ref{appendix:excitation_diagram}).}
    \label{fig:profiles}
\end{figure}

To increase SNR, we average all spaxels with SNR $> 3$ in the transverse direction perpendicular to the flow axis, and further average over 2 spaxels in the Z-direction. Figure~\ref{fig:profiles}a shows the average H$_2$ 1-0S(1) surface brightness distribution inside each of these "slices" as a function of distance along the flow axis. The error bars correspond to the dispersion of values among spaxels in a given slice. 

We then estimate the mean extinction towards each slice from the H$_2$ line ratio 1-0Q(7) / 1-0S(5) (at $\lambda =$ 2.4999$\mu$m and 1.8357$\mu$m respectively). Since these transitions are emitted from the same upper energy level ($v=1, J=7$), their intrinsic ratio is independent of excitation temperature and column density, and is only set  by the ratio of their $A_{ij}/\lambda$ (with $A_{ij}$ the Einstein coefficient), equal to 0.472. Any observed deviation from this value is caused by differential reddening between the wavelengths of the two lines. We adopt the near-infrared extinction law from \citet{wang2019optical}:
\begin{equation}
\label{eq:power_law_extinction}
    A_\lambda = A_{\lambda_{Ref}} \left( \frac{\lambda_{Ref}}{\lambda}\right)^{\alpha}
\end{equation}
with $\alpha = 2.07 \pm 0.03$.
The extinction $A_{2.12\mu m}$ at $\lambda_{Ref} = 2.12\mu$m can then be expressed as: 

\begin{figure}[h!]
    \resizebox{\hsize}{!}{\includegraphics{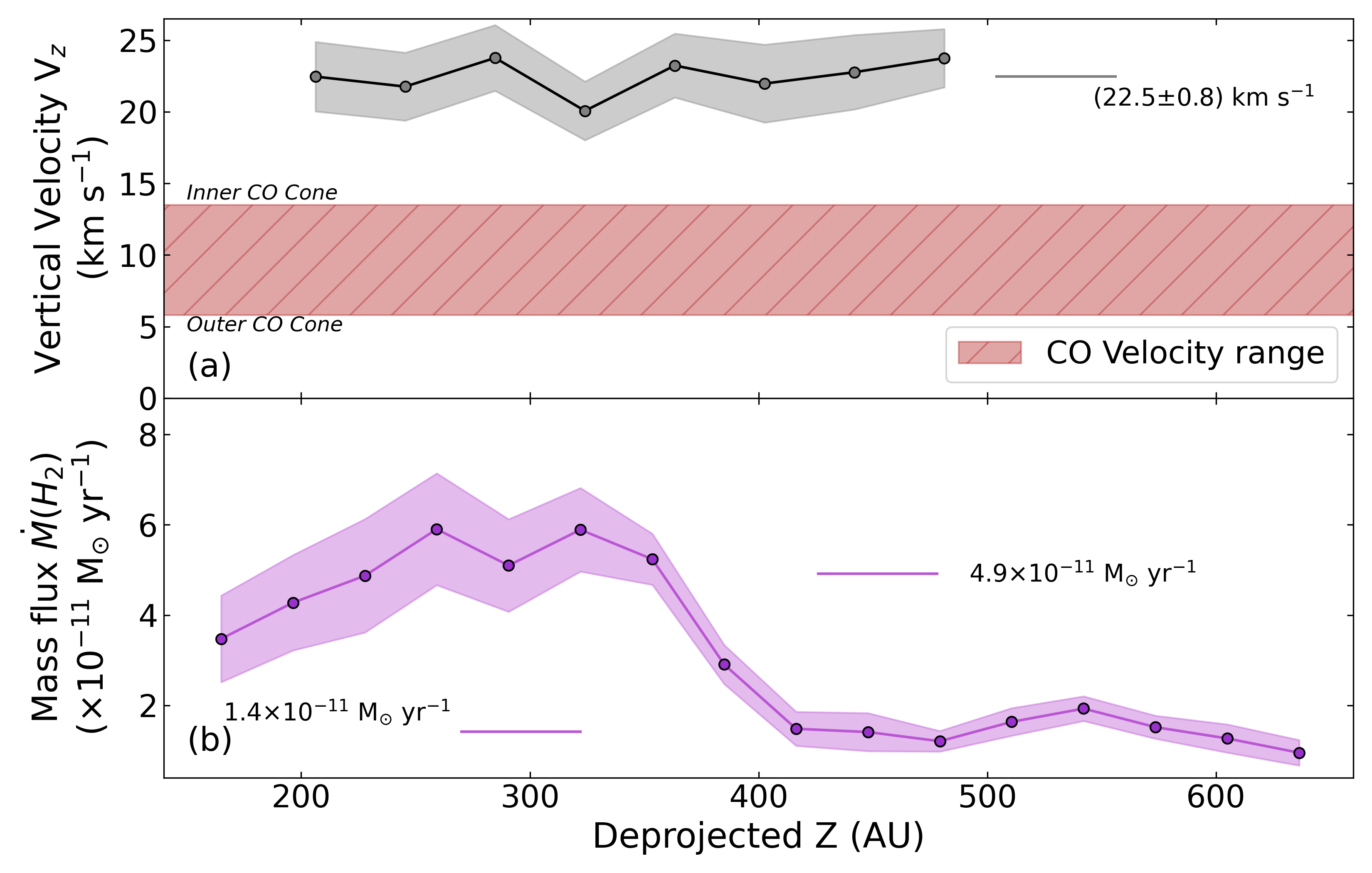}}
    \caption{Vertical velocity and mass-flux of hot H$_2$ along the red-shifted lobe of DG Tau B. (a) $V_z$ as a function of deprojected distance $Z$ from the source, derived from H$_2$ 1-0S(1) SINFONI/VLT observations. For comparison, the shaded red area shows the range of $V_z$ across the CO red-shifted conical flow, from outer to inner streamlines \citepalias{de2022modeling}. (b) Mass flux of hot H$_2$ as a function of $Z$ (see text).}
    \label{fig:mass_flux}
\end{figure}

\begin{equation}
\label{eq:extinction}
    A_{2.12\mu m} = C_{\lambda_{S5}, \lambda_{Q7}} \times 2.5 \log_{10} \left( \frac{1}{0.472} \frac{F(1-0Q(7))}{F(1-0S(5))} \right)
\end{equation}
with $C_{\lambda_{S5}, \lambda_{Q7}} = \left[ \left(\frac{2.12}{1.83}\right)^{2.07} - \left( \frac{2.12}{2.49} \right)^{2.07} \right]^{-1} = 1.565$.

Figure~\ref{fig:profiles}b shows the derived extinction values at 2.12$\mu$m in each slice as a function of distance to the source. A systematic decrease is observed, from a maximum value around 0.9 mag close to the source ($\Delta Z \simeq 150$~au) to negligible values at 600~au. 
Using Equ.~\ref{eq:power_law_extinction}, the corresponding optical extinction in the V-band at 0.55$\mu$m is
$A_V = A_{2.12}\,(2.12/0.55)^{\alpha}$, which gives $A_V = 14.7 \pm 2.3$ mag. 
\mdy{The maximum extinction derived from the H$_2$ line ratio at $Z \simeq 100 - 150$~au in the red lobe is comparable to the average extinction estimated towards the blue lobe from the H{\sc i} lines in Sect.~\ref{sect:BrGamma_emission} ($A_V = 13 \pm 4$ mag). These comparable values on similar spatial scales suggest that there is a large scale circumstellar environment obscuring light in the same way for both lobes of the system.}

This extinction profile cannot be caused by the CO conical outflow. In that case, we would expect 
 values typically 10 times lower than those estimated here (see Appendix \ref{appendix:CO_extinction}). An additional component is required. In Appendix~\ref{appendix:CO_extinction}, we show that a free-falling spherical envelope with a mass infall rate of $4.5 \times 10^{-6} \text{M}_\odot \text{yr}^{-1}$ can reproduce the extinction values along the receding H$_2$ lobe. This value is on the upper end for a Class I source, but very similar to the large envelope infall rate ($5 \times 10^{-6} \text{M}_\odot \text{yr}^{-1}$) derived by \citet{Stark2006} from the modelling of DG Tau B NICMOS/HST images. The broad low-velocity CO outflow evidenced by \citetalias{de2020alma} around the bright inner CO conical flow could also contribute to the observed extinction. 
  
Using the values of $A_{2.12 \mu m}$ derived above, and the extinction law in Equ. \ref{eq:power_law_extinction}, we then build de-reddened H$_2$ excitation diagrams, averaged over each slice. Only line fluxes with SNR $> 3$ are considered. 
No systematic offset between the different vibrational levels $v$ is observed, suggesting that the gas is close to LTE. 
An excitation temperature $T_{ex}$ and \mdy{an average} column density $N({\text{H}_2})$ is then estimated for each slice by fitting a straight line (see Appendix~\ref{appendix:excitation_diagram}). 
The last two panels of Fig.~\ref{fig:profiles} show the derived excitation temperature and column density profiles of hot H$_2$ along the red-shifted lobe. $T_{ex}$ increases slightly with $Z$ with an average value at $T_{ex} \simeq 2200 \pm 180$~K. Since $T_{ex}$ is relatively constant, the $N(\text{H}_2)$ profile follows that of the surface brightness with two plateaus: the first one corresponds to an average value of $N(\text{H}_2) \simeq 2 \times 10^{17} \text{cm}^{-2}$, the second one associated with the fainter part of the cavity, with an average $N(\text{H}_2)$ $\sim$5 times lower. 
Assuming that the H$_2$ layer thickness remains constant along the cavity at $\Delta R \simeq 22$ au, and that it is axisymmetric, the average number density of hot H$_2$ molecules in the layer, $n(\text{H}_2) = N(\text{H}_2) / \pi \Delta R$, is $\sim 200 \text{ cm}^{-3}$ within 350~au from the source and $40 \text{ cm}^{-3}$ further out. \mdy{We stress that this value is only the number density in the form of hot H$_2$ at $\sim 2000$ K; in a PDR layer at $T \simeq 2000 K$, a low H$_2$ abundance of  $\leq 10^{-3}$ is expected \citep{burton90,bron2016} so that the total density of H nuclei $n$(H) could exceed $10^5$~cm$^{-3}$.}

Using the value of $V_z$ estimated in Sect.~\ref{sect:kinematics} and the column density $N(\text{H}_2)$ averaged over each slice (from Fig.~\ref{fig:profiles}d), it is possible to derive the mass flux in the form of hot H$_2$ along the flow axis:
\begin{equation}
    \dot{M}(\text{H}_2) = 2 m_H \times N(\text{H}_2) \times \frac{N_{sp} \Delta X}{2} \times V_Z \sin(i)
\end{equation}
where $N_{sp}$ (given in Table~\ref{tab:table_profiles}) is the number of spaxels inside each slit, of width 2 pixels along the jet axis, and $\Delta X=0.1^{\prime\prime}$ is the spaxel size perpendicular to the jet axis, $m_H$ the hydrogen mass, $V_z$ the deprojected vertical velocity of the H$_2$ red-shifted outflow, and $i = 63\degr$ the inclination of the disk to the line of sight.
Beyond $Z=500$~au, we assume that $V_z$ stays constant at the value derived from the SINFONI observations. The H$_2$ mass flux profile shows two plateaus (Fig.~\ref{fig:mass_flux}b), similarly to the column density profile. The mass flux of hot H$_2$ in the inner part of the red-shifted lobe is $4.9 \times 10^{-11} \text{M}_{\odot} \text{yr}^{-1}$, decreasing by a factor $\sim$ 3 beyond $Z=400$~au at $1.4 \times 10^{-11} \text{M}_{\odot} \text{yr}^{-1}$. \mdy{Again} we stress that this value is only the mass-flux in the form of hot H$_2$ at $\sim 2000$ K; the total mass-flux carried by this layer will be much larger if hydrogen is mostly atomic, eg. if it traces a dense photodissociated layer in the disk wind (see next section).

\section{Discussion}
\label{sec:discussion}
\subsection{Global stratified flow structure}

\begin{figure}[ht!]
\centering
\resizebox{0.99\hsize}{!}{\includegraphics{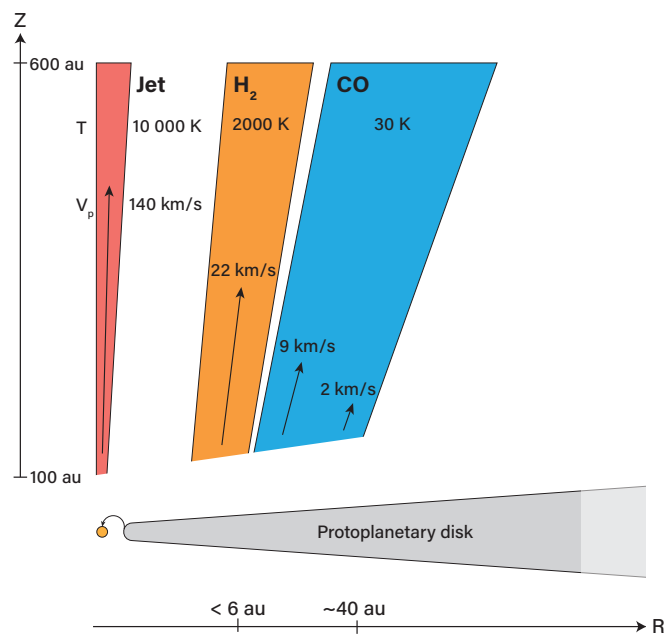}}
    \caption{Schematic representation of the layered transverse structure of the DG Tau B red-shifted lobe. The variation in opening angle, temperature and poloidal velocity between the atomic jet, the H$_2$ and CO red-shifted outflows are illustrated. The different colours describe the temperature gradient in the layers: blue is associated with the coldest component and red with the hottest component. The illustration is not to scale.}
    \label{fig:sketch_slice}
\end{figure}

Fig. \ref{fig:sketch_slice} summarizes the different components of the DG Tau B outflow.
The JWST observations reveal a red-shifted H$_2$ flow nested between the fast axial jet and the bright conical CO outflow and extending inwards the radial stratification in velocity and collimation of the CO outflow mapped by \citetalias{de2022modeling} . The H$_2$ conical flow is slightly more collimated and two times faster on average than the CO outflow. Its temperature ($\simeq 2200$K) and velocity are
intermediate between the atomic jet and CO. 

Small scale H$_2$ outflows have been detected in a few late Class I and high accretion rate Class II sources \citep{Beck2008,takami2007micro,agra2014origin,melnikov2023study} and recently with JWST in TMC1-A \citep{harsono2023jwst} and for 5 young stars by \citet{federman2023}. These outflows all show similar layered structure with wider morphology and lower velocities ($V < 30$ \kms~typically) than the axial jets. 
The blue-shifted H$_2$ lobe in DG Tau \citep{agra2014origin} in particular shows similarities with the DG Tau B H$_2$ flow, although it is wider (half opening angle of 45$^{\circ}$) and much more compact spatially (detected only out to a projected distance of $Z=70$~au). The H$_2$ shell in DG~Tau extends the onion-like velocity structure observed in atomic lines, tracing the axial jet surrounded by a lower-velocity component \citep{agra2011, maurri2014} and appears nested inside a wide CO outflow \citep{gudel2018}. \mdy{Its} intrinsic H$_2$ surface brightnesses are however 100 times brighter on average than in the red H$_2$ lobe of DG Tau B.

We discuss below the current constraints from the DG Tau B JWST observations to the three competing scenarios for the origin of these H$_2$ outflows: disk winds versus swept-up cavities versus shocked disk wind. 

\subsection{MHD Disk wind scenario} 
\label{sect:MHD_wind}

The global layered H$_2$/CO structure observed in DG Tau B with temperature, velocity, collimation increasing towards the axis of the flow are reminiscent of a disk wind scenario \mdy{(as discussed below)}.

In addition, the derived parameters of the H$_2$ cavity (opening angle, $V_z$, temperature and to a lesser extent mass flux) appear constant with distance from the source, similar to the CO structure \citepalias[see][]{de2020alma, de2022modeling}, compatible with an underlying steady ejection process. 

We investigate in the following whether the H$_2$ layer could trace inner streamlines of the same disk wind solution that also accounts for the conical CO flow. A photo-evaporative wind is ruled out for the origin of the conical CO flow by the large mass flux ($2.3 \times 10^{-7} \text{M}_\odot \text{yr}^{-1}$) originating from less than 10~au \citepalias{de2022modeling}, which largely exceeds current model predictions (see for example mass-loss compilations in Fig. 7 of \cite{pascucci2023}. A non-thermal acceleration process must therefore be present. 

We explore below the case of magnetically driven disk winds. In an MHD disk wind of uniform magnetic level arm the terminal poloidal velocity is proportional to the Keplerian velocity at the streamline footpoint radius r$_0$ \citep{blandford-payne1982,anderson2003,ferreira2006jet}. Therefore material from inner streamlines will have a higher velocity than material from more external streamlines, reproducing the onion-like velocity layered structure of the DG Tau B red-shifted outflows. 


The existence of an inner layer of  warm ($T > 2000 K$) molecular gas inside an MHD disk wind, corresponding to the photo-dissociation region, was first predicted by \citet{panoglou2012molecule}. Recent simulations of disk wind outflows by \citet{wang-bai2019}, coupling non-ideal magneto-hydrodynamics and thermochemistry, show a layered structure in temperature, velocity and chemical abundance strikingly similar to the H$_2$/CO radial stratification observed in DG Tau B. In their fiducial model, with parameters comparable to the  DG Tau B case (disk accretion rate $\simeq 10^{-8} \; \text{M}_\odot \text{yr}^{-1}$, stellar mass $ = 1~M_\odot$), the wind launched from $r=1$ to 100 au stays predominantly molecular except for a thin PDR-like layer of a few 1000~K along the innermost streamlines, where photo-dissociation becomes important. 
Models of dense \mdy{photodissociation regions (PDRs)} predict that the layer at 2000~K, bright in H$_2$ ro-vibrational emission, 
would have a low H$_2$ abundance $x_{H_2} \simeq 10^{-3}$ \citep{burton90,bron2016}. 
Taking into account the low H$_2$ abundance, the actual mass-flux along this PDR-like wind layer would then be $\dot{M}({\rm 2000 K})\simeq 1.4-5 \times 10^{-8} \text{M}_{\odot} \text{yr}^{-1}$.

\begin{figure}[ht!]
\resizebox{\hsize}{!}{\includegraphics{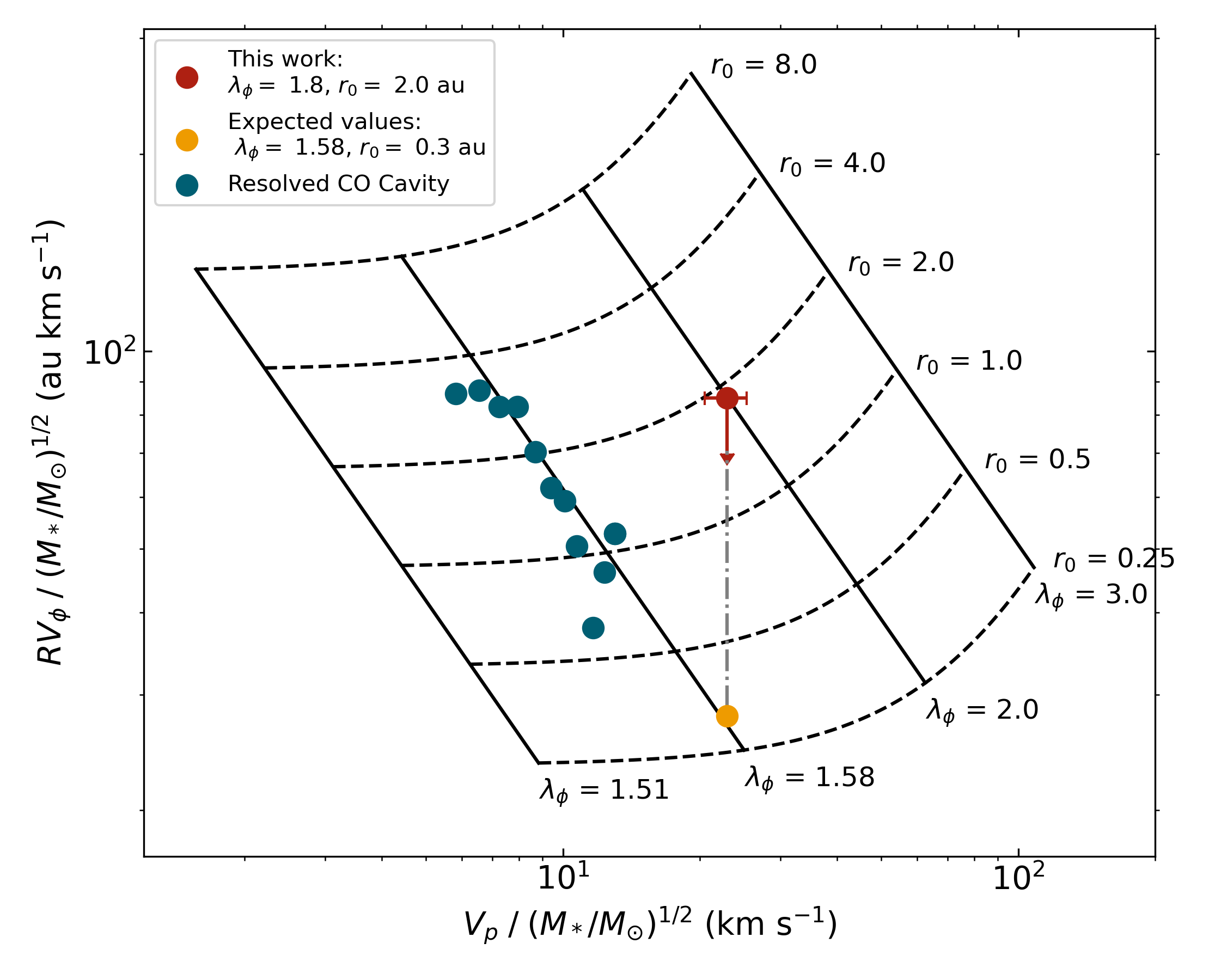}}
    \caption{Relation between the poloidal velocity $V_p$ and the specific angular momentum $J=RV_\phi$ defining the parameter domain ($r_0$,$\lambda_\phi$) in the case of an MHD disk wind, based on \citet{ferreira2006jet}. The values obtained by \citetalias{de2022modeling} for the resolved CO red-shifted cavity are shown in blue. The red dot is associated with the values of $V_p$ and the upper limit on $J$ estimated by the kinematics study of the H$_2$ red-shifted cavity. The yellow point is obtained by considering a constant value of $\lambda_\phi$ between the CO and H$_2$ cavities.}
    \label{fig:anderson_plot}
\end{figure}

The H$_2$ flow kinematics and mass-flux bring further constraints to such MHD disk wind solutions. 
Figure~\ref{fig:anderson_plot} shows the expected relation between the specific angular momentum $J = RV_{\phi}$ and the poloidal velocity $V_p$ 
in the context of a steady self-similar MHD disk wind solution \citep{ferreira2006jet}. As demonstrated in \citetalias{de2022modeling}, the layered kinematics of the conical CO outflow is compatible with such an MHD disk wind solution with a constant magnetic lever arm $\lambda{_\phi}=1.58$ and streamline anchoring radii $r_0$ ranging between 0.7 and 3.4~au (see blue dots in Fig.~\ref{fig:anderson_plot}).
We overplot in Fig.~\ref{fig:anderson_plot} the upper limit on $J$ and $V_p$ estimated for the H$_2$ flow in 
Sect. \ref{sect:kinematics}.
From this diagram, we estimate an upper limit on $\lambda_{H_2} < 1.8$ and on $r_{0,H_2} < 2.0$ au. 
   
If we assume that the magnetic lever arm parameter remains the same between the H$_2$ and CO outflows, then the observed value of $V_{p,H_2} \simeq 23$~\kms implies a launching radius $r_0 = 0.3$ au (yellow dot in Fig.~\ref{fig:anderson_plot}). The specific angular momentum carried by the H$_2$ flow is then expected to be $\simeq$ 24 au~\kms, well below our derived upper limit of $J \le 90$ au~\kms. Given the measured H$_2$ flow radii (cf. Fig.~\ref{fig:widening}),  expected rotation velocities would be $\le 1$~\kms.

In self-similar disk wind solutions, the local ejection efficiency, $\xi$, is defined as $\dot{M}_{acc}(r) \propto r^{\xi}$. If the MHD disk wind dominates the angular momentum extraction from the disk driving accretion, the magnetic lever arm $\lambda$ is related to $\xi$ with $\lambda \simeq 1 + 1/2\xi$ \citep{ferreira1997jet}. With the assumed value of $\lambda = 1.58$ ($\xi \simeq 1$), the relation between mass accretion rate and radius is simplified to $\dot{M}_{acc}(r) \propto r$. Considering the conservation of mass flux across the wind launching region: $\dot{M}_w = \dot{M}_{acc}(r_{0,out}) - \dot{M}_{acc}(r_{0,in})$, the H$_2$ and CO mass fluxes are then linked by a simple expression depending on the respective width of their launching regions:
\begin{equation}
    \frac{\dot{M}_{H_2}}{\dot{M}_{CO}} = \frac{\Delta r_{0,H_2}}{\Delta r_{0,CO}}\text{ ,}
\end{equation}
with $\Delta r_{0,H_2}$ and $\Delta r_{0,CO}$ corresponding  to $(r_{0,out} - r_{0,in})$ for the H$_2$ and CO wind launching regions respectively. This relation allows us to eliminate $\dot{M}_{acc}$ which is poorly constrained in the disk. In the case of the CO conical flow, the kinematic data of \citepalias{de2022modeling} plotted in Fig.~\ref{fig:anderson_plot} give a launching width of $\Delta r_{0,CO} = 3.4 - 0.7 = 2.7$ au. With a mass-flux in the CO conical flow of $\dot{M}_{CO} = 2 \times 10^{-7} \; \text{M}_\odot \text{yr}^{-1}$ \citetalias{de2022modeling}, the total mass flux in the H$_2$ layer at similar distance from the source ($\dot{M}({\rm 2000 K}) = 1.4 \times 10^{-8} \; \text{M}_\odot \text{yr}^{-1}$
assuming a PDR layer with an H$_2$ abundance of $10^{-3}$, see above) can be explained by an H$_2$ wind launching region of width $\Delta r_{0,H_2} = 0.2$ au, i.e. launching radii from \mdy{0.2 to 0.4}~au. This range is consistent with the upper limit on the H$_2$ wind anchoring radius of $r_{0,H_2} \leq 6$~au derived from the large scale morphological study in Sect.~\ref{sect:morphology}. It is also consistent with the H$_2$ ro-vibrational layer tracing MHD wind streamlines interior to the cold CO conical outflow (launched from $r_0 \ge 0.7$~au).
\mdy{However, it remains to be checked that PDR models at the derived disk wind densities (n$_H$ $\geq 10^{5}$ cm$^{-3}$ for H$_2$ abundance $\leq 10^{-3}$ see \S 4.3 above) can reproduce the thermalization of the H$_2$ levels for reasonable G$_0$ values and on the spatial scales observed here. We differ this analysis to a forthcoming paper where we will combine the NIRSpec and MIRI H$_2$ transitions, providing more stringent constraints to the models. The fact that the H$_2$ emission in the blue lobe appears 'knotty' is more difficult to reconcile with an UV irradiated disk wind scenario where the H$_2$ emission may be expected to vary more smoothly, unless different NIR and UV illuminations paths in a clumpy wind are assumed. Alternatively, shocks would be an efficient heating mechanism for the H$_2$ in the disk wind. We discuss this scenario in \S 5.4 below.}

\subsection{Envelope swept-up scenario}

The sharp transition in temperature between the adjacent CO and H$_2$ flows, indicated by the small estimated width of the H$_2$ layer ($\simeq$ 20 au width at $z= 1.5^{\prime\prime}$), could also arise if the H$_2$ layer is tracing a shocked interface with the surrounding envelope.

The popular analytical model developed by \citet{shu91,lee2000} of a radially expanding swept-up cavity created by the interaction of an inner wide-angle wind with an outer static stratified environment is unlikely to reproduce the H$_2$ DG Tau B observations, as already argued by \citetalias{de2022modeling} for the CO cavity. Indeed, in such a 
model, the local shell expansion velocity is proportional to the distance from the source ("Hubble law"), which is not observed here. A similar behavior  of self-similar expansion is predicted for the nested magnetic bubbles model presented in \citet{shang2023} and for the momentum-conserving interaction  with an infalling rotating envelope \citep{lopez-vazquez2019}.

A variant that avoids this first caveat was presented by \citet{liang2020}. Assuming weak mixing between the shocked wide-angle wind and the shocked ambient gas, a stationary shell is obtained, along which a mixing shear-layer with constant speed can develop. A more serious challenge for swept-up models driven by a wide-angle wind is the small radius of the H$_2$ conical cavity in DG Tau B at its base ($< 6$~au). The predicted radius reached by the cavity in the disk mid-plane {coincides with} the centrifugal disk radius \citet{lopez-vazquez2019,liang2020}, which in the case of DG~Tau~B is at least 750~au (see \citetalias{de2020alma}), much larger than the H$_2$ cavity footpoint radius $r_0 < 6$~au observed here.  

\citet{rabenanahary2022} study entrainment of a stratified static core by a narrow jet (instead of a wide-angle wind) and find it can produce stationary conical cavities with a sheared velocity field similar to the structure of the conical H$_2$/CO layer in DG Tau B. The opening angle and final width depends strongly on the ambient density stratification. Therefore,
such a model remains to be computed in the case of more realistic Class 1 environments (including a residual rotating infalling envelope) to see if it could also reproduce the narrow cavity base, large entrained mass-flux in CO and hot H$_2$ along the cavity walls, and large specific angular momentum observed in DG Tau B.

\subsection{Shocked disk-wind scenario}

An alternative possibility would be that the near-IR ro-vibrational H$_2$ is emitted in the shocked interaction layer driven into an outer disk wind by successive jet bow-shocks, or an inner wide-angle wind. Both models have been recently invoked to reproduce the low velocity SiO shells nested inside the outer SO disk wind in HH212 \citep{lee2021_accretion, lee2022_magnetocentrifugal}. This scenario differs from the traditional swept-up scenario discussed before in a fundamental way: the cavity in that case is not tracing entrained envelope material but shocked disk wind (plus jet or inner wide-angle wind) material, and the large measured mass-flux has a much stronger, direct implication for disk evolution. 

\citet{tabone2018} simulated numerically the long-term interaction of a time variable jet with a uniform and vertical disk wind. A conical layer forms at the interface by the stacking of multiple bow-shock wings.
This jet-disk wind interaction scenario was mentioned by DV22 as a possible alternative for the bright CO cone seen in DG Tau B, which is surrounded by a fainter and slower wind. Possible evidence in support of it was the presence of slight brightness enhancements "sub-structures" within the CO cone with spacings very similar to the typical spacing of jet knots along the axis, and with linear shapes (as inferred from their velocity gradients) suggestive of shock fronts that could be driven by the impact of recent bow-shock wings against the outer disk wind. The brighter H$_2$ emission at the base of the \mdy{red-shifted} cone \mdy{($< 350$~au)}, coincident with bright inner CO emission could result from the stronger impact of jet bow-shocks close to the source, where both the densities and shock speeds are expected to be higher. 

The side-to-side brightness asymmetries seen in the present H$_2$ NIRCam images \mdy{on the edges of the redshifted cone} (with some counterparts in CO, cf. Figs.~\ref{fig:NIRCam_images} and \ref{fig:NIRSpec_images}) might be additional possible evidence for a jet-disk wind interaction in this system. These asymetries would be naturally explained if the jet undergoes regular precession of its axis. The  jet precession angle is constrained at less than 1$^{\circ}$. However, \citetalias{de2022modeling} showed that a precession of 0.5$^{\circ}$ of the conical CO flow axis would be sufficient to reproduce the brightness enhancements seen in CO. Jet precession would not lead to direct impact of the jet on the cavity but rather induce asymmetric bow-shock wings impacting the cavity. However, simulations with more realistic disk wind geometry and kinematics are required for a detailed comparison to observations. 

This "shocked disk wind" scenario cannot be fully tested with the near-IR H$_2$ lines alone, as the allowed domain of possible shock conditions is too broad \citep{agra2014origin, kristensen2023}. Stronger constraints will be obtained from the combined analysis of NIRSpec and MIRI data including pure rotational H$_2$ lines, which will be presented in a future publication. 


\section{Conclusions}

We present in this paper JWST observations of the DG Tau B outflows with NIRCam and NIRSpec-IFU complemented by SINFONI/VLT data to constrain the kinematics. We focus our analysis on the red-shifted outflow cavity, which shows in ALMA observations properties compatible with a disk wind origin. Our conclusions can be summarized as follows:

\begin{itemize}

    \item The bipolar jet is detected in several atomic tracers ([C {\sc i}], [S {\sc ii}], [N {\sc i}], [P {\sc ii}], [O {\sc i}], He {\sc i}) but particularly in [Fe {\sc ii}]. The red-shifted jet $PA$ is accurately estimated at $295.0^\circ \pm 0.2^\circ$, in full agreement with the PA of the CO red-shifted cavity derived by \citetalias{de2022modeling}. We derive a jet full-opening angle of $2.9^\circ$ and an upper limit on the jet radius at origin of $r_{0,jet} \le 4$~au. The upper limit on the jet axis precession angle  is $\leq 1^{\circ}$. 

    \item  The H$_2$ images show two lobes separated by a dark lane of decreasing thickness with wavelength. The eastern approaching lobe shows a significant contribution from scattered light. As wavelength increases the peak continuum moves closer to the ALMA continuum peak, showing that the source is not seen directly at wavelengths shorter than 3 $\mu$m.  
    
    \item Estimate of the mass accretion rate onto the star from the dereddened luminosities of H{\sc i} transitions gives $\dot{M}_{acc} \simeq 10^{-10} \text{M}_\odot \text{yr}^{-1}$. This is 3 orders of magnitude smaller than expected from the jet mass-flux, \mdy{suggesting} that only a tiny fraction (0.1\%) of the H{\sc i} line flux reaches us through scattering. This finding clearly demonstrates that deriving mass accretion rates from dereddened H{\sc i} luminosities in embedded sources can be severely in error, in particular for highly inclined systems.
    
    \item  The H$_2$ red-shifted emission shows a narrow conical morphology with a deprojected semi-opening angle $9.4^\circ$. Extrapolating at the origin gives an H$_2$ radius on the disk mid-plane of $r_0 \leq 6$~au. The H$_2$ emission appears to coincide with the CO cavity inner edge. The atomic jet, the H$_2$ conical wind and the CO conical outflow are nested within each other and form a layered outflow structure. 
    
    \item From the SINFONI observations, we derive a constant vertical velocity $V_Z = 22.5$ \kms ~for H$_2$, a factor two larger than the average CO velocities. An upper limit on the H$_2$ specific angular momentum $J < 90$ au \kms is also derived.
   
    \item  A constant H$_2$ excitation temperature $T \simeq 2200$ K is derived inside the red-shifted cavity. The \mdy{average} column density of hot H$_2$ is $2 \times 10^{17} \text{cm}^{-2}$ close to the source, dropping to $5 \times 10^{16} \text{cm}^{-2}$ beyond $Z=400$~au. The mass flux in the form of hot H$_2$ shows two plateaus at $4.9 \times 10^{-11} \text{M}_\odot \text{yr}^{-1}$ close to the source and $1.4\times 10^{-11} \text{M}_\odot \text{yr}^{-1}$ beyond. If ro-vibrational H$_2$ is tracing a dense photodissociated layer, the actual mass-flux would be a thousand times larger.
      
\end{itemize}

The global layered H$_2$/CO structure observed in DG Tau B with temperature, velocity and collimation increasing towards the axis of the flow is suggestive of an MHD disk wind scenario. The  hot H$_2$ ($T \simeq 2000$~K) could trace the inner photodissociation layer as predicted by recent simulations. 
Assuming the same wind magnetic lever arm as for the rotating CO conical outflow ($\xi \simeq 1$, \citetalias{de2022modeling}), an H$_2$ launching region at disk radii \mdy{0.2-0.4}~au would account for both the poloidal velocity and the total mass-flux. 
Alternatively, the large jump in temperature between the H$_2$ and CO layers together with the small thickness of the H$_2$ cavity and the striking conical geometry might be suggestive of a shocked interface driven by successive jet bow-shocks into an outer disk wind \citep{tabone2018} or the envelope \citep{rabenanahary2022}. Dedicated simulations including a more realistic disk wind or infalling envelope will be necessary to test these alternatives.
Additional constraints on the H$_2$ \mdy{excitation processes and} cavity formation mechanism will be possible by analyzing the MIRI pure rotational mid-IR H$_2$ transitions, which will be presented in a future publication. The present work clearly demonstrates the unique power of JWST for studies of disk winds in young stars.

\begin{acknowledgements}
This work is based on observations made with the NASA/ESA/CSA James Webb Space Telescope. The data were obtained from the Mikulski Archive for Space Telescopes at the Space Telescope Science Institute, which is operated by the Association of Universities for Research in Astronomy, Inc., under NASA contract NAS 5-03127 for JWST. These observations are associated with program \#1644 (PI: C. Dougados). This work strongly benefited from the Core2disk-III residential program of Institut Pascal at Universit\'e Paris-Saclay, supported by the program “Investissements d’avenir” ANR-11-IDEX-0003-01. This work was supported by the "Programme National de Physique Stellaire" (PNPS) and the Programme National “Physique et Chimie du Milieu Interstellaire” (PCMI) of CNRS/INSU co-funded by CEA and CNES. \mdy{LP acknowledges the project PRIN MUR 2022 FOSSILS (Chemical origins: linking the fossil composition of the Solar System with the chemistry of protoplanetary disks, Prot. 2022JC2Y93).}
\end{acknowledgements}

\bibliographystyle{aa}
\bibliography{references}

\begin{appendix}

\section{Correction of 1/f correlated noise in NIRSpec and NIRCam data}
\label{appendix:noise_bgd}
All JWST instruments, i.e. NIRISS, MIRI, and in particular NIRSpec and NIRCam, are connected to a pixel reading device called SIDECAR ASIC. As a result, all the detectors on all the instruments suffer from correlated 1/f read-out noise \citep{rauscher2011}. It is quite complicated to fully correct the contribution of this noise in the images due to the stochastic mechanisms at its origin, but we propose a simple method that significantly reduces its impact on both NIRCam and NIRSpec datasets.

The most obvious property of 1/f noise in images is the appearance of "banding". These bands stand out from the standard dispersion of the image background. For example, in the case of NIRSpec, the frames of the NRS1 and NRS2 detectors are covered by vertical "bands". The method used here applies to both NIRSpec and NIRCam observations and works on the output files from step 1 of the JWST pipeline (\texttt{Detector1}) named with the suffix "\_rate". Applying this method does not modify the file structure in any way. It is then possible to restart the rest of the pipeline using these corrected files as input files for the "Spec2" or "Image2" stage.

\begin{figure}[h]
\begin{minipage}{\textwidth}
    \centering
    \includegraphics[width=\textwidth]{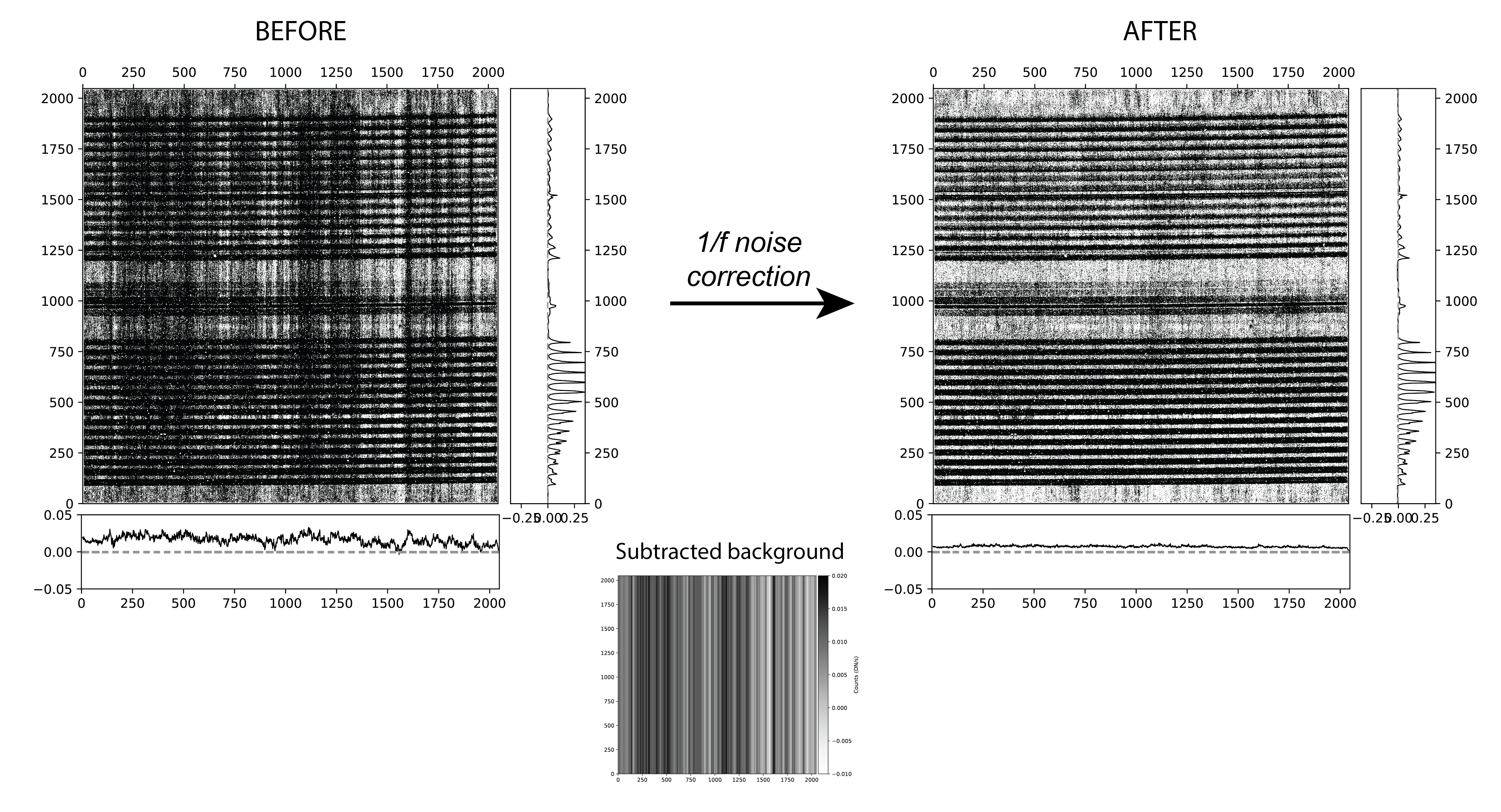}
    \caption{Illustration of the 1/f noise correction. Example of an image from NIRSpec NRS2 detector before (left) and after (right) correction for 1/f correlated noise. The box to the right of each image plots the median values of each horizontal slit in the image. The boxes at the bottom show the median values of each vertical slit in the image. The smaller image in the middle corresponds to the map of mean values subtracted from the initial image.}
    \label{fig:noise_correction}
\end{minipage}
\end{figure}

\newpage

To correct the banding effect, an average value is estimated for each slit in the image. This average value is estimated by applying a $\sigma$-clipping method. The $\sigma$ value is estimated from the histogram of the image pixel distribution. A clipping value of 1.5$\sigma$ has been found optimum and was applied to all the images presented in this article. 

Fig.~\ref{fig:noise_correction} shows an example of the results obtained for the NIRSpec IFU NRS2 detector in the F100LP/G140H configuration. The left image shows a raw image output by \texttt{Detector1} before correction, with the characteristic "bands" of 1/f noise. The second figure shows the image obtained after noise correction. The subtracted flux values used are shown in the smaller middle image. It can be seen that the bands are for the most part relatively well subtracted, increasing the SNR of the spectra present in the image. We can also see that the horizontal dispersion of the image is reduced by the correction, without altering the average  value of the vertical slits. 

The routine is publicly available at: https://github.com/delabrov/JWST-Background-Noise-Removal




\clearpage
\newpage

\clearpage
\newpage

\section{SINFONI precise wavelength calibration method using atmospheric lines}
\label{appendix:wvs_calibration}

\begin{figure}[h]
    \resizebox{\hsize}{!}{\includegraphics{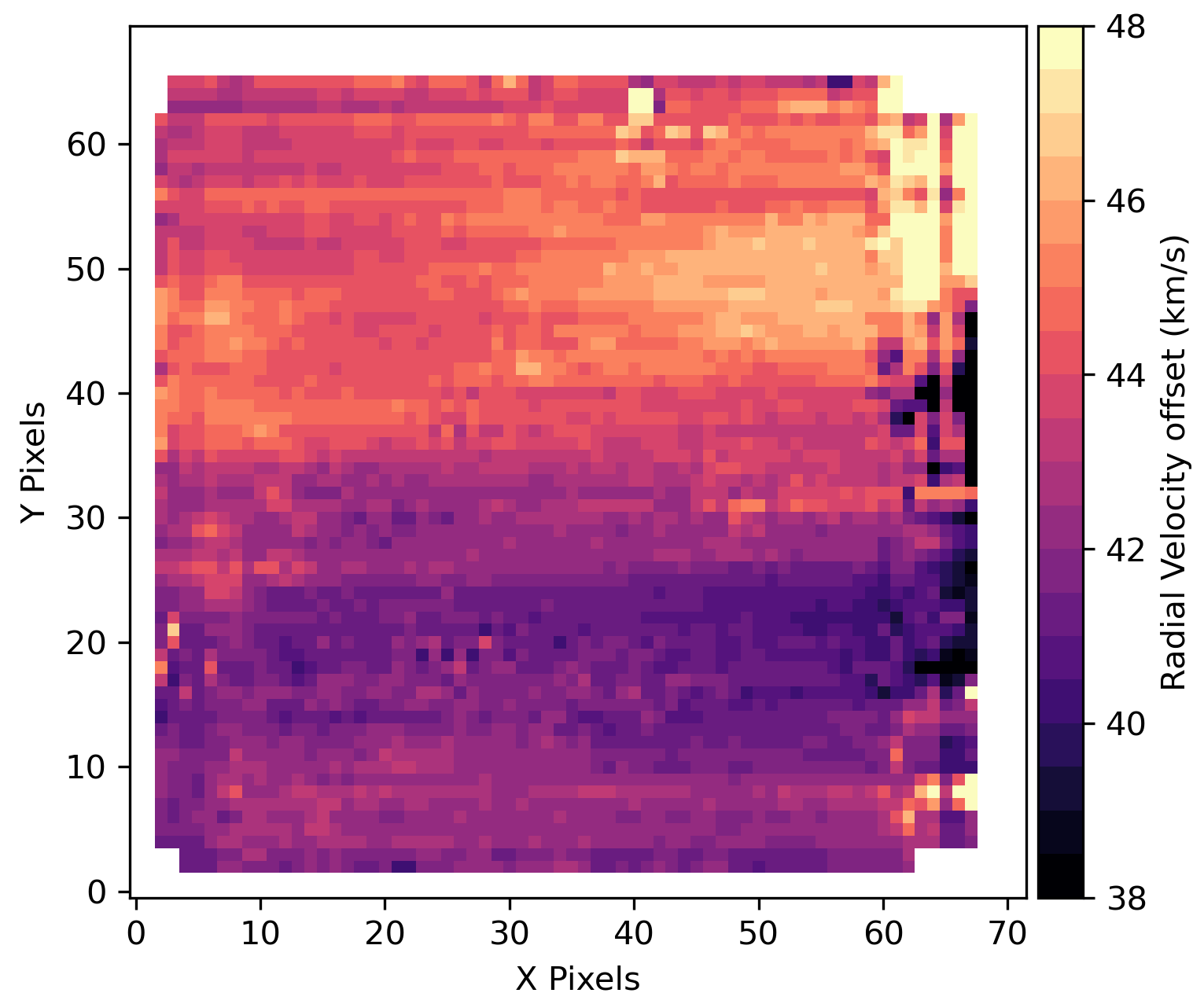}}
    \caption{Map of radial velocity offsets in the SINFONI data cube. The values are estimated via the cross-correlation method over a spectral range between 2.10 and 2.17$\mu$m.}
    \label{fig:RV_offsets_map}
\end{figure}

In the SINFONI observations, after processing with the reduction pipeline, a systematic offset in radial velocity ($RV$) of $\sim 40$\kms is observed in the OH telluric emission lines in all the spaxels. Since the aim of the SINFONI observations is to study the kinematics of the H$_2$ ejecta, we perform a spectral recalibration of the datacube 
in order to constrain the H$_2$ radial velocities more precisely.  

The method used is based on the cross-correlation between the individual spectra of the data cube and a theoretical atmospheric spectrum containing the telluric lines emitted in the spectral band of our observations. This atmospheric model is obtained from ESO's \textit{Sky Model Calculator}\footnote{https://www.eso.org/observing/etc/bin/gen/form?INS.MODE=\\swspectr+INS.NAME=SKYCALC} tool by specifying the date and conditions of the  observations. This analysis is applied to the data cube
reduced with the pipeline but deactivating sky correction so that OH sky emission lines are preserved. The cross-correlation routine used on our data is taken from the TexTRIS python module \citep{bonnefoy2014}, a package tailored to the  processing of raw SINFONI data. 

\begin{figure}[h]
    \centering
    \resizebox{\hsize}{!}{\includegraphics{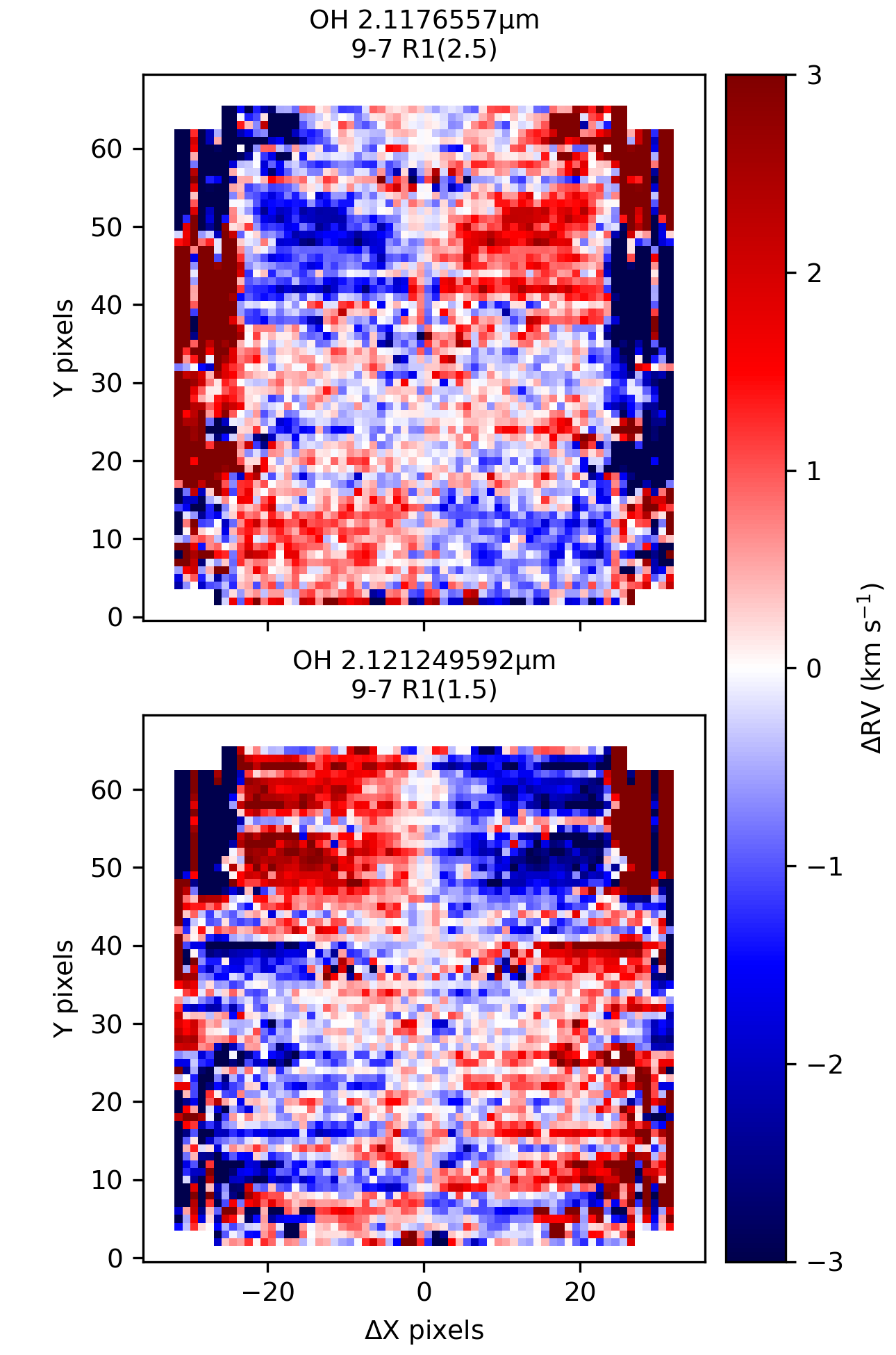}}
    \caption{Top panel: Radial velocity differences $\Delta RV$ map of the 9-7 R1(2.5) atmospheric OH line. Bottom panel: Same map for the 9-7 R1(1.5) atmospheric OH line. The X $=0$ axis defines the red-shifted outflow axis in this figure. $\Delta RV$ are constructed by computing the difference between RV values at symmetric X distances from the outflow axis. The red-shifted outflow is located in the bottom part of these maps (Y $ < 30$).}
    \label{fig:dRV_OH}
\end{figure}

The cross-correlation between the current spectrum and the reference atmospheric spectrum is computed at each spaxel position of the datacube. The cross-correlation is sampled every 0.2 \kms. Its maximum gives the estimated radial velocity correction RV to be applied at the current spaxel position.  The TexTRIS routine returns the SNR associated with the cross-correlation curve \citep{houlle2021}, which makes it possible to estimate offset errors via the relation:

\begin{equation}
    err(RV) = \frac{\sigma}{\text{SNR}}
\end{equation}

$\sigma$ the width of the cross-correlation profile estimated by a Gaussian fit. The derived RV offsets map is shown in Fig.~\ref{fig:RV_offsets_map}. The values vary between +36 and +50 \kms. We observe a vertical gradient, corresponding to the Uneven Slit Illumination Effect (USIE) \citep{marconi2003}. This effect occurs when strong gradients of illumination across the SINFONI slitlet direction are present \citep{erkal2021}. In our case, the pattern reflects the non-uniform illumination due to the DG Tau B continuum emission. Horizontal gradients are also visible along each slitlet direction. These may be due to systematic errors in the spectrum extraction from the detector or to horizontal variations in illumination, which generates a second USIE. 

To improve the SNR on the RV correction map, we fit the horizontal gradients with a polynomial function of order between 1 and 3, depending on  the variation observed. The resulting fit values are used to build the RV correction map. This RV correction map has been applied to produce the H$_2$ centroid velocity map presented in Fig.~\ref{fig:velocity_map}.

The average resulting fit RMS over the entire frame is 0.35 \kms, which corresponds to $\sim$1/100 of the spectral sampling. This corresponds to a typical relative error between spectra on the same SINFONI slitlet direction.


To estimate an absolute error of the RV offsets, we perform the cross-correlation at each spaxel position over several small spectral intervals. This allows us to estimate the variation of the $RV$ offset as a function of the central wavelength of the correlation interval. The  derived $\Delta \lambda$ offsets are constant with wavelength, which results in $RV$ offsets decreasing linearly with wavelength. By averaging the RMS values of $\Delta \lambda$ over all the spaxels of the cube, we estimate an absolute error on the recalibration of $3.5 \times 10^{-6} \mu$m, which corresponds to a $RV$ offset error equal to $\sim 0.5$~\kms at 2.12$\mu$m. Combining relative and absolute errors we estimate a final uncertainty of 0.6~\kms (1 $\sigma$) in the global velocity calibration at each spaxel, averaged over all OH lines used for the correlation. 

 However, this method derives average RV offset values over a wavelength range. 
By analysing the residual radial velocities of two atmospheric OH transitions adjacent to the 1-0S(1) H$_2$ line, we show that this procedure does not fully correct the offsets at a given wavelength. 

Figure~\ref{fig:dRV_OH} shows the $RV$ offsets difference maps for two OH transitions close to 2.12$\mu$m. The $\Delta RV$ values are computed as differences between the two $RV$ values at symmetrical distances with respect to the H$_2$ red-shifted outflow axis. Residual horizontal velocity gradients are still present that limit our ability to detect rotation in the DG Tau B H$_2$ flow.
 In the region where the H$_2$ red-shifted cavity is located ($\text{Y} < 30$), there are $\Delta RV$ gradients in opposite directions between the two maps, on average around 1~\kms and going up to 3~\kms. Therefore, we estimate a conservative 3$\sigma$ upper limit on $\Delta RV$ of $\sim 3$~\kms~ in the DG Tau B red-shifted H$_2$ cavity.

\section{H$_2$ red-shifted cavity thickness}
\label{appendix:cavity_thickness}

\begin{figure}[h]
    \resizebox{\hsize}{!}{\includegraphics{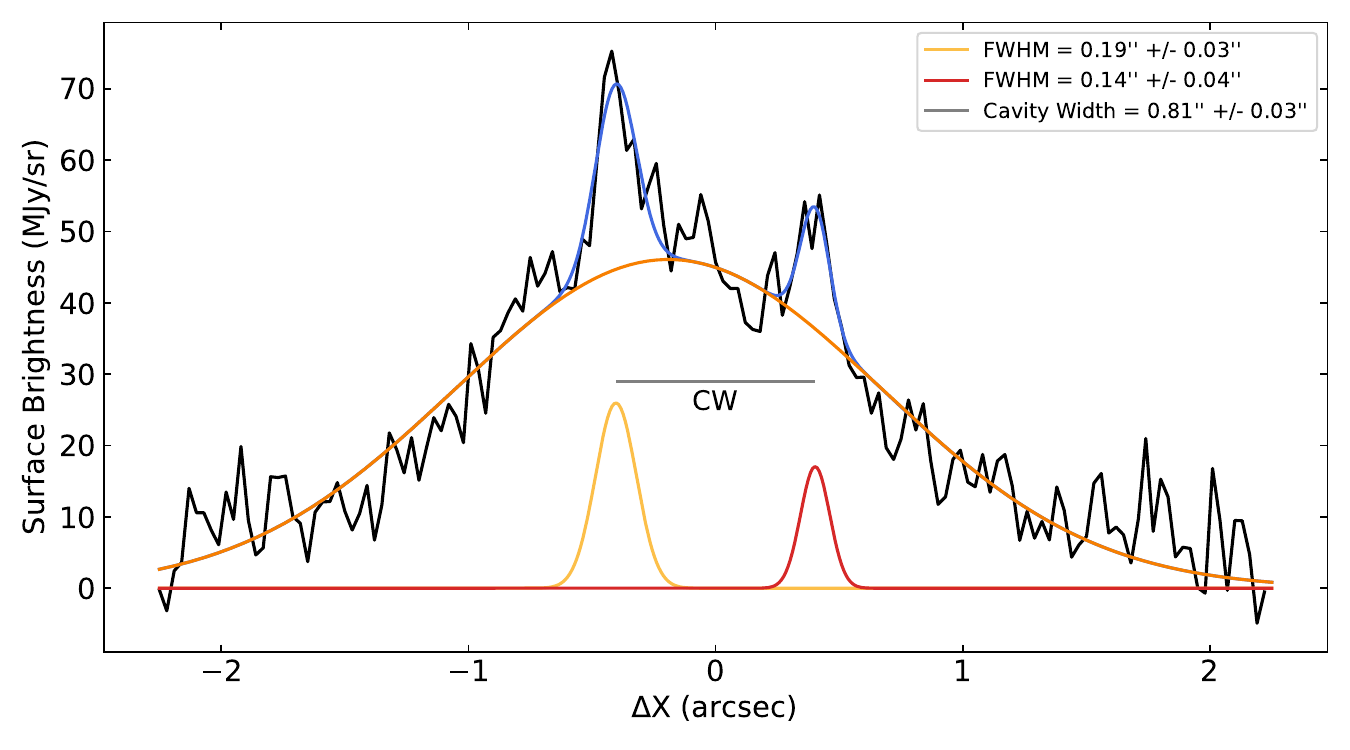}}
    \caption{Measure of the H$_2$ cavity thickness from the NIRCam images. The black curve shows the transverse intensity profile perpendicular to the axis of the red-shifted H$_2$ cavity measured on the \mdy{NIRCam} F212N image at $\Delta Z = 1.5^{\prime\prime}$. The blue curve corresponds to the fit of the profile by a combination of 3 Gaussians. The red and yellow curves are the individual Gaussians fitting the two intensity peaks associated with the bright edges of the cavity. The orange curve corresponds to the fit of the wide baseline component. The grey line, called "CW" for Cavity Width, shows the separation between the two Gaussian center positions.}
    \label{fig:relative_thickness}
\end{figure}

The edges of the H$_2$ conical emission in red-shifted lobe appear marginally resolved in the NIRCam F212N images. To estimate their thickness, we construct an intensity profile perpendicular to the cavity axis at $Z=1.5^{\prime\prime}$ from the central source. In order to increase the SNR, we summed the profiles over 5 pixels along the flow axis. We checked that this averaging process does not overestimate the thickness by more than 10\% due to the conical opening of the cavity. The profile obtained is shown in Fig.~\ref{fig:relative_thickness}. The profile was fit by a function composed of three Gaussians in order to match the wide extended component and the two peaks associated with the bright edges. We derive a H$_2$ cavity thickness of $0.16^{\prime\prime}\pm 0.04^{\prime\prime}$ (deconvolved quadratically from the FWHM of the PSF\footnote{https://jwst-docs.stsci.edu/jwst-near-infrared-camera/nircam-performance/nircam-point-spread-functions}, equivalent to 0.072$^{\prime\prime}$ in the F212N filter.), corresponding to 22$\pm$5 au. 
Unfortunately, it was not possible to estimate the thickness of the H$_2$ edges at larger distances from the source due to significantly lower SNR in the transverse intensity profiles.

\section{Accretion rate onto the star from de-reddening of H{\sc i} lines}
\label{appendix:accretion_rate}

The accretion rate is estimated using 4 H{\sc i} transitions, those with the best SNR in the spectra: Pa$\delta$ 7-3 at 1.00$\mu$m, Pa$\gamma$ 6-3 at 1.09$\mu$m, Pa$\beta$ 5-3 at 1.28$\mu$m and Br$\gamma$ at 2.16$\mu$m. The fluxes considered for estimating the accretion luminosities are obtained in a circular aperture of diameter 1.2$^{\prime\prime}$ centred in the blue lobe and containing the central source position. This aperture encloses all H{\sc i} Br$\gamma$ flux \mdy{above the 3$\sigma$ detection limit}. The observed fluxes are listed in Table~\ref{tab:accretion_rate}, together with the parameters $a$ and $b$ of the empirical relations in \citet{alcala2017x} of the shape: $\log (L_{acc} / L_\odot) = a\log (L_{HI} / L_\odot) + b$, where $L_{HI}$ is the de-reddened luminosity of each line. 
{By using 4 H{\sc i} transitions at different wavelengths, and the extinction curve of \citet{wang2019optical}, we can find the value of $A_V$ that gives the same $L_{acc}$ for the 4 derredened line fluxes.} 
The accretion rate is then deduced from the accretion luminosity by the relation $\dot{M}_{acc} = L_{acc} R_* / (G M_*)$ considering the DG Tau B parameters as follows: $R_* = 2 R_{\odot}$ and $M_* = 1.1 M_{\odot}$ \citepalias{de2020alma}. 

\begin{table}[h!]
    \centering
    \caption{H{\sc i} transitions used in accretion rate estimation. Observed flux values are given in $10^{-16}$ $\text{erg} ~ \text{s}^{-1} \text{cm}^{-2}$. The coefficients $a$ and $b$ appear in the relations $\log (L_{\rm acc} / L_\odot) = a\log (L_{\rm HI} / L_\odot) + b$ as determined by \citet{alcala2017x} (see text).}
    \begin{tabular}{c|c|c|c}
        H{\sc i} Transition & Flux & $a$ & $b$\\
        \hline
        \hline
        Pa$\delta$-1.00$\mu$m & 1.71$\pm$0.18 & 1.22$\pm$0.09 & 3.74$\pm$0.43\\
        Pa$\gamma$-1.09$\mu$m & 3.97$\pm$0.13 & 1.24$\pm$0.06 & 3.58$\pm$0.27\\
        Pa$\beta$-1.28$\mu$m & 10.1$\pm$0.38 & 1.06$\pm$0.07 & 2.76$\pm$0.34\\
        Br$\gamma$-2.16$\mu$m & 11.8$\pm$0.28 & 1.19$\pm$0.10 & 4.02$\pm$0.51\\
    \end{tabular}
    \label{tab:accretion_rate}
\end{table}

\begin{figure}[h!]
\centering
    \resizebox{\hsize}{!}{\includegraphics{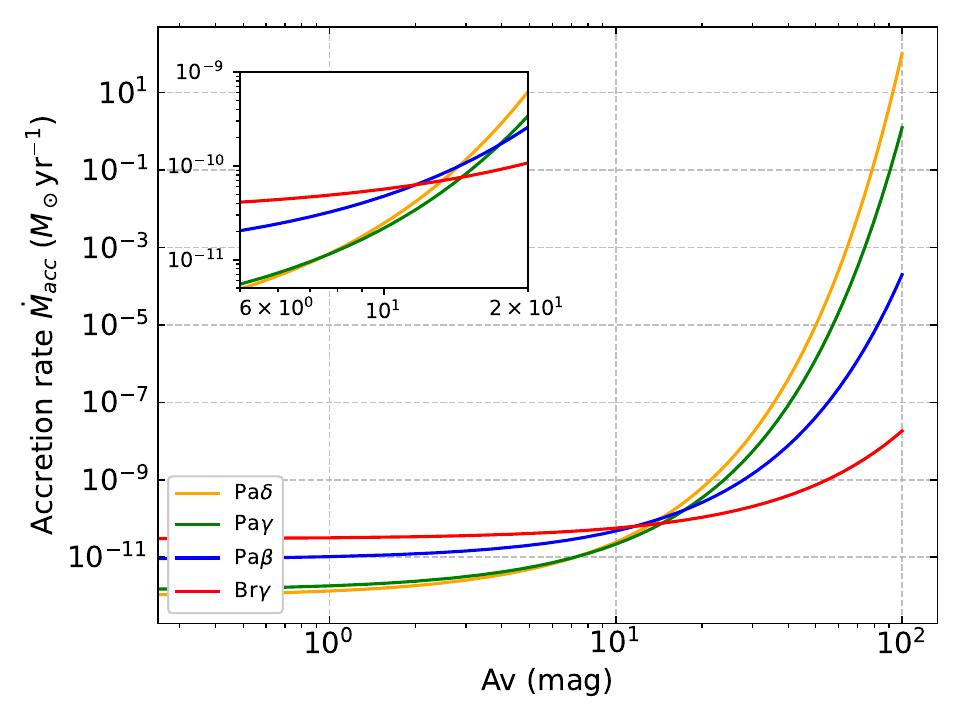}}
    \caption{Accretion rate $\dot{M}_{acc}$ as a function of extinction $A_V$ reproducing the observed fluxes of the atomic transitions H{\sc i} Pa$\delta$, Pa$\gamma$, Pa$\beta$ and Br$\gamma$. The curves are derived from the empirical relations between $L_{acc}$ and $L_{HI}$ of \citet{alcala2017x}. The intersection of the four curves gives the best estimate of $\dot{M}_{acc}$ and $A_V$ towards DG Tau B.}
    \label{fig:MaccVsAv}
\end{figure}

Figure~\ref{fig:MaccVsAv} shows the estimated accretion rates from the 4 H{\sc i} lines considered as a function of the extinction $A_V$. The curves intersect at approximately the same values: $A_v = 13 \pm 4$ for $\dot{M}_{acc} = (1.0 \pm 0.5) \times 10^{-10} \text{M}_{\odot} \text{yr}^{-1}$. The errors correspond to the average dispersion of the values for each intersection of two curves. 

\section{Extinction caused by the CO conical outflow and an envelope in free-fall}
\label{appendix:CO_extinction}

From the analysis of the excitation diagrams in Sect. \ref{sect:excitation_conditions} we estimate the extinction between us and the H$_2$ emission along the red-shifted lobe axis. The aim here is to compare to the extinction due to the CO conical outflow, and to an envelope in free-fall. 

\subsection{CO conical outflow}

\citet{de2020alma} estimates a constant total mass flux along Z in the CO red-shifted conical outflow of $2.2\pm 0.7 \times 10^{-7} M_{\odot} yr^{-1}$. From this value and considering the average CO ejection velocity, we can estimate the mass in a slice of thickness $\Delta Z$ as : 

\begin{equation}
    \begin{split}
    M_{tot} & = \dot{M}_{tot} \times t_{cross} = \dot{M}_{tot} \times \left( \frac{\Delta Z_{real}}{V_{Z,real}} \right) \\
    & \simeq \pi (R^2_{out} - R^2_{in}) \times \rho \times \Delta Z_{real}
    \end{split}
\end{equation}

$V_{Z,real}$ is the CO deprojected velocity and $R_{out}$ and $R_{in}$ are respectively the radii associated with the edges of the CO conical flow at a given Z position. Figure~\ref{fig:schemaAvCO} illustrates the considered geometry and the definition of $R_{in}$ and $R_{out}$. The subscript "$real$" means that the quantities are deprojected, considering the angle of inclination $i = 63^{\circ}$ of the DG Tau B flow axis with respect to the line of sight, as opposed to projected quantities annotated with "$proj$". From these quantities we can derive the hydrogen column density $N_H$ as:

\begin{equation}
    \begin{split}
    N_H & = \frac{\rho (R_{out} - R_{in})}{\mu m_H} \\
    & = \frac{M_{tot}}{\pi (R_{out} + R_{in}) \times \mu m_H \times \Delta Z_{real}} \\
    & = \frac{\dot{M}_{tot}}{\pi (R_{out} + R_{in}) \times \mu m_H \times V_{Z,real}}
    \end{split}
\end{equation}

And the projected values of $Z$ and the radii are defined as:

\begin{figure}[h]
\centering
    \resizebox{0.7\hsize}{!}{\includegraphics{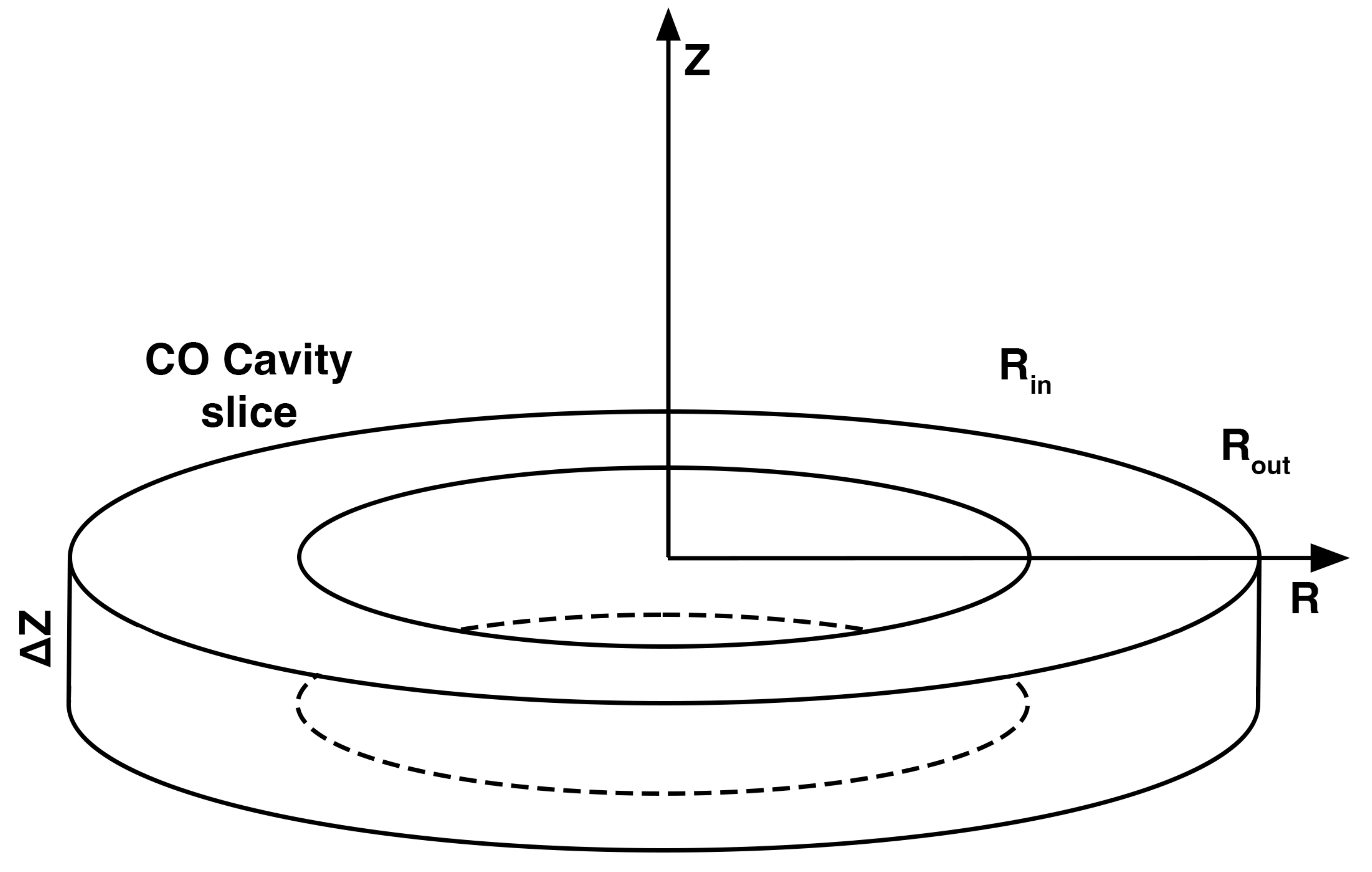}}
    \caption{Illustration of a slice of the CO red-shifted conical outflow perpendicular to its axis. If the slice is sufficiently thin, it can be approximated as a hollow cylinder of thickness $\Delta R = R_{out} - R_{in}$.}
    \label{fig:schemaCOslice}
\end{figure}

\begin{figure}[h]
    \centering
    \resizebox{0.8\hsize}{!}{\includegraphics{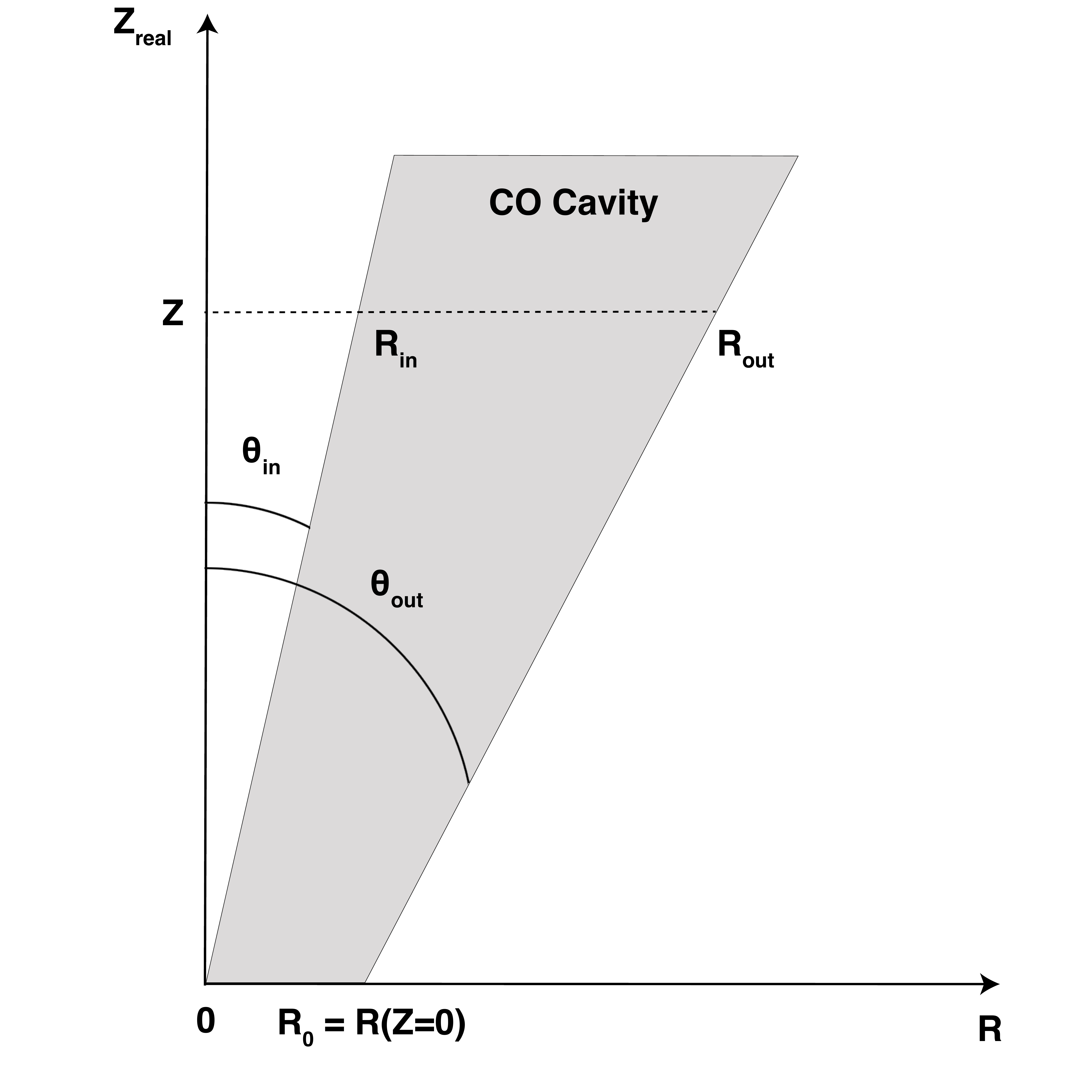}}
    \caption{Cutout in the ($R,Z$) plane of the CO conical red-shifted outflow defined by its two opening angles $\theta_{in}$ and $\theta_{out}$ and having the parameter $R_0$ as its external anchoring radius in the disk. These parameters are as derived in \citetalias{de2022modeling}.}
    \label{fig:schemaAvCO}
\end{figure}

\begin{figure}[h]
    \resizebox{\hsize}{!}{\includegraphics{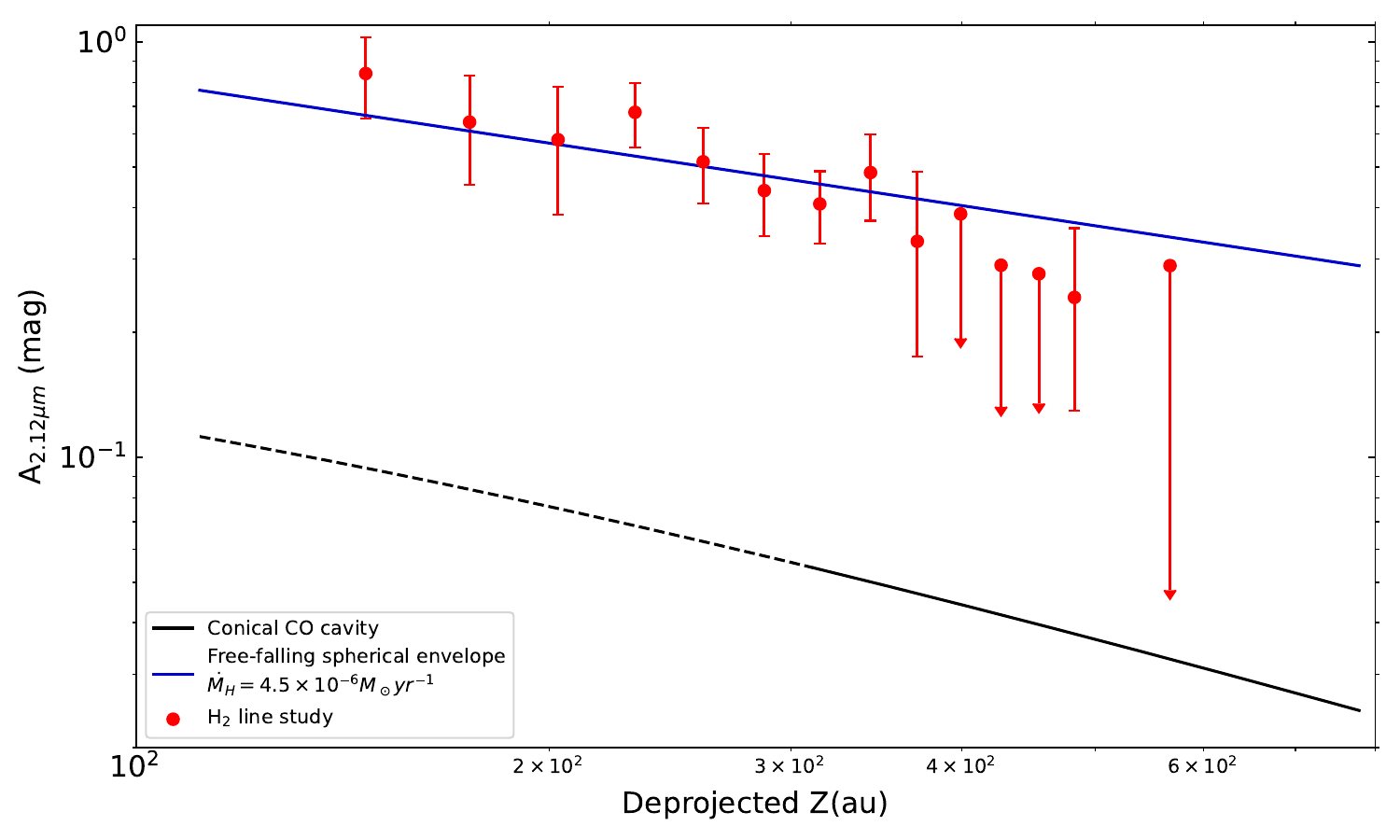}}
    \caption{Predicted reddening variation along the DG Tau B redshifted lobe for the conical CO outflow and envelope. The red points are the extinction values estimated from the H$_2$ ro-vibrational transitions in Sect.~\ref{sect:excitation_conditions}. The black curve shows the expected extinction at 2.12$\mu$m as a function of distance from the source due to the CO conical red-shifted outflow using the mass flux estimated in \citetalias{de2020alma}. The black dashed curve represents the extrapolation of the model below 300~au assuming a constant CO mass flux. The blue line corresponds to the extinction profile for a spherical envelope in free fall with an infall rate of $4.5 \times 10^{-6} \text{M}_\odot \text{yr}^{-1}$.}
    \label{fig:CO_extinction}
\end{figure}

\begin{equation}
    Z_{real} = Z_{proj} / \sin{i}
\end{equation}
\begin{align}
    &R_{in} = \tan{\theta_{in}} \times Z_{real} \quad \mathrm{and} \\ 
    &R_{out} = R_{out}(Z=0) + \tan{\theta_{out}} \times Z_{real}
\end{align}

$R_{in}$ corresponds to the extrapolated internal radius of the CO conical outflow, defined from the internal deprojected opening angle $\theta_{in}$ equal to $12.8^{\circ}$ and $R_{out}$ being the external radius of the conical outflow, using the deprojected external opening angle $\theta_{out}$ of $16.5^{\circ}$ ($R_{in}(Z=0)=0$ and $R_{out}(Z=0)= 40$~au, values taken from Fig.~9 of \citetalias{de2022modeling}).

From this column density profile $N_H(Z)$, we can estimate the extinction associated with the CO conical flow using the \citet{guver2009} relation: $N_H(\text{cm}^{-2}) = (2.21 \pm 0.09) \times 10^{21} A_V$. The extinction profile obtained is shown in Fig. \ref{fig:CO_extinction}. The profile is inversely proportional to $Z$ (due to the fact that $R \propto Z$ since the cavity is considered to be conical). Extinctions are given at 2.12$\mu$m, which is more consistent since we are working in the near infrared. The conversion between $A_V$ and $A_{2.12\mu m}$ is made via the power law given  in Equ. \ref{eq:power_law_extinction}, from \citet{wang2019optical}. 

\subsection{Envelope in free-fall}

In the same way, we  estimate the extinction caused by a spherical envelope free-falling towards the star. The envelope mass infall rate is given by: 
\begin{equation}
    \dot{M}_{ff} = 4 \pi R^2 \rho(R) \times V_{ff}(R)
\end{equation}
where the gas free-fall velocity is given by:
\begin{equation}
    V_{ff} = \sqrt{\frac{2 G M_*}{R}}.
\end{equation}
The column density of atomic hydrogen is then expressed as: 
\begin{equation}
    N_H(R) \sim \frac{2 R \times \rho(R)}{\mu m_H} = \frac{\dot{M}_{ff}}{2 \pi \mu m_H} \frac{1}{\sqrt{2 G M_* R}}
\end{equation}
\begin{equation}
    N_H(R) \sim 6.2 \times 10^{20} \times \left( \frac{\dot{M}_{ff}}{10^{-7} \text{M}_\odot \text{yr}^{-1}} \right) \times \left( \frac{R}{100 \text{ au}} \right)^{-1/2} \text{ cm}^{-2}
\end{equation}
Using the relation $N_H(\text{cm}^{-2}) = (2.21 \pm 0.09) \times 10^{21} A_V$ \citep{guver2009}, we derive the following relation:
\begin{equation}
\label{eq:Av_infall_envelope}
    A_V \sim 0.29 \times \left( \frac{\dot{M}_{ff}}{10^{-7} \text{M}_\odot \text{yr}^{-1}} \right) \times \left( \frac{R}{100 \text{ au}} \right)^{-1/2} \text{ mag}
\end{equation}
Finally, we convert to $A_{2.12\mu m} \simeq A_V/16.3$ (see Eq. \ref{eq:power_law_extinction}).
Figure~\ref{fig:CO_extinction} plots the profile that best reproduces the extinction values derived from the H$_2$ line ratios, the derived infall rate is $\dot{M}_{ff} = 4.5 \times 10^{-6} \text{M}_\odot \text{yr}^{-1}$. 


\clearpage
\section{Excitation Diagrams}
\label{appendix:excitation_diagram}

\begin{figure}[h]
\begin{minipage}{0.99\textwidth}
\centering

\begin{minipage}{0.5\hsize}
  \centering
  \includegraphics[width=\hsize]{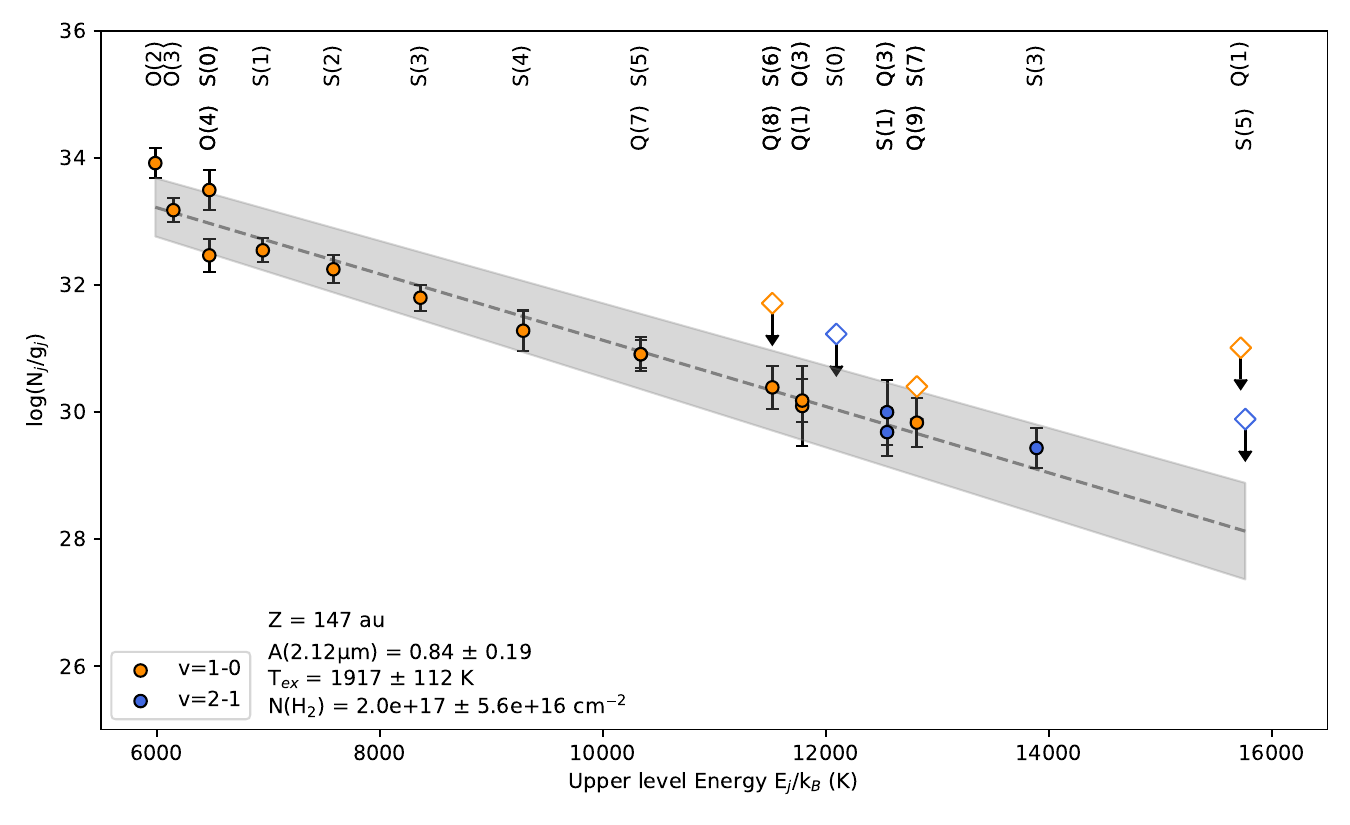}
\end{minipage}%
\begin{minipage}{.5\hsize}
  \centering
  \includegraphics[width=\hsize]{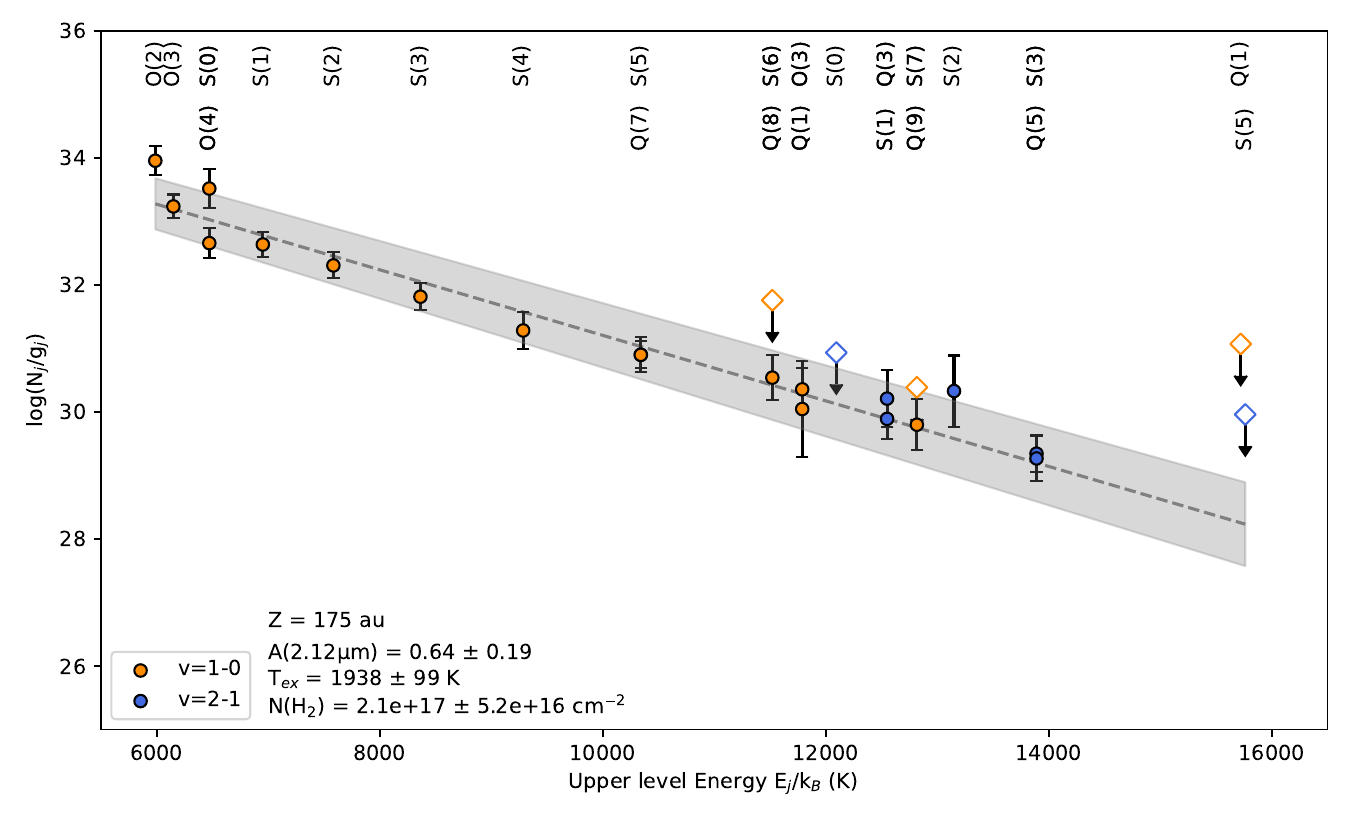}
\end{minipage}

\begin{minipage}{.5\hsize}
  \centering
  \includegraphics[width=\hsize]{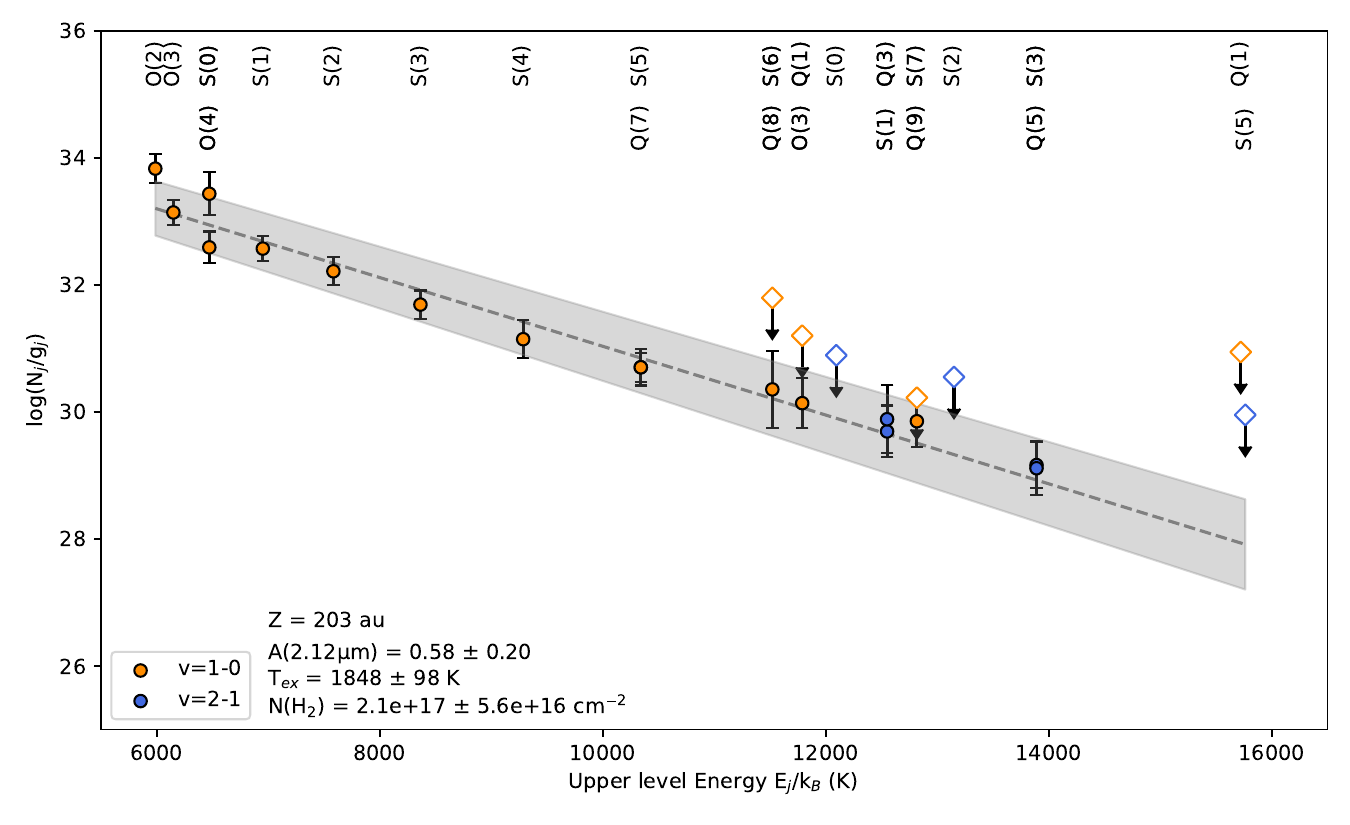}
\end{minipage}%
\begin{minipage}{.5\hsize}
  \centering
  \includegraphics[width=\hsize]{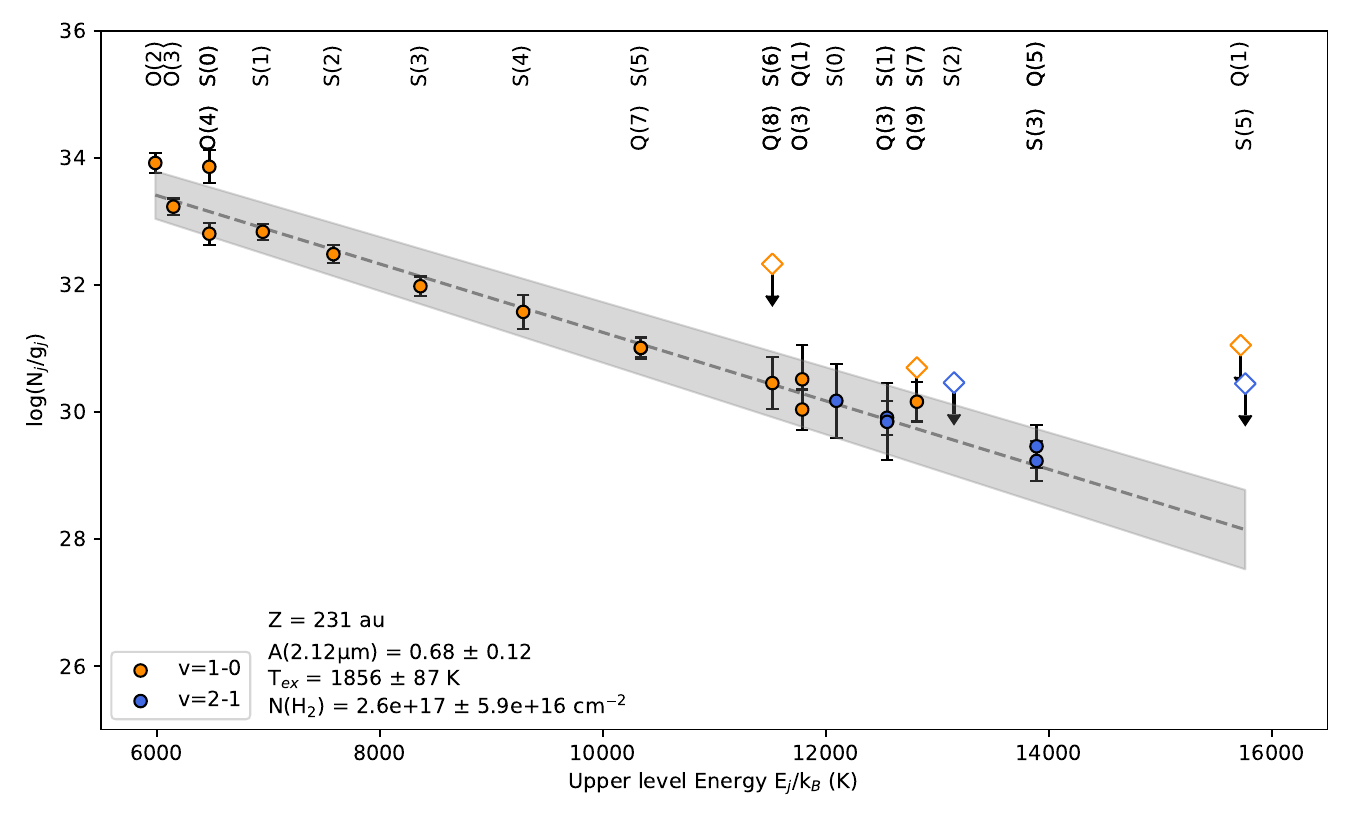}
\end{minipage}

\begin{minipage}{.5\hsize}
  \centering
  \includegraphics[width=\hsize]{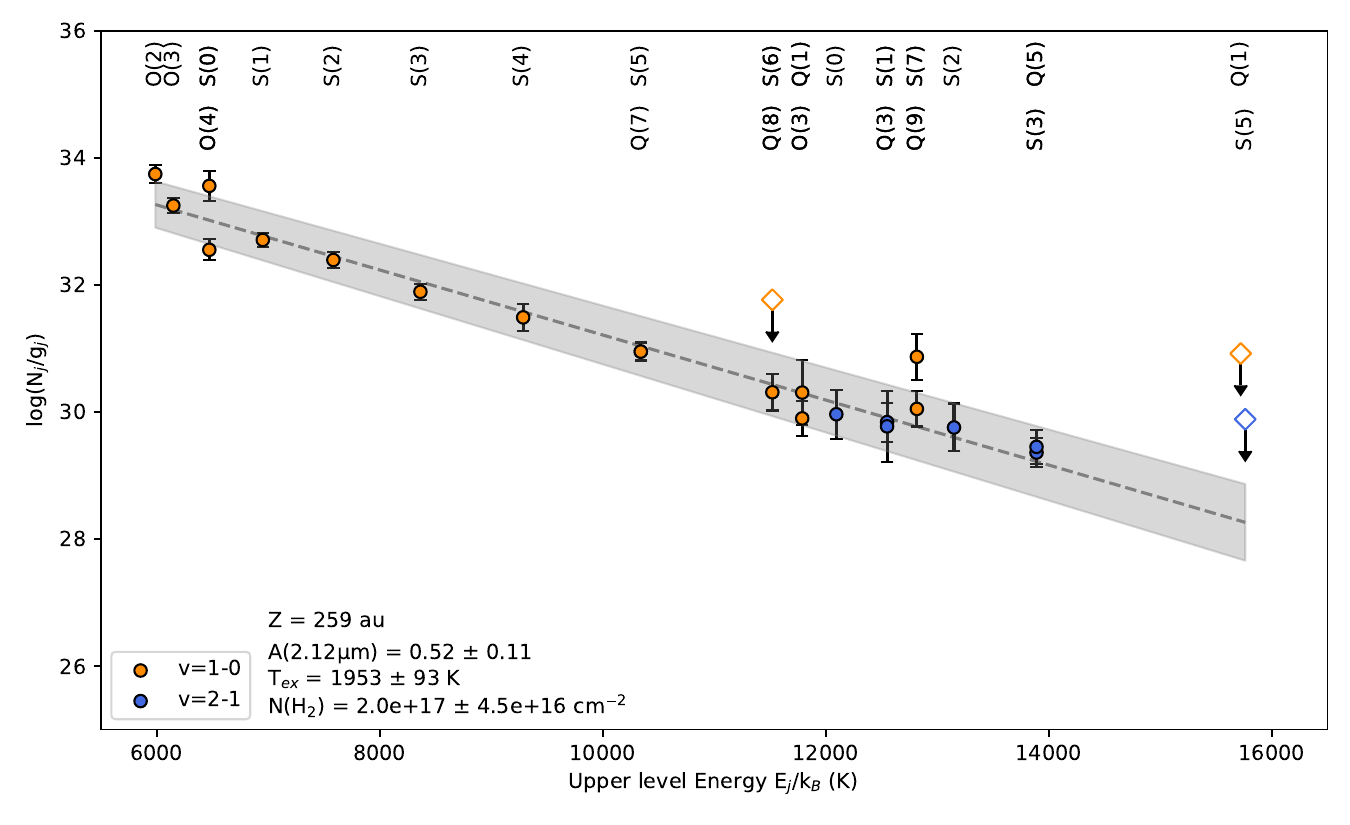}
\end{minipage}%
\begin{minipage}{.5\hsize}
  \centering
  \includegraphics[width=\hsize]{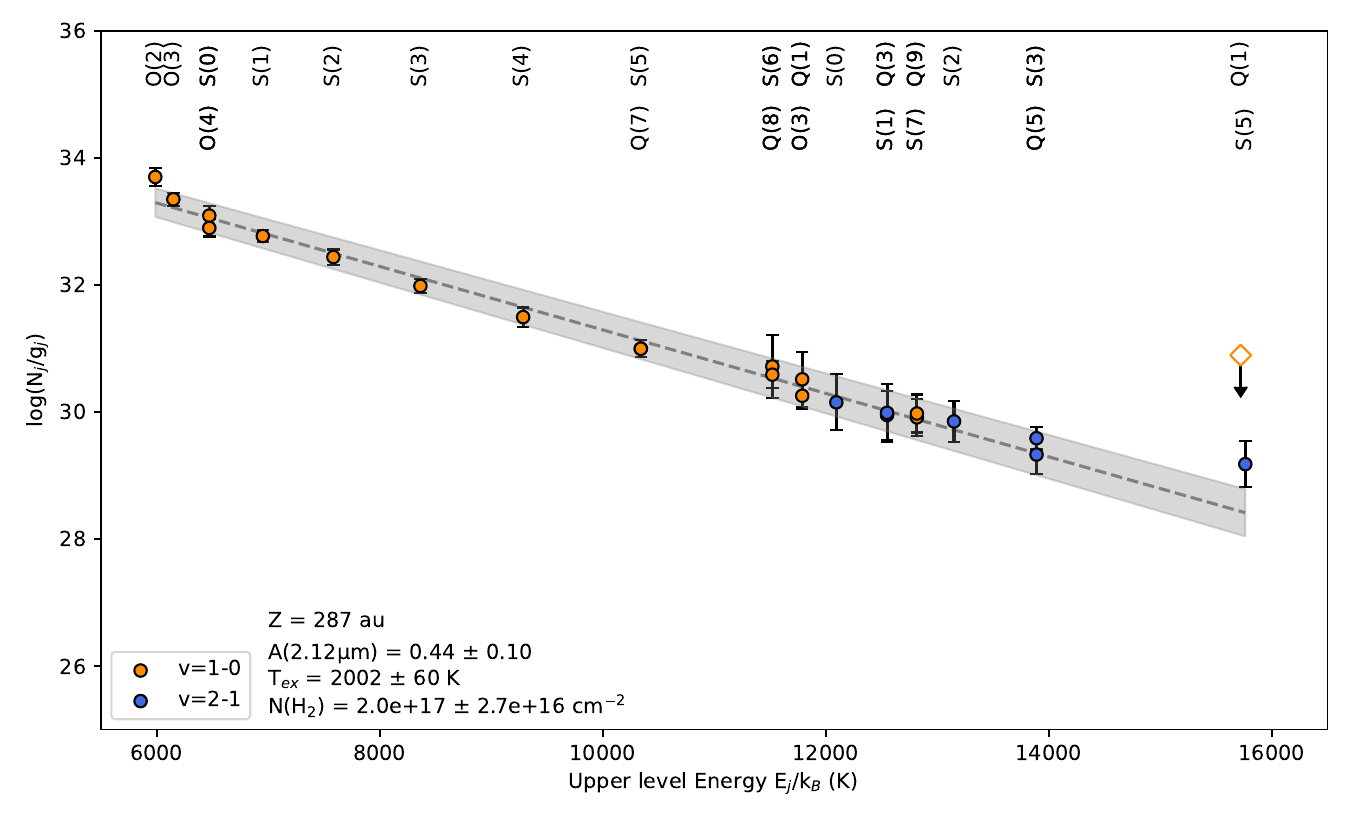}
\end{minipage}

\caption{H$_2$ excitation diagrams at various deprojected distances $Z$ along the red-shifted lobe. They are obtained in slits of width $\Delta Z = 0.2^{\prime\prime}$ ($= 28$~au). Rotational transitions are labeled at the top of each diagram and were de-reddened using the mean extinction $A_{2.12\mu\text{m}}$ estimated from the 1-0Q(7)/1-0S(5) line ratio. $v=1$ levels are plotted in orange and $v=2$ in blue. \mdy{The opened diamond symbols represent the 3$\sigma$ upper limits of undetected H$_2$ transitions. Error bars are at 1$\sigma$.} The grey dashed line is the best fit to the points and the grey area shows the 1$\sigma$ range. The text at the bottom of each panel gives respectively the deprojected distance $Z$, the mean extinction $A_{2.12\mu\text{m}}$, the excitation temperature $T_{ex}$ and the hot H$_2$ column density $N(\text{H}_2)$ derived from the fit parameters.}

\end{minipage}
\end{figure}

\clearpage
\newpage

\begin{figure}[h]
\begin{minipage}{0.99\textwidth}
\centering
\begin{minipage}{.5\hsize}
  \centering
  \includegraphics[width=\hsize]{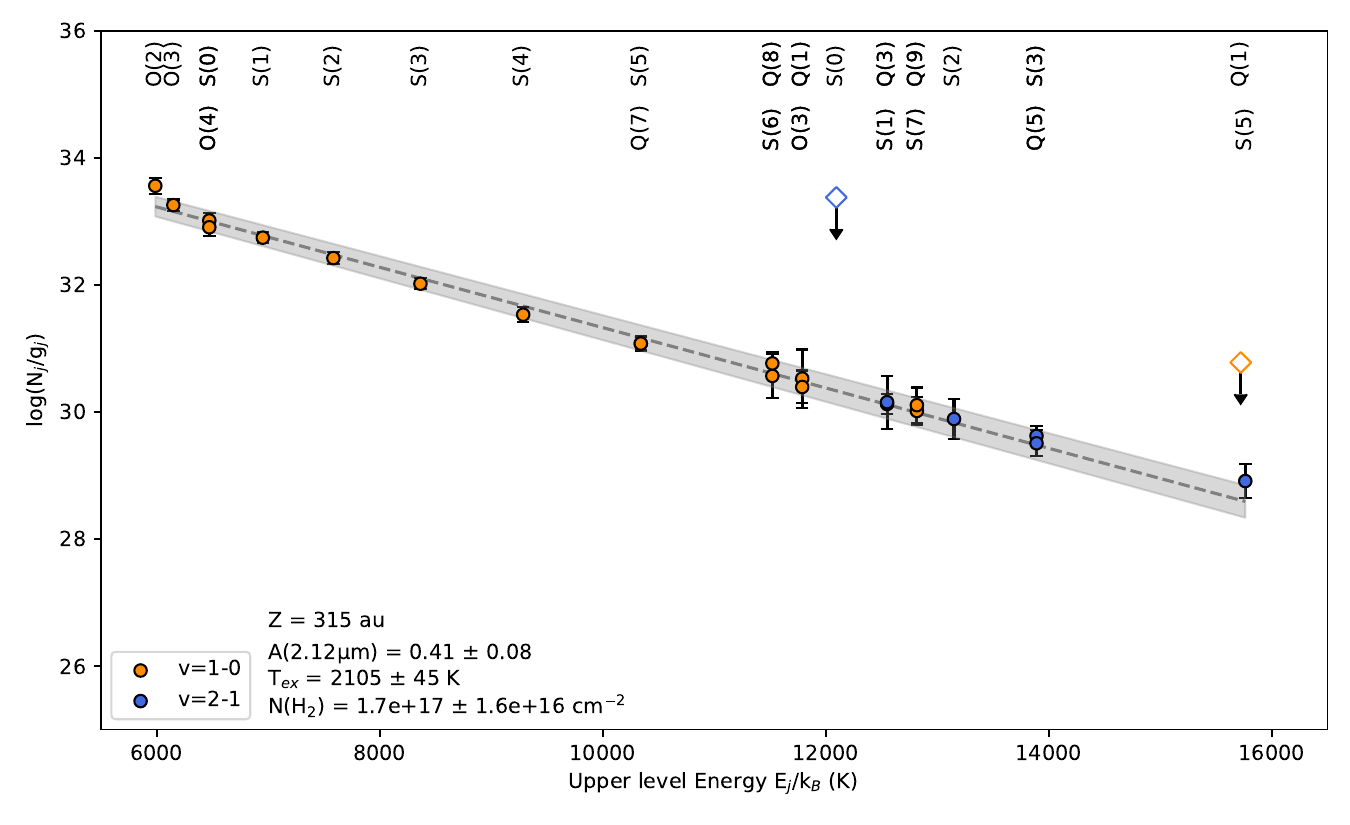}
\end{minipage}%
\begin{minipage}{.5\hsize}
  \centering
  \includegraphics[width=\hsize]{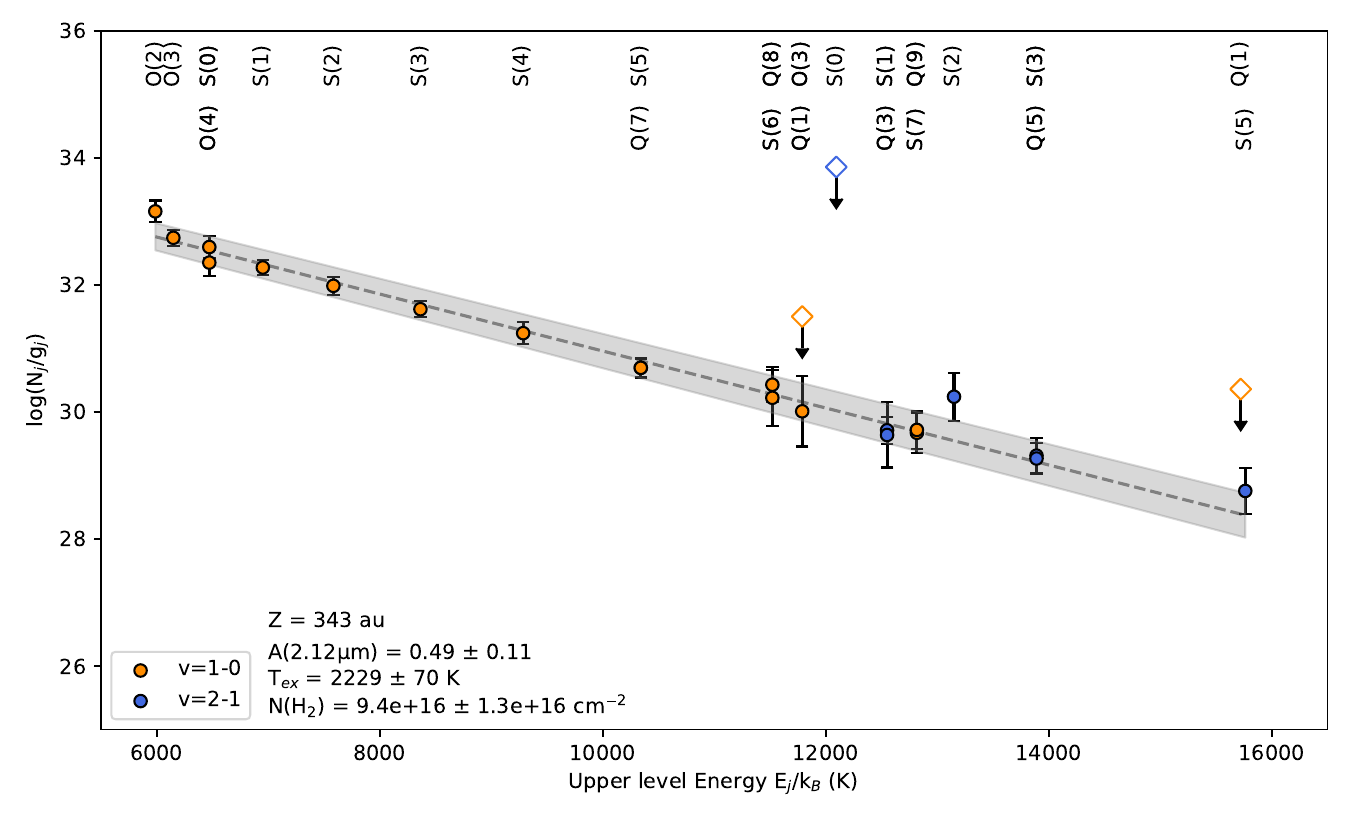}
\end{minipage}

\begin{minipage}{.5\hsize}
  \centering
  \includegraphics[width=\hsize]{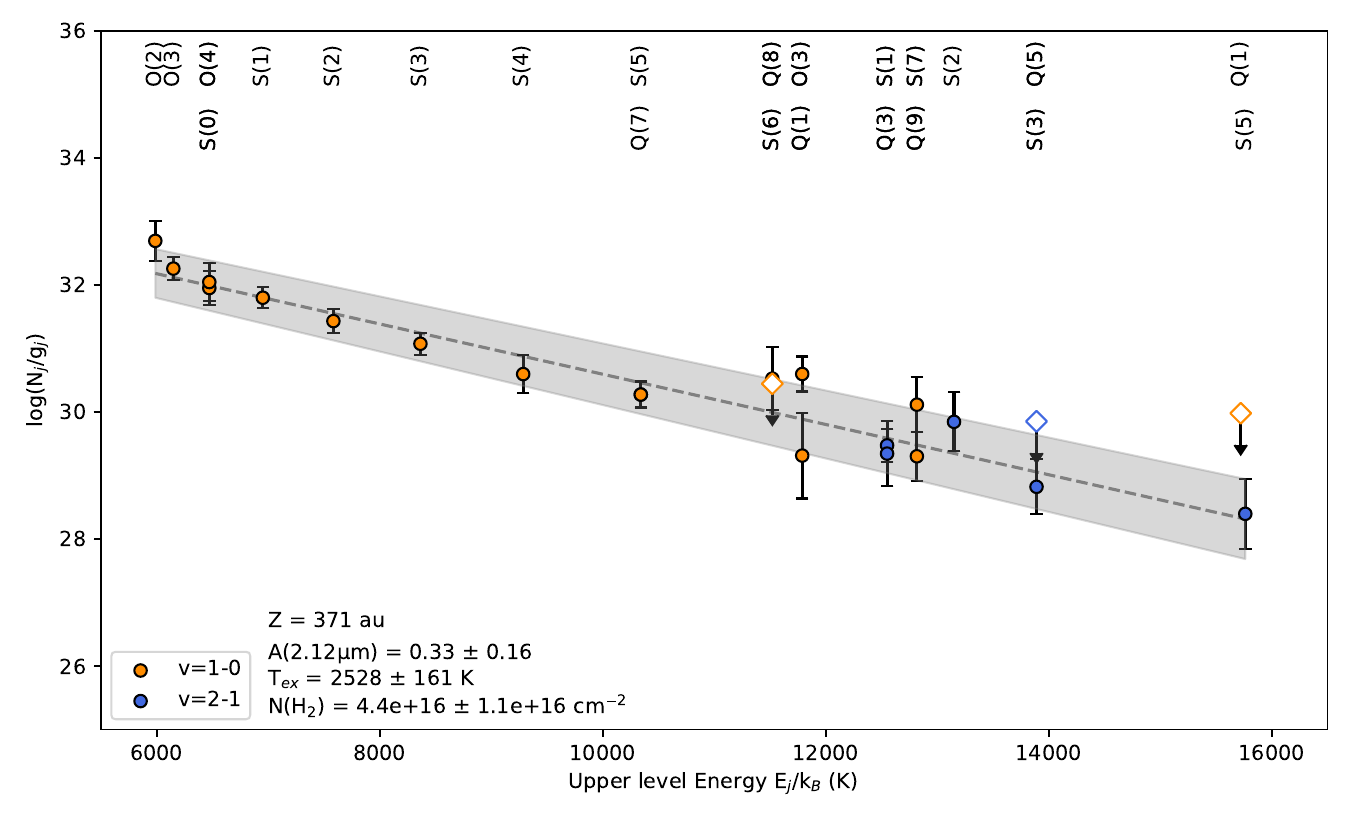}
\end{minipage}%
\begin{minipage}{.5\hsize}
  \centering
  \includegraphics[width=\hsize]{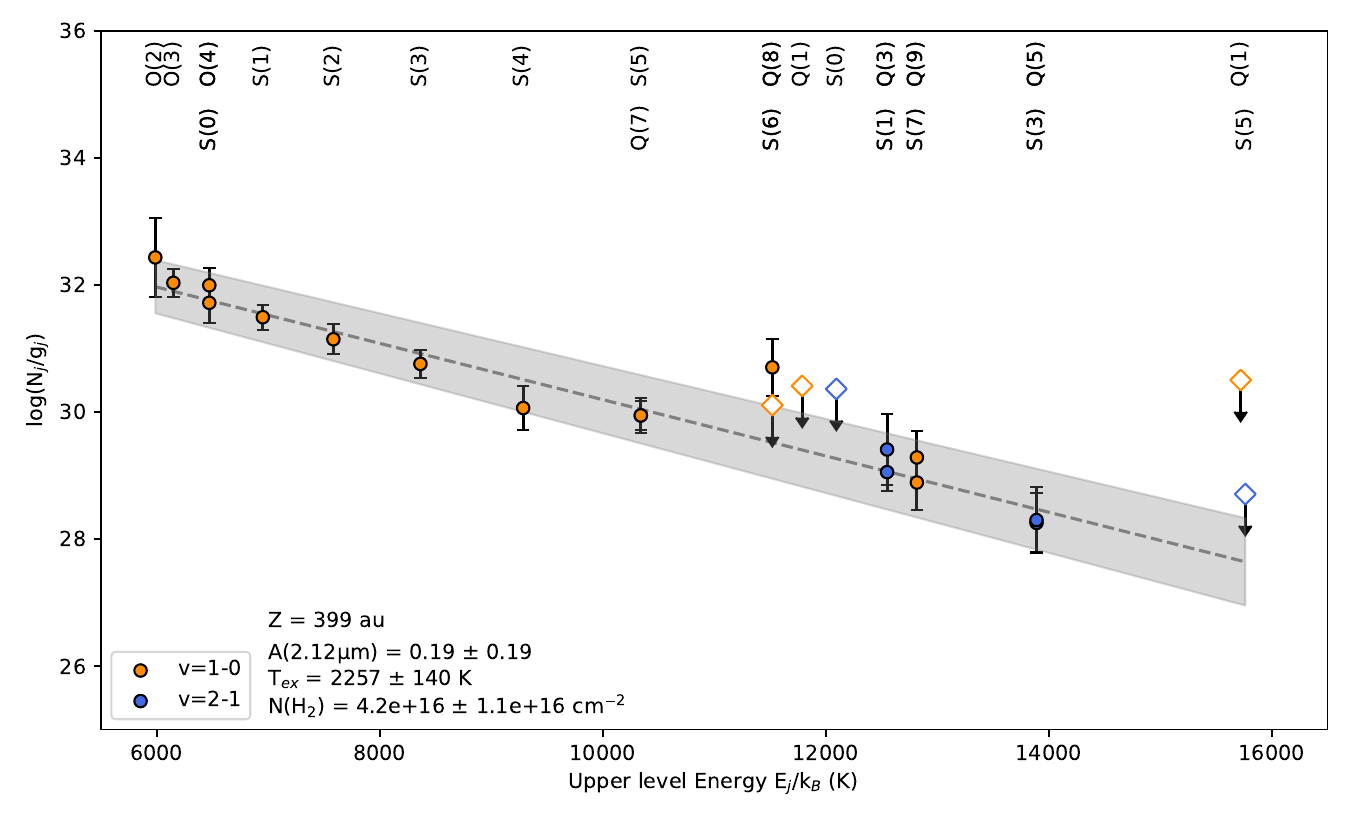}
\end{minipage}

\begin{minipage}{.5\hsize}
  \centering
  \includegraphics[width=\hsize]{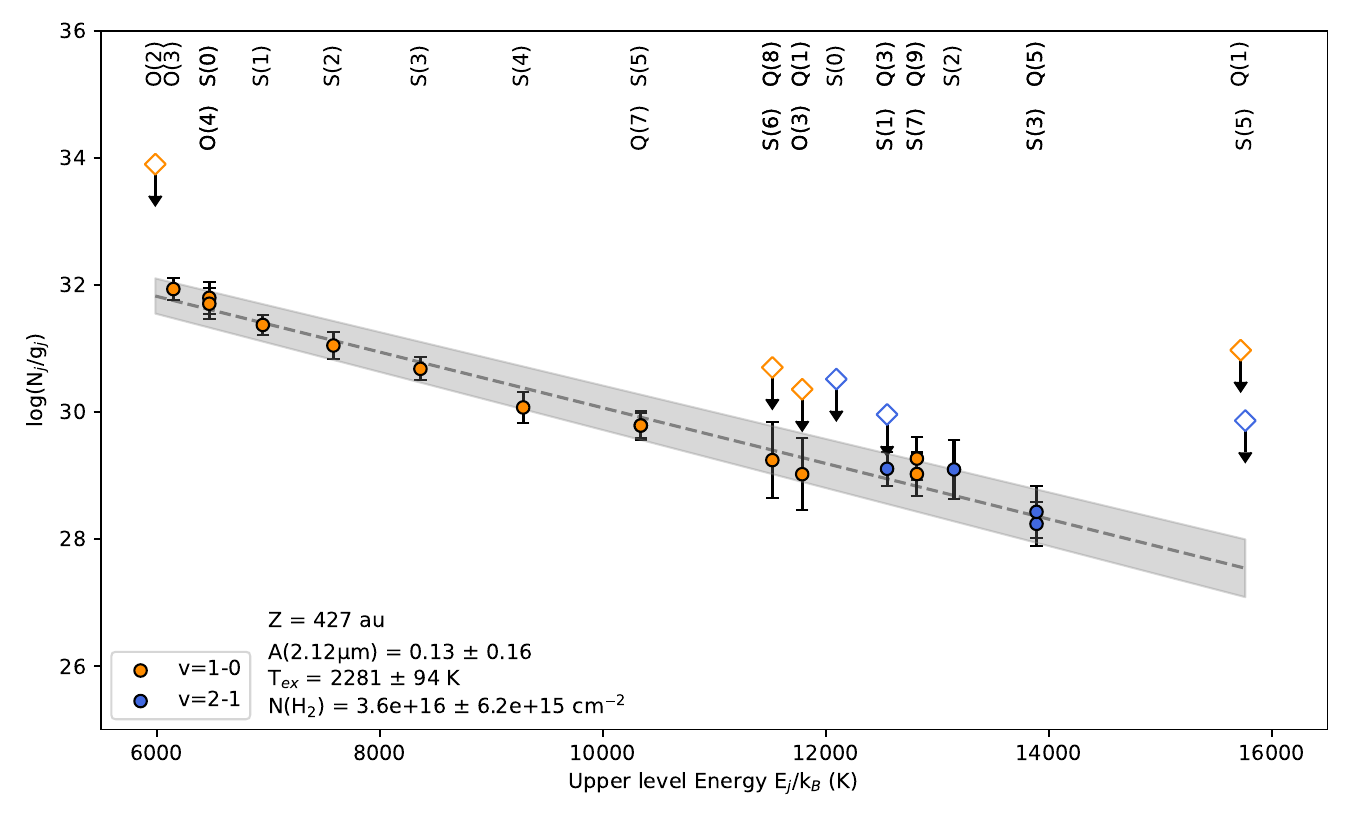}
\end{minipage}%
\begin{minipage}{.5\hsize}
  \centering
  \includegraphics[width=\hsize]{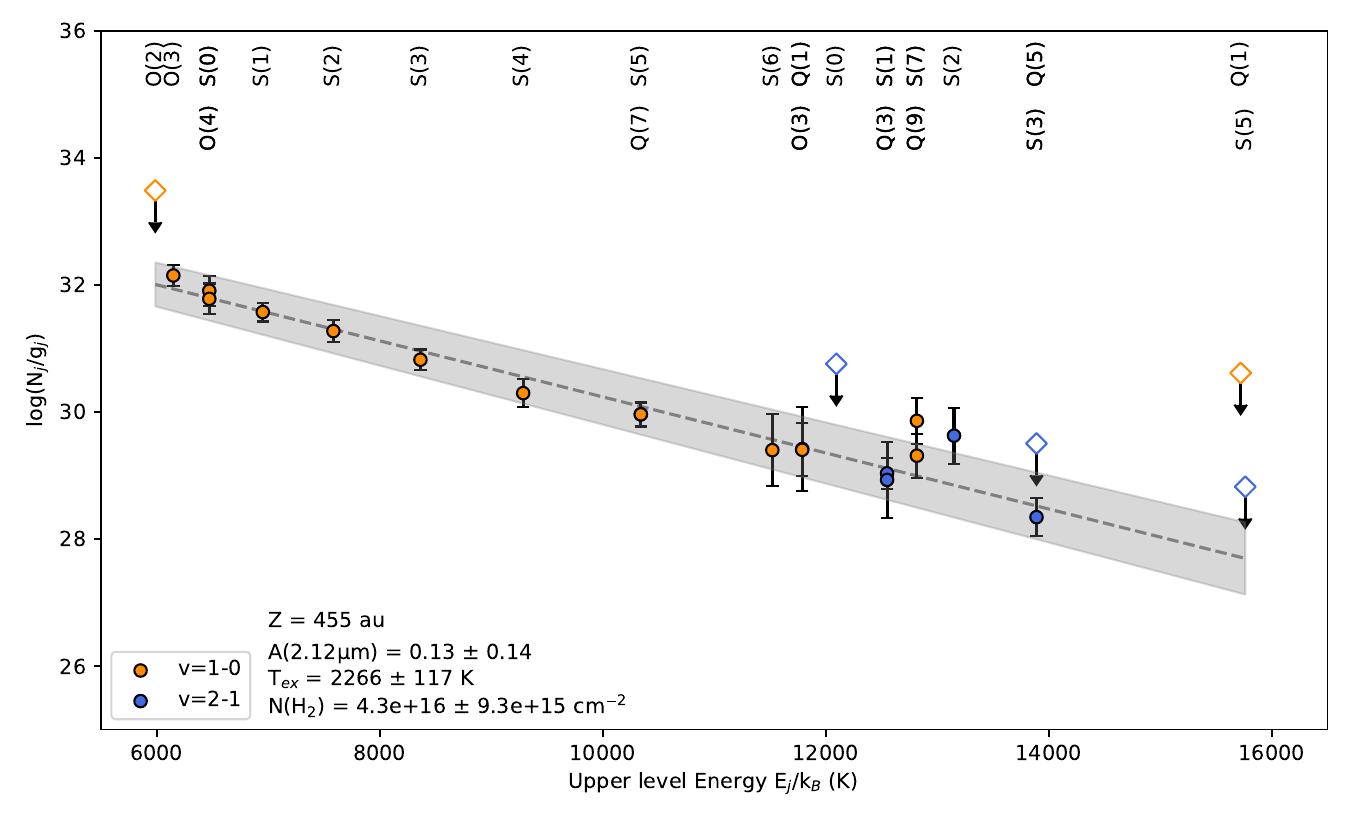}
\end{minipage}

\caption{continued.}
\end{minipage}
\end{figure}

\clearpage
\newpage

\begin{figure}[h]
\begin{minipage}{0.99\textwidth}
\centering

\begin{minipage}{.5\hsize}
  \centering
  \includegraphics[width=\hsize]{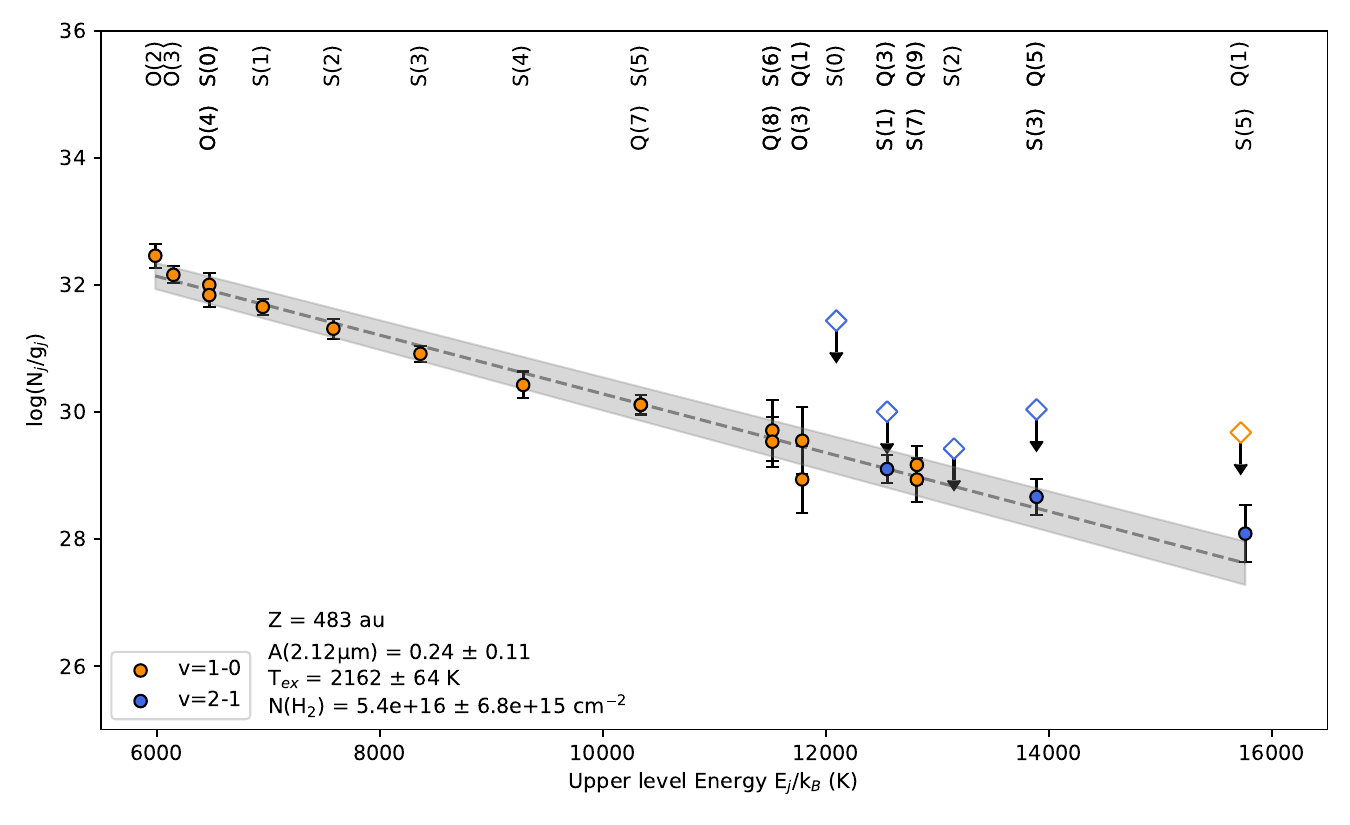}
\end{minipage}%
\begin{minipage}{.5\hsize}
  \centering
  \includegraphics[width=\hsize]{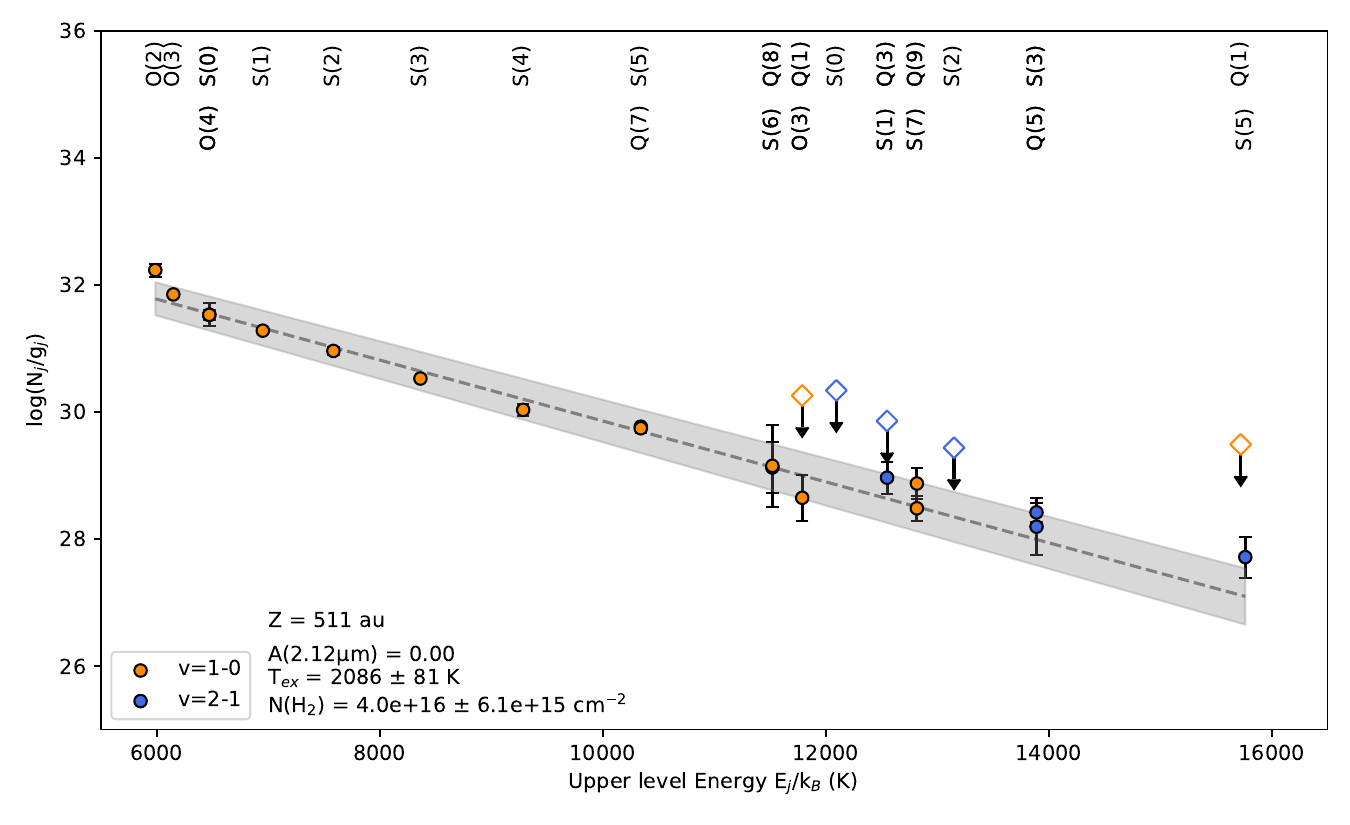}
\end{minipage}

\begin{minipage}{.5\hsize}
  \centering
  \includegraphics[width=\hsize]{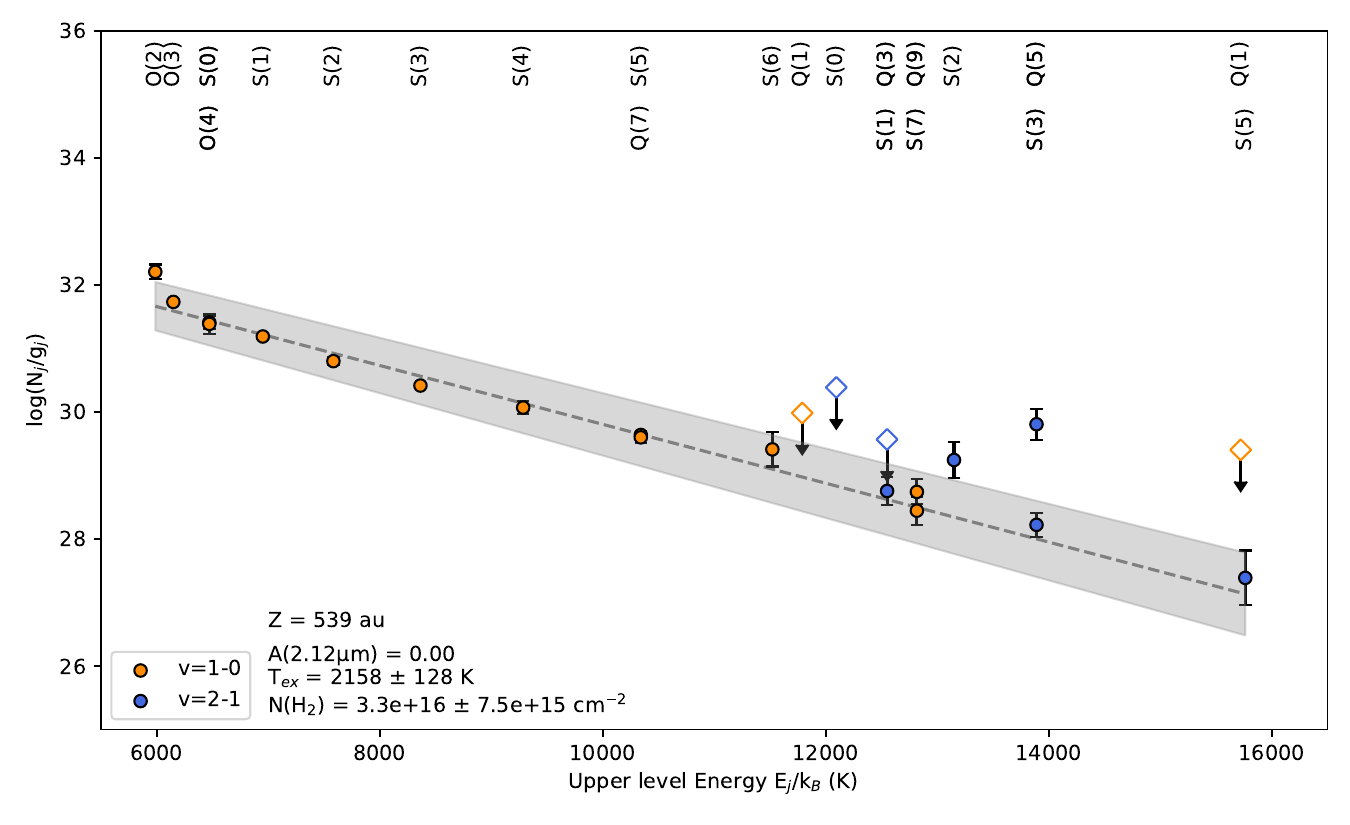}
\end{minipage}%
\begin{minipage}{.5\hsize}
  \centering
  \includegraphics[width=\hsize]{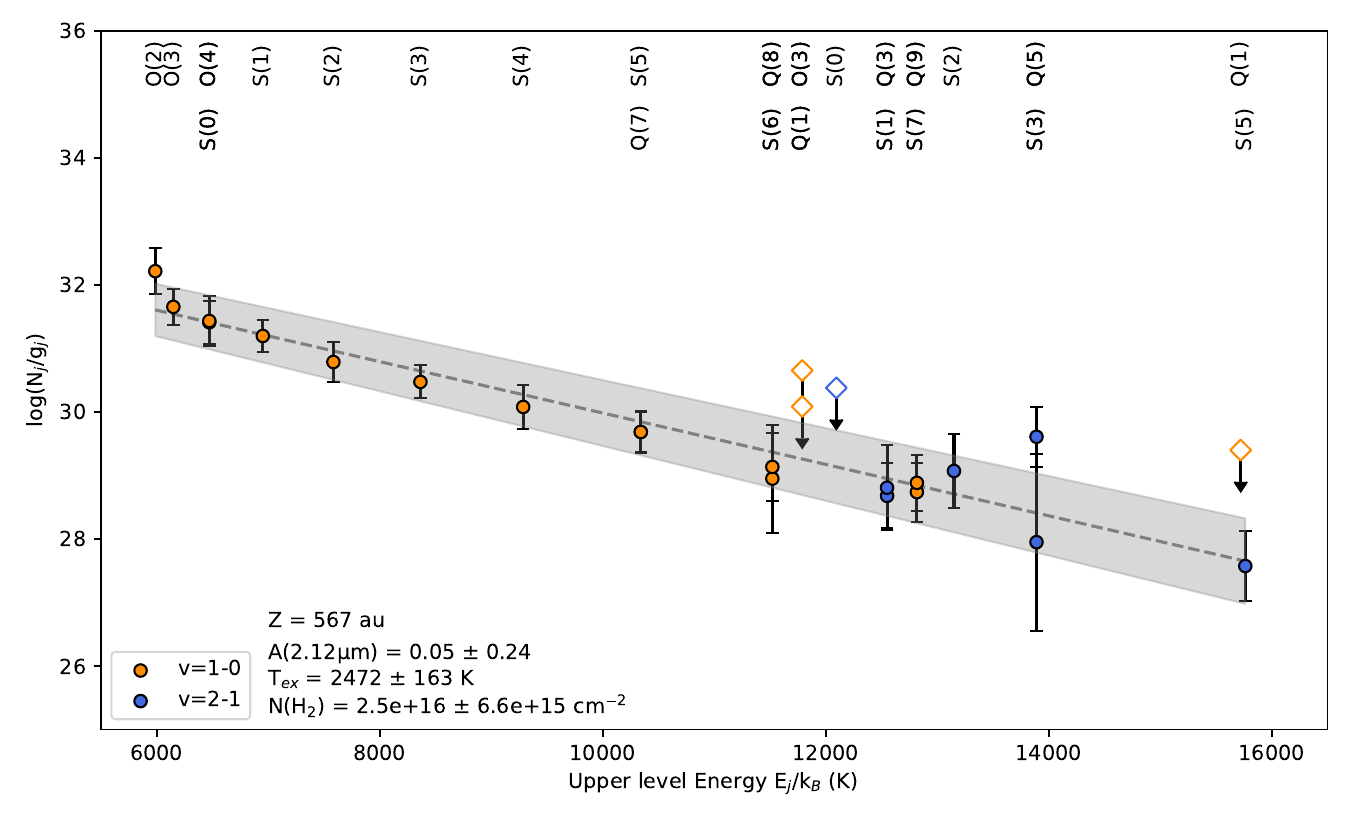}
\end{minipage}
\caption{continued.}
\end{minipage}
\end{figure}

\begin{table*}[]
    \centering
    \caption{Quantities along the H$_2$ red-shifted cavity of DG Tau B.}
    \begin{tabular}{c|cccccc}
        \hline
        \noalign{\smallskip}
         $Z$ (au) & N$_{sp}^a$ & F$^b$ ($\times 10^{-5}$ erg s$^{-1}$cm$^{-2}$sr$^{-1}$) & $A_{2.12\mu m}^c$ & $T_{ex}^d$ (K) & $N(\text{H}_2)^d$ ($\times 10^{17}$cm$^{-2}$) & $\dot{M}(\text{H}_2)^e$ ($\times 10^{-11}\text{M}_\odot yr^{-1}$) \\
         \noalign{\smallskip}
         \hline
         \noalign{\smallskip}
         147 & 17 & 3.42$\pm$0.07 & 0.84$\pm$0.18 & 1952$\pm$113 & 1.83$\pm$0.51 & 3.5$\pm$0.9\\
         175 & 20 & 4.49$\pm$0.09 & 0.64$\pm$0.19 & 1968$\pm$100 & 1.92$\pm$0.47 & 4.3$\pm$1.1\\
         203 & 22 & 4.44$\pm$0.09 & 0.58$\pm$0.20 & 1874$\pm$99  & 1.99$\pm$0.51 & 4.9$\pm$1.3\\
         231 & 22 & 5.30$\pm$0.12 & 0.68$\pm$0.12 & 1886$\pm$83  & 2.41$\pm$0.50 & 5.9$\pm$1.2\\
         259 & 23 & 5.41$\pm$0.09 & 0.52$\pm$0.11 & 1943$\pm$83  & 1.99$\pm$0.40 & 5.1$\pm$1.0\\
         287 & 27 & 6.18$\pm$0.06 & 0.44$\pm$0.10 & 2005$\pm$68  & 1.96$\pm$0.31 & 5.9$\pm$0.9\\
         315 & 28 & 6.20$\pm$0.08 & 0.41$\pm$0.08 & 2106$\pm$51  & 1.68$\pm$0.18 & 5.2$\pm$0.6\\
         343 & 28 & 3.61$\pm$0.08 & 0.49$\pm$0.11 & 2238$\pm$78  & 0.93$\pm$0.14 & 2.9$\pm$0.4\\
         371 & 29 & 2.59$\pm$0.06 & 0.33$\pm$0.16 & 2472$\pm$162 & 0.46$\pm$0.12 & 1.5$\pm$0.4\\
         399 & 30 & 2.16$\pm$0.05 & 0.19$\pm$0.19 & 2259$\pm$162 & 0.42$\pm$0.13 & 1.4$\pm$0.4\\
         427 & 31 & 2.02$\pm$0.04 & 0.13$\pm$0.16 & 2284$\pm$103 & 0.35$\pm$0.07 & 1.2$\pm$0.2\\
         455 & 32 & 2.47$\pm$0.05 & 0.13$\pm$0.14 & 2203$\pm$96  & 0.46$\pm$0.09 & 1.6$\pm$0.3\\
         483 & 33 & 2.43$\pm$0.06 & 0.24$\pm$0.11 & 2176$\pm$73  & 0.52$\pm$0.07 & 1.9$\pm$0.3\\
         511 & 34 & 2.08$\pm$0.05 & $\sim$0       & 2086$\pm$90  & 0.40$\pm$0.07 & 1.5$\pm$0.3\\
         539 & 34 & 1.90$\pm$0.05 & $\sim$0       & 2159$\pm$142 & 0.33$\pm$0.08 & 1.3$\pm$0.3\\
         567 & 34 & 1.84$\pm$0.07 & 0.05$\pm$0.24 & 2486$\pm$187 & 0.25$\pm$0.07 & 0.9$\pm$0.3\\
         \noalign{\smallskip}
         \hline
    \end{tabular}
    \label{tab:table_profiles}

\begin{minipage}{\hsize}
\vspace{0.1cm}
\vspace{0.1cm}
\small a. Number of spaxels per slit. All spaxels follow an SNR $>3$ criterion on the 1-0S(1) H$_2$ transition, allowing them to be selected. b. Mean surface brightness over the slit of the 1-0S(1) H$_2$ transition, not corrected from extinction. The error bars correspond to the full range of values in each slit. c. Average extinction at $2.12\mu$m inside the slits using the 1-0Q(7)/1-0S(5) line ratio. d. Excitation temperature $T_{ex}$ and column density $N(\text{H}_2)$ derived from excitation diagram fits. e. H$_2$ mass flux estimated using the $N(\text{H}_2)$ values and considering the vertical velocity $V_Z$ ($22.5$~\kms) derived in Sect.~\ref{sect:kinematics}.
\end{minipage}
    
\end{table*}

\end{appendix}

\end{document}